\documentclass[journal]{IEEEtran} 
\usepackage[table,xcdraw]{xcolor}
\usepackage{cite, color}
\usepackage{algorithm}
\usepackage{algorithmicx} 
\usepackage[noend]{algcompatible}
\usepackage{amsmath,bm}
\usepackage[pdftex]{graphicx}
\usepackage{multirow}
\usepackage[table,xcdraw]{xcolor}
\usepackage{xcolor}
\usepackage{array}
\usepackage{multirow}
\usepackage[justification=centering]{caption}
\usepackage{epstopdf}
\usepackage{float}
\abovecaptionskip 
\belowcaptionskip
\usepackage{comment, balance}
\usepackage[flushleft]{threeparttable}
\usepackage{booktabs} 
\usepackage{graphicx}
\usepackage{subcaption}
\usepackage{multirow}
\usepackage{titlesec}
\usepackage{url}
\usepackage{tikz}

\usepackage{breakurl}
\usepackage[breaklinks]{hyperref}
\usepackage{amssymb}
\usepackage{pifont}
\newcommand{\cmark}{\ding{51}}%
\newcommand{\xmark}{\ding{55}}%

\makeatletter
\renewcommand{\paragraph}
{\@startsection
 {paragraph}
    {4}
  {\parindent}                    
    {0ex plus 0.1ex minus 0.1ex}    
    {0ex}  
    {\normalfont\normalsize\itshape}}
\makeatother

\ifCLASSINFOpdf
\else
\fi
\usepackage{tabularx}
\newcolumntype{Y}{>{\centering\arraybackslash}X}
\newcolumntype{U}{>{\arraybackslash}X}
\usepackage{algorithm,algpseudocode}
\newcommand{\argmax}{\mathop{\mathrm{arg\,max}}}

\begin{document}
\title{Pervasive AI for IoT applications: A Survey on Resource-efficient Distributed Artificial Intelligence}
\author{Emna Baccour$^\ddagger$, Naram Mhaisen$^\star$, Alaa Awad Abdellatif$^\ast$,~\IEEEmembership{Member,~IEEE}, Aiman Erbad$^\ddagger$,~\IEEEmembership{Senior,~IEEE}, Amr Mohamed$^\ast$,~\IEEEmembership{Senior,~IEEE}, Mounir Hamdi$^\ddagger$,~\IEEEmembership{Fellow,~IEEE}, and Mohsen Guizani$^\diamond$,~\IEEEmembership{Fellow,~IEEE}. 
\thanks{$\ddagger$ E. Baccour, A. Erbad, and M. Hamdi are with the Division of Information and Computing Technology, College of Science and Engineering, Hamad Bin Khalifa University, Qatar Foundation.}
\thanks{$\ast$ A.A. Abdellatif and A. Mohamed are with the
Department of Computer Science and Engineering, College of Engineering, Qatar University, Qatar.}
\thanks{$\star$ N. Mhaisen is now with the Faculty of Electrical Engineering, Mathematics and Computer Science. Delft University of Technology, The Netherlands. This work was done while he was a research fellow at Qatar University.}
\thanks{$\diamond$ M. Guizani is with Mohamed Bin Zayed University of Artificial Intelligence (MBZUAI), Abu Dhabi, UAE.}
\vspace{-0.6cm}
}

\maketitle

\begin{abstract}
Artificial intelligence (AI) has witnessed a substantial breakthrough in a variety of Internet of Things (IoT) applications and services, spanning from recommendation systems and speech processing applications to robotics control and military surveillance. This is driven by the easier access to sensory data and the enormous scale of pervasive/ubiquitous devices that generate zettabytes of real-time data streams. Designing accurate models using such data streams, to 
revolutionize the decision-taking process, inaugurates pervasive computing as a worthy paradigm for a better quality-of-life (e.g., smart homes and self-driving cars.). The confluence of pervasive computing and artificial intelligence, namely Pervasive AI, expanded the role of ubiquitous IoT systems from mainly data collection to executing distributed computations with a promising alternative to centralized learning, presenting various challenges, including privacy 
and latency requirements. In this context, an intelligent resource scheduling should be envisaged among IoT devices (e.g., smartphones, smart vehicles) and infrastructure (e.g., edge nodes and base stations) to avoid communication and computation overheads and ensure maximum performance. In this paper, we conduct a comprehensive survey of the recent techniques and strategies developed to overcome these resource challenges in pervasive AI systems. Specifically, we first present an overview of the pervasive computing, its architecture, and its intersection with artificial intelligence. We then review the background, applications and performance metrics of AI, particularly Deep Learning (DL) and reinforcement learning, running in a ubiquitous system.  Next, we provide a deep literature review of communication-efficient techniques, from both algorithmic and system perspectives, of distributed training and inference across the combination of IoT devices, edge devices and cloud servers. Finally, we discuss our future vision and research challenges. 
\end{abstract}

\IEEEpeerreviewmaketitle

\section{Introduction}
\IEEEPARstart{D}{riven} by the recent development and prevalence of computing power,  
Internet of Things (IoT) systems, and big data, a booming era of AI has emerged, covering a wide spectrum of applications including natural language processing \cite{NaturalLanguage}, speech recognition \cite{speechRecognition}, computer vision \cite{ResNet}, and robotics \cite{robotics}. Owing to these breakthroughs, AI has achieved unprecedented improvements in multiple sectors of academia, industry, and daily services in order to improve the humans' productivity and lifestyle. As an example, multiple intelligent IoT applications have been designed such as self-driving cars, disease mapping services, smart home appliances, manufacturing robots, and surveillance systems. In this context, studies estimate that AI will have higher impact on the global Gross Domestic Product (GDP) by 2030, accounting for \$ 13 trillion additional gains compared to 2018 \cite{AIEconomic}. 
\begin{figure*}[!h]
\centering
	\frame{\includegraphics[scale=0.55]{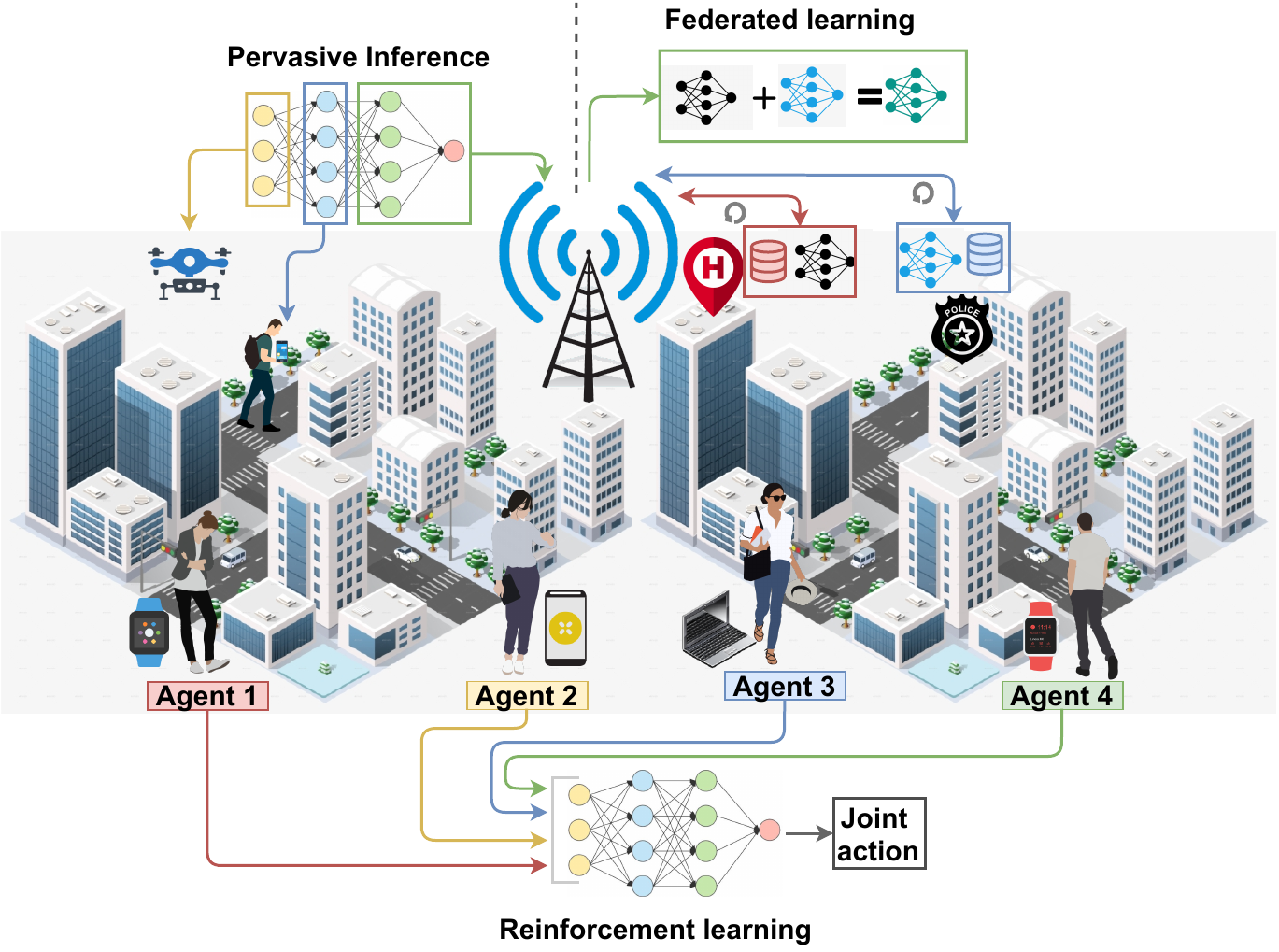}}
	\caption{A scenario illustrating examples of pervasive AI techniques.}
	\label{intro}
\end{figure*}

The popularity of AI is also related to the abundance of storage and computing devices, ranging from server clusters in the cloud to personal phones and computers, further, to  wearables and IoT units. In fact, the unprecedented amount of data generated by the massive number of ubiquitous devices opens up an attractive opportunity to provide intelligent IoT services that can transform all aspects of our modern life and fuel the continuous advancement of AI. Statistics forecast that, by 2025, the number of devices connected to the internet will reach more than 500 billion \cite{cisco} owing to the maturity of their sensing capabilities and affordable prices.  Furthermore, reports revealed that these devices will generate enormous data reaching more than 79 ZB by 2025 and will increase the economic gains up to 11 trillion by the same year \cite{Survey30}. 

With the rapid evolution of AI and the enormous bulks of data generated by pervasive devices,  conventional wisdom resorts to  centralized cloud servers for analytics. In fact, the high performance of AI systems applied to multiple fields comes at the expense of a huge memory requirement and an intensive computational load to perform both training and inference phases, which requires powerful servers.
However, this approach is no longer sustainable as it introduces several challenges: (1) the appearance of a new breed of services and the advent of delay-sensitive technologies spanning from self-driving cars to Virtual and Augmented Reality (VR/AR), make the cloud-approaches inadequate for AI tasks due to the long transmission delays. More precisely, the aforementioned applications are real-time and cannot allow any additional latency or connectivity loss. For example, autonomous cars sending camera frames to remote servers need to receive prompt inferences to detect potential obstacles and apply brakes \cite{autonomousCars,autonomousCars2}. 
Sending data to cloud servers may not satisfy the latency requirements of the real-time applications. Particularly, experiments in \cite{Amazon} demonstrated that executing a computer vision task on a camera frame  offloaded to an Amazon server takes more than 200 ms. (2) In addition to latency, privacy presents a major concern for cloud-based AI approaches. In fact, end-users are typically reluctant to upload their private data to cloud servers (e.g., photos or audios), as they can be highly exposed to cyber risks, malicious attacks, or disclosures. Among the most popular breaches reported in the 21st century, we can cite the Marriott attack revealed in 2018 and affecting 500 million customers and Equifax breach recorded in 2017 and affecting 147 million users \cite{breach}. (3) A tremendous number of AI tasks, involving unstructured and bandwidth-intensive data, needs to be transferred across the Wide Area Network (WAN), which poses huge pressure on the network infrastructure having varying quality. (4) In the same context, offloading the data to remote servers encounters also scalability issues, as the access to the cloud can become a bottleneck when the number of data sources increases, particularly if some devices import irrelevant and noisy inputs. (5) Nowadays, Explainable AI (XAI) \cite{XAI1} has become extremely popular, aiming to enhance the transparency of learning and detect prediction errors. However, consigning AI tasks to the cloud makes the whole process a black-box vis-a-vis the end-user, and prevents model decomposability and debugging. 

Pushing AI to the network edge has been introduced as a viable solution to face latency, privacy, and scalability challenges described earlier. As such, the large amount of computational tasks can be handled by edge devices without exchanging the related data with the remote servers, which guarantees agile IoT services owing to the physical proximity of computing devices to the data sources \cite{EdgeComputing2019}. In the case when the AI tasks can only be executed at the cloud datacenters, the edge devices  can be used to pre-process the data and polish it from noisy inputs in order to reduce the transmission load \cite{elsevierbook}. Furthermore, the edge network can play the role of a firewall that enhances the privacy by discarding sensitive information prior to data transfer. A variety of edge devices can be candidate for executing different AI tasks with different computation requirements, ranging from edge servers provisioned with GPUs, to smart-phones with strong processors and even small IoT wearable with Raspberry Pi computing. These edge devices have been continuously improving to fit for deep AI models.
In spite of this technological advancement, a large range of pervasive devices used in countless fields of our daily life still suffers from limited power and memory, such as smart home IoT appliances, sensors, and gaming gears.

Given the limited resources of edge-devices, computing the full AI model in one device may be infeasible, particularly when the task requires high computational load, e.g., Deep Neural Networks (DNN). A promising solution is to opt for pervasive computing, where different data storage and processing capacities existing everywhere (e.g., distributed cloud datacenters, edge servers, and IoT devices.) cooperate to accomplish AI tasks that need large memory and intensive computation. This marriage of pervasive computing and AI has given rise to a new research area, which garnered considerable attention from both academia and industry. The new research area comprises different concepts (e.g., federated learning, active learning, etc.) that suggest to distribute AI tasks into pervasive devices for multiple objectives. In this paper, we propose to gather all existing concepts having different terminologies under the same umbrella that we named  \textit{“Pervasive AI"}. Indeed, we define the pervasive AI as \textit{“The intelligent and efficient distribution of AI tasks and models over/amongst any types of devices with heterogeneous capabilities in order to execute sophisticated global missions”}.  
The pervasive AI concepts are firstly introduced to solve the described challenges of centralized approaches (e.g., on-cloud or on-device computation):
(1) To preserve privacy and reduce the huge overhead of data collection and the complexity of training an astronomical dataset, \textit{Federated Learning (FL)} is proposed, where raw data are stored in their source entities and the model is trained collaboratively. Particularly, each entity computes a local model using its collected data, then sends the results to a fusion server to aggregate the global model. Such an approach suggests the distribution of data and the assembly of the trained AI models. (2) To cope with the limited resources provided by edge devices and simultaneously avoid latency overheads caused by cloud transmissions, the inference task is distributed among ubiquitous devices located at the proximity of the source. The basic idea is to divide the trained model into segments and subsequently, each segment is assigned to a participant. Each participant shares the output to the next one until generating the final prediction.  In other words, the \textit{Pervasive Inference} covers the distribution of the established model resulting from the training phase. (3) Some AI techniques are inherently distributed such as Multi-Agent Reinforcement Learning (MARL) or Multi-agent Bandits (MAB), where agents cooperate to build and improve a policy in real-time enabling them to take on-the-fly decisions/actions based on the environment status. In this case, the distribution covers the online creation and update of the Reinforcement Learning (RL) policy. A scenario illustrating some pervasive AI techniques is presented in Fig. \ref{intro}.

The pervasive AI exploits the on-device computation capacities to collaboratively achieve learning tasks. This requires 
a careful scheduling to wisely use the available resources without resorting to remote computing. Yet, some intensive AI tasks can only be performed by involving the cloud servers, which results in higher communication costs. Therefore, leveraging the small and ubiquitous resources and managing the enormous communication overheads present a major bottleneck for the pervasive AI.
\subsection{Our scope}
In this survey, we focus on the confluence of the two emerging paradigms: pervasive computing and artificial intelligence, which we name \textit{Pervasive AI}. The pervasive AI is a promising research field, in which the system design is highly correlated to the resource constraints of the ubiquitous participants (e.g., memory, computation, bandwidth, and energy.) and the communication overheads between them.  More specifically, the size of some deep AI models, their computational requirements and their energy consumption may exceed the available memory (e.g., RAM) or the power supply capacity of some devices, which restricts them from participating in the collaborative system. Furthermore, the process of decentralized training or inference may involve a big number of participants that potentially communicate over wireless links, which creates new challenges related to channels capacities and conditions, the delay performance, and the privacy aspect. Therefore, the pervasive AI should rely on various parameters, including the optimal AI partitioning, the efficient design of architectures and algorithms managing the distributed learning, and the smart selection and scheduling of pervasive participants supported by efficient communication protocols.
Not only that, all the on-device constraints should be taken into consideration such as the memory, the computation, the energy, not to mention the privacy requirements of the system. Finally, the load of real-time inferences (e.g., an area that needs 24/7 surveillance), the pace of data collection (e.g., weather monitoring) and the dynamics of the studied environment should also be considered as they highly impact the number of selected participants and the parallelization strategies. In this paper, we survey the aforementioned challenges in deploying pervasive AI models and algorithms. Particularly, we provide a deep study of resource-efficient distributed learning for the training phase, the inference tasks, and real-time training and decision process. We start by identifying the motives behind establishing a pervasive AI system for IoT applications and the corresponding communication and resource challenges.
\subsection{Contributions and structure of the paper}
The contributions of this paper are summarized as follows:
\begin{itemize}
    \item We present an overview of the pervasive computing and introduce its architecture and potential participants. 
    \item We provide a brief background of artificial intelligence, particularly deep learning and reinforcement learning. We, also, describe the frameworks that support AI tasks and the metrics that assess their performance. Furthermore, we present multiple IoT applications, in which pervasive AI can be useful.
    \item For each phase of the AI (i.e., training and inference), we profile the communication and computation models and review the state-of-the-art. A comparison between different existing works, lessons learned, in addition to recent use cases, are also provided. 
    \item We conclude by an elaborative discussion of our future vision and we identify some open challenges that may arouse new promising research ideas.
\end{itemize}
\begin{figure*}[h]
\centering
\hspace{-9 mm}
	\includegraphics[scale=0.5]{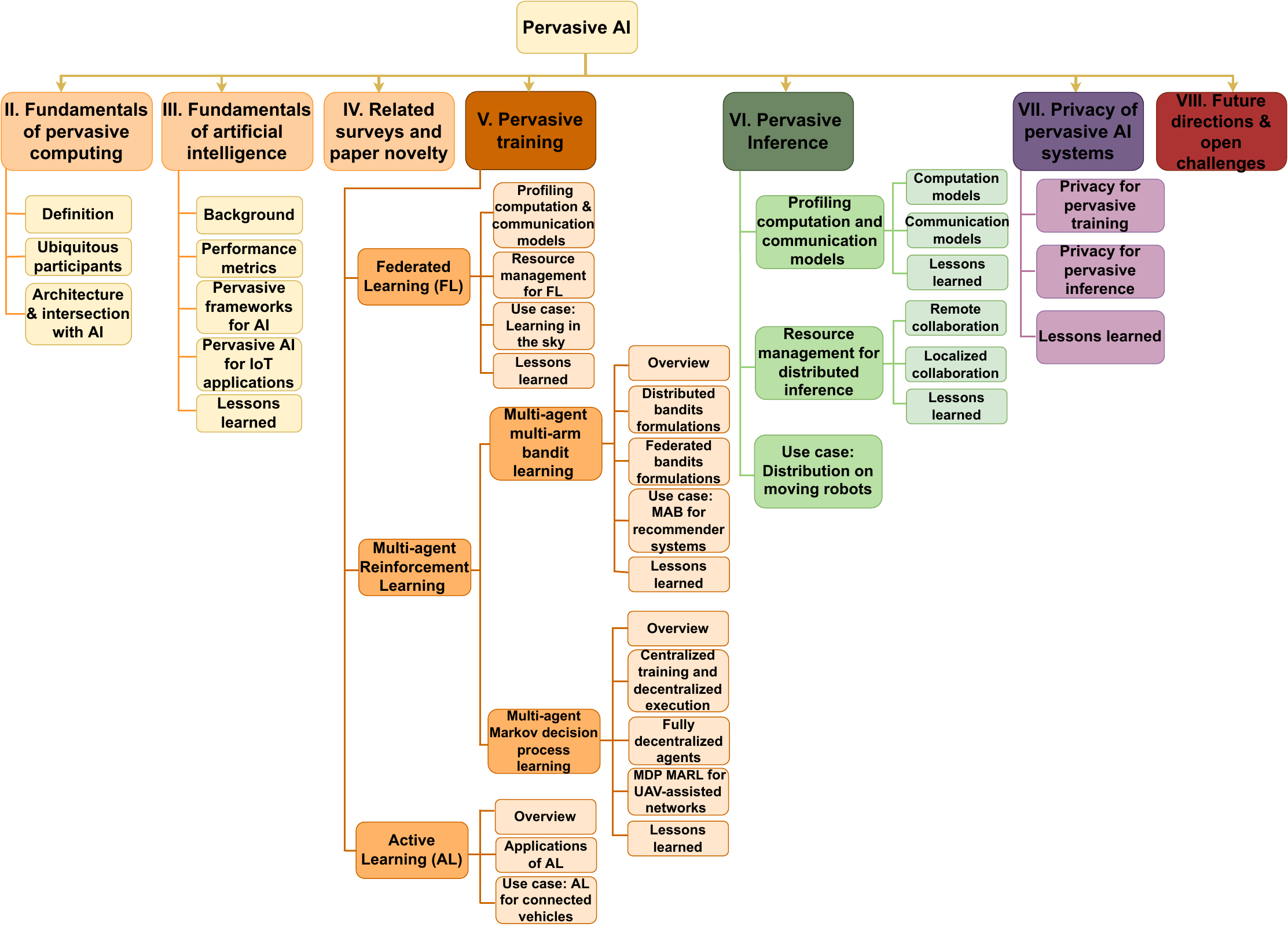}
	\caption{Pervasive AI survey roadmap.}
	\label{journalSkeleton}
\end{figure*}

The rest of this paper is organized as follows:  Sections \ref{pervasive_systems} and \ref{AI} present the fundamentals of pervasive computing and artificial intelligence, respectively. In section \ref{related_surveys}, we introduce the related surveys and we highlight the novelty of our paper. Section \ref{Pervasive_training} presents the related studies that investigated the potential of \emph{federated learning} schemes in different domains. Moreover, it highlights the use of FL within UAV swarms for cooperative target recognition as a case study.   
Then, we investigate diverse \emph{reinforcement learning} schemes, and \emph{active learning}. 
Specifically, we focus on the state-of-art algorithms that study the trade-off between the utilized communication resources and the performance, which allows us next to evaluate the strengths and weaknesses of the discussed approaches. In section \ref{PI}, we present a deep study of the \emph{pervasive inference}. Particularly, we review the state-of-the-art approaches adopting different splitting strategies and managing the existing pervasive resources to distribute the inference. Next, we compare the performance of these works and discuss the learned lessons and potential use cases. 
In section \ref{privacy_AI}, we discuss the privacy and security issues associated with pervasive AI and the corresponding mitigation strategies proposed in the literature.
We discuss the future vision and open challenges, in section \ref{future}. Finally, we conclude in section \ref{conclusion}. More details about the road map of the paper are illustrated in Fig. \ref{journalSkeleton}. 
\section{Fundamentals of pervasive computing}\label{pervasive_systems}
\subsection{Definition}
The pervasive computing \cite{pervasive,pervasive2}, named also ubiquitous computing, is the growing trend to embed computational capabilities in all devices in order to enable them to communicate efficiently and accomplish any computing task, while minimizing their resource consumptions e.g. battery, memory, cpu time, etc. The pervasive computing can occur in any device, at any format, in any place and any time. More specifically, it can span from resource-constrained devices to highly performant servers and can involve cloud datacenters, mobile edge computing servers, mobile devices, wearable computers, embedded systems, laptops, tablets, pair of intelligent glasses, and even a refrigerator or a TV. These ubiquitous devices are constantly connected and available for any task. In other words, we are not talking anymore about devices acting on a passive data. Instead, the pervasive systems are able to collect, process, communicate any data type or size, understand its surroundings, adapt to the input context, and enhance humans’ experiences and lifestyles.

\subsection{Ubiquitous participants}
The pervasive systems are characterized by highly heterogeneous devices (see Fig. \ref{participants}), where the critical challenge is to design a scalable infrastructure able to dynamically discover different components, manage their interconnection and interaction, interpret their context, and adapt rapidly to the deployment of new software and user interfaces.  A pervasive system can be composed of:
\begin{figure}[!h]
\centering
	\includegraphics[scale=0.5]{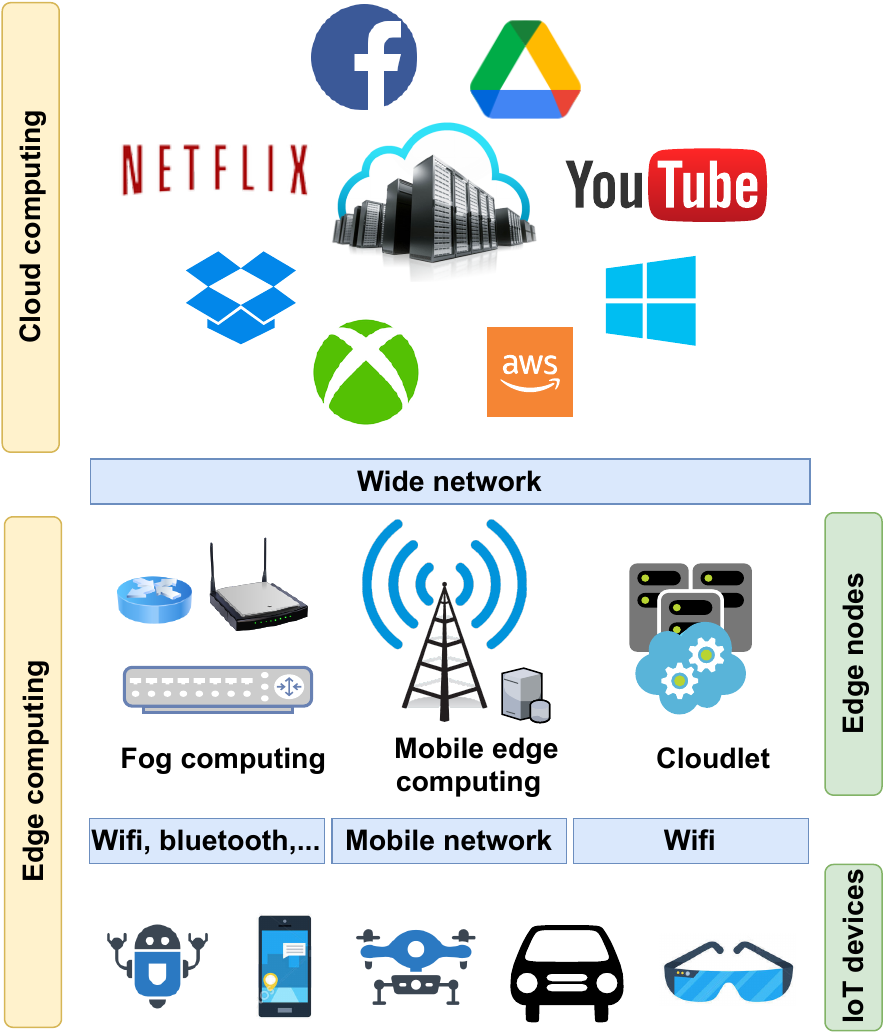}
	\caption{Ubiquitous participants.}
	\label{participants}
\end{figure}
\subsubsection{Data center and cloud servers}
Cloud computing \cite{vital,QoE,RLopra,ptnet,GlobecomFatima} is defined as delivering on-demand services from storage, management, advertising, and computation to artificial intelligence and natural language processing, following different pricing models, such as pay-as-you-go  and subscription-based billing.
Hence, instead of owning computing servers, companies, operators, and end-users can exploit the high-performance facilities offered by the cloud service provider. In this way, they can benefit from better computational capacities, while reducing the cost of owning and maintaining a computation infrastructure, and paying only for their requested services. Cloud computing underpins a broad number of services, including data storage, cloud back-up of photos, video streaming services, and online gaming. 
\subsubsection{Mobile Edge Computing (MEC) servers}
Edge computing is introduced as a solution to bring cloud facilities at the vicinity of users in order to minimize the services perceived latency, relieve the data transmission, and ease the cloud congestion. In another word, the edge computing has become an essential complement to the cloud and even a substitute in some scenarios. Services and computing capabilities equipped at the edge of cellular networks are called Mobile Edge Computing (MEC) facilities \cite{CE-D2D,MEC,CE-D2D2}. Deploying MEC servers within the edge Base Stations (BSs) allows providing location and context awareness, deploying new services quickly and flexibly, and enhancing the Quality of Service (QoS).
\subsubsection{Cloudlet devices}
Cloudlets \cite{cloudlet} are the network components that connect cloud computing to mobile computing  (e.g., computers cluster). This network part presents the middle layer of the three-tier hierarchical architecture composed of mobile devices, micro-clouds, and cloud data centers. The role of cloudlets is to define the algorithms and implement the functionalities that support low latency edge-cloud tasks offloading.
\subsubsection{Fog devices}
The fog \cite{fog} and cloud computing share the same set of services provided to end-users, such as storage, networking, computing, and artificial intelligence. However, the cloud architecture is composed of  fully distributed large-scale data centers. Meanwhile, fog services focus on IoT devices in a specific geographical area and target applications requiring real-time response such as live streaming, interactive applications, and online collective gaming.  Examples include phones, wearable health monitoring devices, connected vehicles, etc.
\subsubsection{Edge devices}
In most of the studies, the interpretation of edge devices (i.e., edge nodes and IoT devices) is still ambiguous \cite{BILAL201894}, which means the difference between end or IoT devices and edge nodes is still unclear. Yet, common consensus defines the end-devices/IoT as ubiquitous gadgets that are embedded with processing capacities, sensors, and software, serving to connect and exchange data with other systems over different communication networks.
Meanwhile, the edge nodes are defined as devices in higher levels including fog nodes, MEC servers, and cloudlets. The edge nodes are expected to possess high storage and computation capacities and to offer high-quality networking and processing service with a lower response time compared to the cloud remote servers.  

Driven by the expansion and pervasiveness of the computing devices, we believe that the heterogeneity of ubiquitous systems will increase in the future. These devices have to interact seamlessly and coherently, despite their difference in terms of software and hardware capacities. 
\subsection{Architecture and intersection with AI}
\subsubsection{Architecture}
\begin{figure}[H]
\centering
	\includegraphics[scale=0.645]{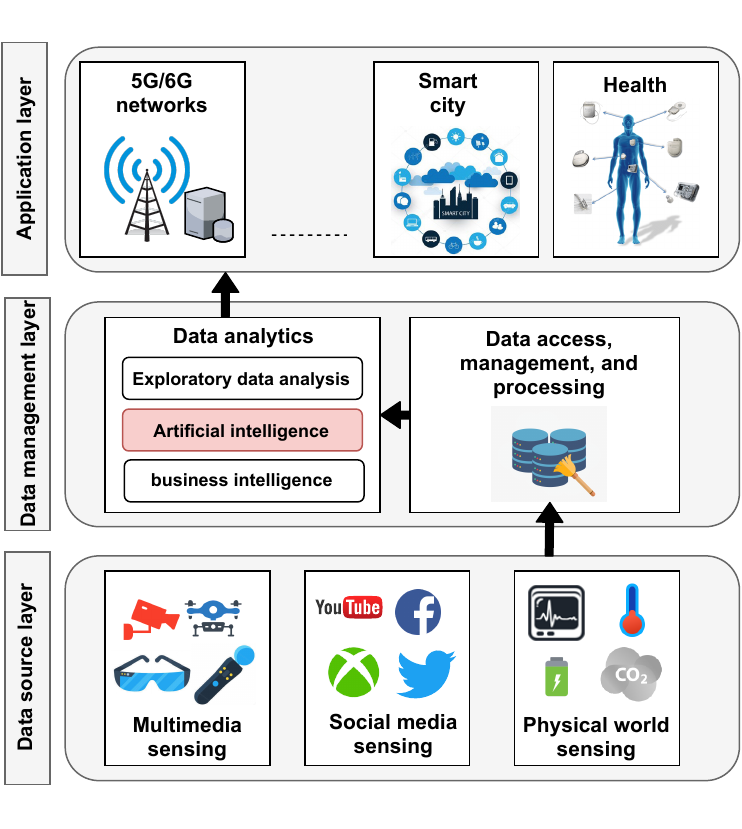}
	\caption{Pervasive architecture.}
	\label{Architecture}
\end{figure}
Fig. \ref{Architecture} illustrates the hierarchical architecture of a pervasive system \cite{pervasive2}, which is composed of three layers:
\begin{itemize}
    \item Data source layer: the data is collected from different monitored sources generating information of physical world or human activities, multimedia data such as images and audio, and social media information.
    \item Data management layer: this layer involves the storage and integration of heterogeneous data incoming from pervasive sources, the cleaning and pre-processing that tailor the context of the system, and the data analytics that convert the raw information into useful and personalized insights using multiple approaches, such as business and artificial intelligence.
    \item Application layer: to this end, the insights generated from the previous layer are used to offer multiple intelligent applications, such as health advisor and smart home applications. 
\end{itemize}

In our paper, we focus only on the data management layer, specifically the data analytics using artificial intelligence. The data source layer is thoroughly discussed in \cite{Survey31}, whereas the application layer can be found in \cite{Survey30}.
\subsection{Intersection with AI}
\begin{figure}[H]
\centering
	\includegraphics[scale=0.6]{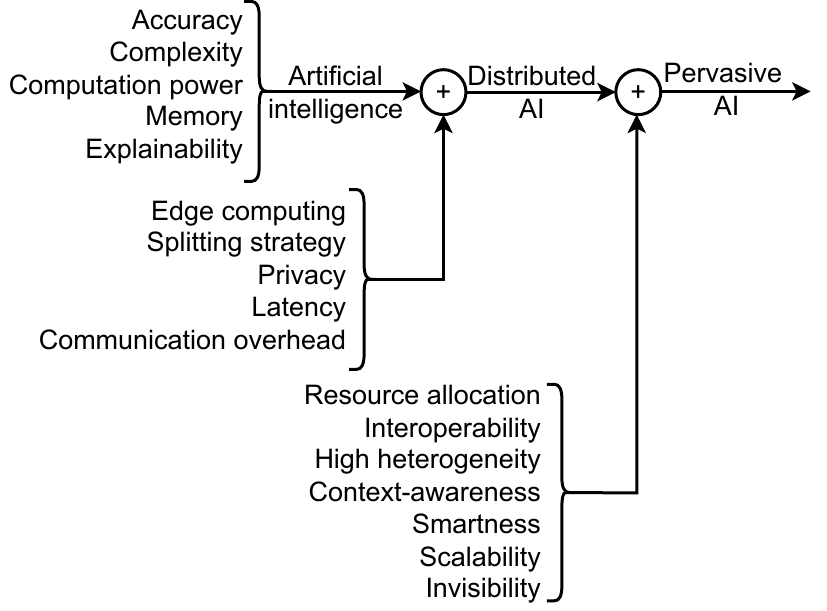}
	\caption{AI forms and considerations.}
	\label{distVSPervasiveV4}
\end{figure}
 The artificial intelligence techniques are centralized by design and most of the challenges revolve around the accuracy, complexity, computation power, memory, and explainability. With the evolution and migration to edge computing characterized by scarce resources, solving these issues as well as facing the new challenges related to privacy, become crucial. This called for AI distribution, where the training and the inference (e.g., data, models, policies.) are split into smaller parts in order to reduce local computation and memory overheads, while considering the privacy and latency constraints. However, the nascent IoT applications deployed in large-scale IoT devices have driven the distributed computation towards further dispersion, which urged the support of interoperability, high heterogeneity, scalability, context-awareness, smart resource allocation, coordination, and invisibility, which are the characteristics of pervasive computing. To this end, the intersection between AI and pervasive computing came to the light paving the way to introduce \textit{"Pervasive AI"}. As shown in Fig. \ref{distVSPervasiveV4} , Pervasive AI is a special class of distributed AI, where the decentralization of AI models is managed using intelligent techniques that take into consideration the IoT resource constraints, their heterogeneity, the application context, etc. 

\section{Fundamentals of Artificial Intelligence}\label{AI}
Since approaches and techniques reviewed in this survey rely on artificial intelligence and deep neural networks, we start first by providing a brief background of deep learning. A deeper and detailed review of AI can be found in the reference book in \cite{DeepLearning}.
\subsection{Background}
\begin{figure}[!h]
\centering
	\includegraphics[scale=0.56]{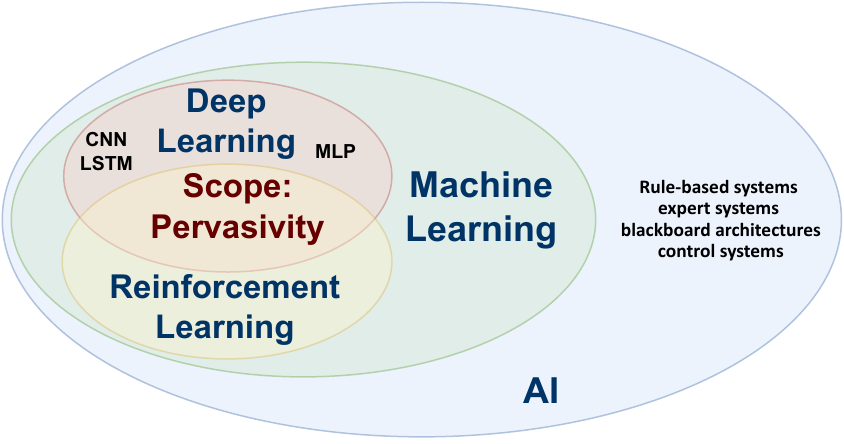}
	\caption{Relation between AI,
machine learning, deep learning, and reinforcement learning. This survey mainly focuses on
pervasive deep and reinforcement learning.}
	\label{DL}
\end{figure}

Even though AI has recently gained enormous attention, it is not a new term and it was initially coined in 1956. 
Multiple techniques and procedures fall under AI broad umbrella, such as rule-based systems, expert systems, control systems, and  well-known machine learning algorithms. Machine learning generally includes three categories, which are supervised, unsupervised and reinforcement learning. An important branch of machine learning is deep learning that can be supervised or unsupervised and it is based on simulating the biological nervous system and performing the learning through subsequent layers transformation. As most of the pervasive applications are led by deep learning techniques and recently reinforcement learning, the crossover between the above-mentioned domains (shown in Fig. \ref{DL}) defines the scope of this paper.
\begin{figure*}[!h]
\centering
	\includegraphics[scale=0.7]{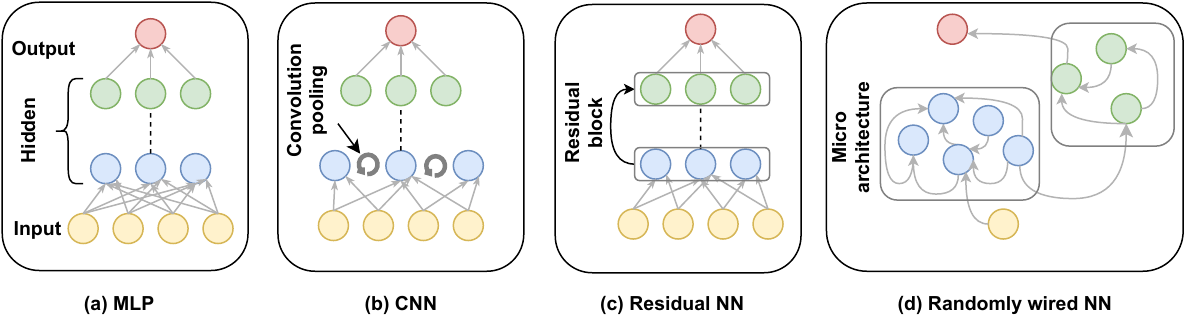}
	\caption{ Examples of NN structures: (a) Multilayer Perceptron (MLP), (b) Convolutional Neural Network (CNN), (c) Residual Neural Network, (d) Randomly Wired Neural Network.}
	\label{DNN_types}
\end{figure*}
\subsubsection{Deep learning and Deep Neural Networks}
In the following, we briefly present an overview of the most common deep learning networks. 

Neural networks consist of a first input layer, one or multiple hidden layers, and a last output layer, as shown in Fig. \ref{DNN_types}. When the neural network contains a high number of sequential layers, it can be called Deep Neural Network (DNN). The DNN layers include smaller units, namely neurons.
Most commonly, the output of one layer is the input of the next layer and the output of the final layer is either a classification or a feature. The correctness of the prediction is assessed by the loss function that calculates the error between the true and predicted values. 

\begin{figure}[h]
\centering
	\includegraphics[scale=0.65]{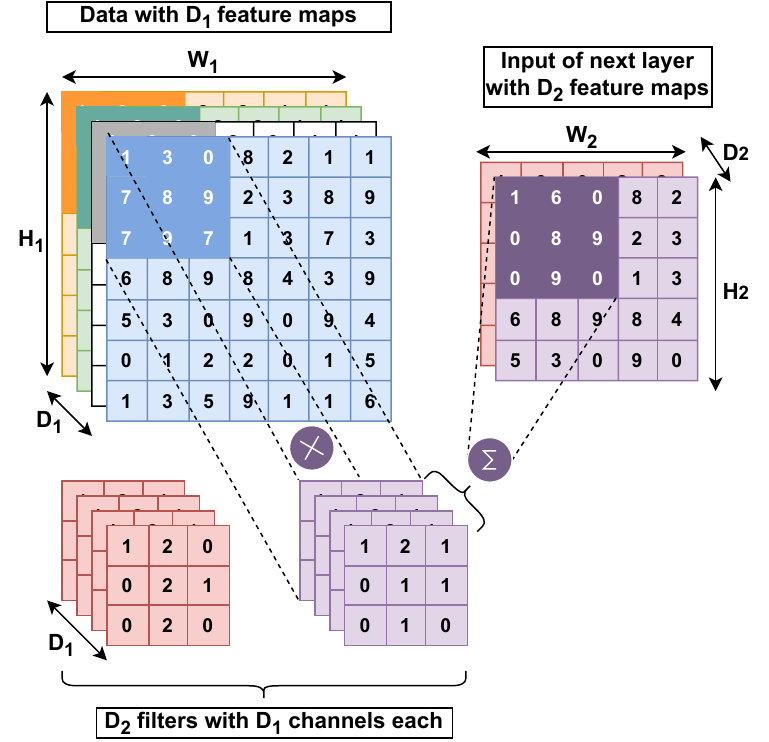}
	\caption{Convolutional task.}
	\label{Conv}
\end{figure}

The DNN networks have various structures. Hence, we introduce the fundamentals of the most known types as follows: 
\paragraph{Multilayer Perceptron (MLP)}
If the output of one layer is fed forward to the subsequent layer, the Neural Network (NN) is termed as the Feed Forward NN (FNN). The baseline FNN is called MLP or Vanilla.  As shown in Fig. \ref{DNN_types} (a), each layer is Fully connected (Fc) to the next one and the output is sent to the next layer’s perceptron without any additional computation or recursion other than the activation function.
\paragraph{Convolutional Neural Networks (CNN)}\label{CNN}
Processing vision-based tasks (e.g., image data), using MLP, potentially requires a deep model with a huge number of perceptrons, as for each data pixel a perceptron is assigned, which makes the network  hard to train and scale. One of the successors of MLP is CNN that is introduced to solve this problem by defining additional pre-processing layers, (i.e., convolutional (conv) and pooling layers), as shown in Fig. \ref{DNN_types} (b). 
Furthermore, the convolutional layer includes a set of learning parameters, namely filters that have the same number of channels as the data feature maps with smaller dimensions. Each filter channel passes through the length and width of the corresponding input feature map and calculates the inner product to the data. The summation of all the outputs produces one feature map. Finally, the number of output feature maps equals the number of filters, as illustrated in Fig. \ref{Conv}. The main difference between the Fc and the conv layers is that each neuron in Fc networks is connected to the entire input, which is not the case of CNN that is connected to only a subset of the input. The second basic component of the CNN network is the pooling task, which has an objective to reduce the spatial size of the input feature maps and minimize the computation time. 

A milestone for CNN applied to computer vision problems is the design of AlexNet \cite{AlexNet} and VGG \cite{VGG}.

\paragraph{Deep Residual Networks}\label{DRN}
Following the victory of AlexNet and VGG, the deep residual networks have achieved a new breakthrough in the computer vision challenges during the recent years. Particularly, the residual networks paved the way for the deep learning community to train up to hundreds and even thousands of layers, while achieving high performance.
ResNet \cite{ResNet} is the-state-of-the-art variant of the residual network. This model uses the so called shortcut/skip connections that skip multiple nodes and feed the intermediate output to a destination layer (see Fig. \ref{DNN_types} (c)), which serves as a memory to the model. A similar idea is applied in the Long Short Term  Memory (LSTM) networks \cite{LSTM}, where a forget gate is added to control the information that will be fed to the next time step. LSTM belongs to the Recurrent Neural Networks (RNN) family.

\paragraph{Randomly Wired Networks}
The aforementioned networks focus more on connecting operations such as convolutional tasks through wise and sequential paths. Unlike previous DNNs, the randomly wired networks \cite{Randomly_wired} arbitrarily connect the same operations throughout the sequential micro-architectures, as shown in Fig. \ref{DNN_types} (d). Still, some decisions are required to design a random DNN, such as the number of stages to down-sample feature maps using Maxpooling and the number of nodes to deploy in each stage. The edge of the randomly wired networks over the other models is that the training is faster, the number of weights is reduced and the memory footprint is optimized.\\

Fig. \ref{DNN_types} presents the NN structures introduced in this section and serving to understand the following sections. Other state-of-the-art structures achieved unprecedented performance in multiple deep learning applications \cite{Survey32}, including Recurrent Neural Networks (RNNs) \cite{RNN}, Auto-Encoders (AEs) \cite{AEs}, and Generative Adversarial Networks (GANs) \cite{GAN}; however, detailed overview of all models falls outside the scope of this paper.

\subsubsection{Reinforcement Learning (RL)} 
Reinforcement learning, also known as sequential decision making, refers to techniques that update the model/policy at each time step, i.e., when receiving each new instance of data. 
The advantage of RL is that it is adaptable, as it does not have any knowledge or assumption about the data distribution. In this way, if the trend of data drifts or morphs, the policy or the model can adapt to the changes on the fly.

\paragraph{Bandit learning}
The bandit problem represents the simplest RL formulation, where an agent interacts with an environment by performing actions at discrete time steps. Each of these actions results in a feedback signal that is referred to as reward, which describes the goodness of that action. Consider a website that wants to maximize the engagement and relevance of articles presented to users. When a new user arrives, the website needs to decide on an article header to show and observe whether or not the user interacts with this article. In this example, the selected action is the article to display, and the reward is binary, $1$ if clicked, $0$ otherwise.
Note that a critical assumption in bandits is that actions do not have any effect on the agent other than causing a sample of a reward signal. In cases where actions may transform the environment from a well-described state to another, a Markov Decision Process (MDP) is required to model the problem and the formulation is known as MDP reinforcement learning. 

\begin{figure}[!h]
\centering
	\includegraphics[scale=0.6]{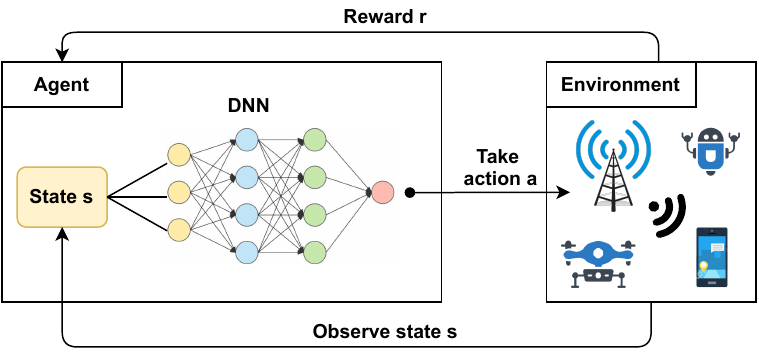}
	\caption{Deep Reinforcement Learning (DRL) design.}
	\label{DRL}
\end{figure}

\paragraph{Markov Decision Process (MDP)-based Learning}
This RL concept is based on learning how to map MDP's states to actions in order to maximize the long-term reward signal. The  RL-agent is not apprised  which  action  to  choose; instead, it  discovers  the actions that  achieve  the  highest  reward by trying  different combinations and receiving immediate gains and penalties, which can be modeled as MDP process. Different from the bandit learning, the RL chosen  action does not impact only the direct reward, but  also  all subsequent situations and related  rewards.  
Deep Reinforcement Learning (DRL) \cite{DRL, 9207771} combines reinforcement learning and the deep learning, as illustrated in Fig. \ref{DRL}. The DRL is well-suited, and even indispensable, when the environment is highly dynamic and dimensional and the number of states is large or continuous. 

Variants of DRL include the deep policy gradient RL \cite{policy_gradient}, the Deep Q-Networks (DQN) \cite{DQN}, Distributed Proximal Policy Optimization (DPPO) \cite{PPO}, and Asynchronous Advantage Actor-Critic \cite{AAAC}. 
\subsection{Performance metrics}
The assessment of the DNN performance depends on the proximity-aware IoT application where deep learning is used. For example, for object detection, face authentication, or self-driving car, the accuracy is of an ultrahigh importance. Yet, some performance metrics are general and not specific to any application, including latency, memory footprint, and energy consumption. An overview of different performance metrics is presented as follows:
\subsubsection{Latency}
The latency, typically measured in milliseconds, is defined as the required time to perform the whole inference/training process, which includes the data pre-processing, data transmission, the classification process or the model training, and the post processing. Real-time applications led by artificial intelligence (e.g., autonomous vehicles and AR/VR gaming) have usually stringent latency constraints, of around 100 ms \cite{Survey8}. Hence, the near-processing is advantageous for fast inference response.
The latency metric is affected by different factors, such as the size of the DNN model, the computational capacity of the host device, and the transmission efficiency.
\subsubsection{Energy efficiency}
Unlike the cloud and edge servers, the IoT devices are battery-limited (e.g., commercial drones.). Moreover, the communication and computation overhead caused by the deep model training/inference incurs huge energy consumption. Hence, the energy efficiency, typically measured in nanojoules, is of a large importance in the context of edge AI and it primarily depends on the size of the DNN and the capabilities of the computing device.
\subsubsection{Computation and memory footprint} To perform DNN training/inference, significant cycles are executed for memory data transfer to/from computational array, which makes it a highly intensive and challenging task. For example, VGG 16 and AlexNet require respectively 512 MB and 217 MB of memory to store more than 138 M and 60 M of weights in addition to the model complexity or Multiply-ACCumulate operations (MACC) which is equal to 154.7 G and 7.27 G  \cite{survey_DNN}. Such amounts of memory and computational tasks, typically measured in Megabyte and number of multiplications respectively, are infeasible to be executed in power and resource constrained devices with a real-time response. 

\subsubsection{Communication Overhead} The communication overhead impacts the performance of the system, when the DNN computation is offloaded to the cloud or other edge participants. Hence, it is indispensable to minimize this overhead, particularly in costly network infrastructures. The data overhead, typically measured in Megabyte, depends on the input and how the model is designed, i.e., types and configuration of the layers that determine the output size, in addition to the communication technology.  Furthermore, the fault-tolerance should be guaranteed to deal with communication failures efficiently.
\subsubsection{Privacy} IoT devices produce and offload a massive amount of data every second, which can result in serious  privacy vulnerabilities and security attacks such as  white-box attacks \cite{white-box}, data poisoning \cite{data_poisoning}, and membership attacks \cite{membership}. Guaranteeing the robustness and privacy of the DNN system has become a primary concern for the deep learning community. The traditional wise resorts to data encryption, pre-processing, and watermarking. Yet, all these solutions can be neutralized using model stealing attacks. Hence, more sophisticated defenses need to be designed to secure the DNN training and execution, through data distribution. The robustness of a privacy mechanism is judged by its ability to protect the data from attacks while maintaining the accuracy performance.

To design an efficient deep learning network or select the adequate one for the targeted application, a large number of hyperparameters need to be considered. Therefore, understanding the trade-off between these parameters (e.g., latency, accuracy, energy, privacy, and memory.) is essential before designing the model.  Recently, automated machine learning frameworks responsible for DNN selection and parameters tuning, have been introduced, such as Talos \cite{talos}.
\subsection{Pervasive frameworks for AI}
Several hardware and software libraries are publicly available for pervasive devices, particularly resource-limited ones, to enable DNN training and inference. As a first example, Google TensorFlow \cite{Tensorflow} is an open source deep learning framework released in 2015 to execute DNN tasks on heterogeneous distributed systems based on their estimated computational and communication capacities, which was optimized later to be adequate for resource constrained devices (e.g., Raspberry Pi) and GPU execution.
Another lightweight deep learning framework developed by Facebook is Caffe2 \cite{caffe2} that provides a straightforward way to experiment heterogeneous deep learning models on low-power devices.
Core ML \cite{CoreML} and DeepLearningKit \cite{ DeepLearningKit} are two machine learning frameworks commercialized by Apple to support pre-trained models on iPhone/iPad devices. More specifically, Core ML was designed to leverage the CPU/GPU endowed with the end-device for deep learning applications such as natural language and image processing, while DeepLearningKit supports more complex networks such as CNNs and it is coined to utilize the GPU more efficiently for iOS based applications.

Since pervasive AI is still in its early stages, only few frameworks are dedicated specifically for distributed learning. One of these deep learning frameworks is MXNet \cite{mxnet}, which is used for pervasive training. MXNet 
uses KVStore\footnote{www.kvstore.io} to synchronize parameters shared among participants during the learning process. To monitor the utilization of pervasive resources, Ganglia \cite{Ganglia} is designed to identify memory, CPU, and network requirements of the training and track the hardware usage for each participant. As for the inference phase, authors in \cite{hardInf} designed a hardware prototype targeting distributed deep learning for on-device prediction.

\subsection{Pervasive AI for IoT Applications}
Deep learning methods have brought substantial breakthroughs in a broad range of IoT applications, spanning from signal and natural language processing to image and motion recognition.
In this section, we review the accomplishments of deep learning in different domains where pervasive computing is needed, including intelligent vehicles and robots, smart homes and cities, and virtual reality/augmented reality.
\subsubsection{Intelligent vehicles, robots, and drones}
Recently, DNNs have been widely used to lead a variety of mobile platforms such as drones, robots, and vehicles, in order to achieve critical tasks. 
In this context, applications such as driving assistance, autonomous driving, and mobility mapping have become more reliable and commonly used in intelligent mobile systems. As an example, in \cite{self_driving}, the captured image from the vehicle front facing camera is used to decide the steering angle and keep the car in the middle of the lane.
The ever-improving online learning is broadly exploited for UAVs/robots guidance, including the work in \cite{drone1} where drones learn how to navigate and avoid obstacles while searching target objects. Several start-ups are also using DL for their self-driving systems, such as prime-air UAVs of Amazon used to deliver packages \cite{prime_air}, and Uber self-navigating cars \cite{Uber}.
\subsubsection{Smart homes and cities}
The concept of a smart home covers a large range of applications, that contribute to enhance the productivity, convenience, and life quality of the house occupants. Nowadays, many smart appliances are able to connect to the internet and offer intelligent services, such as smart air conditioners, smart televisions, and lighting control systems. Most of these appliances require the deployment of wireless controllers and sensors in walls, floors, and corners to collect data for motion recognition DL services. Speech/voice DL recognition services are also involved for a better home control, where a Well-known example is Amazon Alexa \cite{alexa}. 

Compared to smart homes, smart city services are more relevant to the deep learning community as the data collected from different ubiquitous participants is huge and highly heterogeneous, which allows high-quality analysis. Examples involve waste classification \cite{waste_classification}, energy consumption and smart grid \cite{smart_grid}, and parking control \cite{car_parking}.


\subsubsection{Virtual Reality (VR) and Augmented Reality (AR)}
VR is designed to create an artificial environment, where users are placed into a 3D  experience while AR can be defined as a VR that inserts artificial objects into the real environment. Popular examples of  applications using AR/VR include the tactile internet and holographic telepresence \cite{holographic}, and multi-players VR games. The latency of the virtual reality systems is measured in terms of “motion-to-photons” metric, which is defined as the delay starting from moving the headset to updating the display according to the movement. This motion-to-photons latency should be in the range of tens to hundreds of milliseconds \cite{VR}. Offloading the VR/AR computation to the remote cloud servers may incur higher latencies exceeding the required constraints. Hence, on-device computation is indispensable to achieve real-time performance.
\subsection{Lessons learned}
In this section, we reviewed state-of-the-art deep learning and reinforcement learning techniques, examined their performance metrics, and presented some of their applications that may require pervasive deployment. In this context, multiple conclusions can be stated:
\begin{itemize}
    \item The AI proximity-aware IoT applications have different requirements and each one has its distinctive performance keys. For example, VR/AR is highly sensitive to delays and cannot tolerate any motion sickness. Meanwhile, the applications relying on UAVs and moving robots have stringent requirements in terms of energy to accomplish their missions. For the surveillance applications, the accuracy is paramount.
    However, such requirements come with other costs. More specifically, lower delays and energy consumption can be achieved using small DL networks that generate fast inference and can be deployed locally. On the other hand, high accuracy requires deep networks that incur higher memory and computation utilization and consequently higher communication overheads for remote execution. 
    Therefore, understanding the requirements of the targeted application and the trade-off between different hyper-parameters is crucial for selecting the adequate AI model and the processing device.
    \item The  common  characteristic for  most of AI applications, particularly for IoT applications that require real-time data collection, is the need for prompt response and fast analytics that should not be piled for later processing. Hence, centralized solutions such as cloud-based data analytics are not feasible anymore, due to the communication overheads. Pervasive computation has emerged as a solution that enables the deployment of AI at the proximity of the data source for latency-sensitive applications, and in collaboration with high-performance servers for better computational resources.
    \item Understanding the application requirements and the pervasive environment and wisely selecting the data shape and the adopted AI technique, is critical for determining the distribution mode. More specifically, the privacy constraints and the size of the data open the doors for federated learning where each entity trains its data locally.  The low latency requirements and the limited resources imposed by some pervasive systems, push for the partitioning of  inference where the AI model is split into smaller segments.  Finally, the dynamics of the system, the unavailability of labeled data and the inherently decentralized architectures call for the reinforcement learning where agents are distributed.
\end{itemize}
After understanding the motivations for \textit{pervasive AI} and the requirements of the IoT applications and their related AI models, we present  different distribution modes and their communication and computation models in the subsequent sections. We review, first, the pervasive training including federated learning, multi-agent RL, and active learning, and then we survey the pervasive inference. However, we start by presenting the related surveys and highlighting our paper novelty. 
\section{Related surveys and paper novelty}\label{related_surveys}
\begin{table*}[!h]
\footnotesize
\centering
\tabcolsep=0.09cm
\caption{Comparison with existing surveys.}
\label{tab:Related_works}
\begin{tabular}{|c|c|c|c|c|c|c|c|c|c|c|c|}
\hline
\textbf{Refs} &\textbf{Summary}  & \multicolumn{2}{c|}{\textbf{AI/pervasivity}} & \multicolumn{3}{c|}{\textbf{Scope}} & \multicolumn{3}{c|}{\textbf{AI technique}} & \multicolumn{2}{c|}{\textbf{Topic}} \\ \hline
 &  & \begin{tabular}[c]{@{}c@{}}AI on pervasive\\ networks\end{tabular} &  \begin{tabular}[c]{@{}c@{}}AI for pervasive\\ networks\end{tabular} & cloud & \begin{tabular}[c]{@{}c@{}}edge\\ servers\end{tabular} & IoT & DI & FL & MARL & \begin{tabular}[c]{@{}c@{}}Deployment:\\ hardware,\\ software\\ techniques, \\protocols.\end{tabular} & \begin{tabular}[c]{@{}c@{}}Management: \\ communication, \\ resource allocation, \\ and algorithms\end{tabular} \\ \hline
\ {}\begin{tabular}[c]{@{}c@{}}\cite{Survey6,Survey13} \\ (2020-2021)\end{tabular} & \begin{tabular}[c]{@{}c@{}}Deep Learning \\ applications for the \\ Mobile Edge \\ computing networks\end{tabular} & \xmark& \begin{tabular}[c]{@{}c@{}}\cmark\\ 5G,\\ wireless \\ networks\end{tabular}& \cmark& \cmark & \cmark& \cmark& \cmark& \xmark& \cmark& \xmark\\ \hline
\ {}\begin{tabular}[c]{@{}c@{}}\cite{Survey8}\\ (2019)\end{tabular} & \begin{tabular}[c]{@{}c@{}}Efficient usage of IoT\\  hardware and software\\  for AI applications\end{tabular} & \cmark& \xmark& \xmark& \xmark& \cmark& \xmark& \xmark& \xmark& \cmark& \xmark\\ \hline
\ {}\begin{tabular}[c]{@{}c@{}}\cite{Survey1,Survey2,Survey3,Survey4,Survey28}\\ (2019-2020)\end{tabular} & \begin{tabular}[c]{@{}c@{}}Enabling AI on \\ edge networks\end{tabular} & \cmark & \cmark& \xmark& \cmark& \cmark& \cmark& \cmark& \xmark& \cmark& \xmark\\ \hline
\ {}\begin{tabular}[c]{@{}c@{}}\cite{Survey7} \\ (2018)\end{tabular}& \begin{tabular}[c]{@{}c@{}}Enabling AI on \\ edge networks\end{tabular} & \cmark& \xmark& \xmark& \cmark& \cmark& \cmark& \xmark& \xmark& \cmark& \xmark\\ \hline
\ {}\begin{tabular}[c]{@{}c@{}}\cite{Survey22,Survey23,Survey27}\\ (2018-2020)\end{tabular} & \begin{tabular}[c]{@{}c@{}}Decision making in \\multi-agent \\ systems and related \\applications\end{tabular} & \xmark& \xmark& \xmark& \xmark& \xmark& \xmark& \xmark& \cmark& \xmark& \xmark\\ \hline
\ {}\begin{tabular}[c]{@{}c@{}}\cite{Survey11}\\ (2020)\end{tabular} & \begin{tabular}[c]{@{}c@{}}Deep RL \\ for IoT systems\end{tabular} & \cmark& \cmark& \xmark& \xmark& \cmark& \xmark& \xmark& \xmark& \cmark& \cmark\\ \hline
\ {}\begin{tabular}[c]{@{}c@{}}\cite{Survey24} \\ (2020)\end{tabular}& \begin{tabular}[c]{@{}c@{}}Deep RL for \\wireless networks\end{tabular} & \cmark& \begin{tabular}[c]{@{}c@{}}\cmark\\ wireless \\ networks\end{tabular} & \xmark& \cmark& \cmark& \xmark& \xmark& \cmark& \cmark& \xmark\\ \hline
\ {}\begin{tabular}[c]{@{}c@{}}\cite{surveyV14} \\ (2020)\end{tabular}& \begin{tabular}[c]{@{}c@{}}Distributed ML\end{tabular} & \cmark& \begin{tabular}[c]{@{}c@{}}\xmark\end{tabular}& \xmark& \xmark& \xmark& \xmark& \xmark& \xmark& \xmark& \cmark\\ \hline
\ {}\begin{tabular}[c]{@{}c@{}}\cite{Survey5} \\ (2019)\end{tabular}& \begin{tabular}[c]{@{}c@{}}Communication for ML \\ and \\ML for communication\end{tabular} & \cmark& \begin{tabular}[c]{@{}c@{}}\cmark\\  wireless \\ networks\end{tabular}& \xmark& \cmark& \cmark& \xmark& \cmark& \xmark& \xmark& \cmark\\ \hline
\ {}\begin{tabular}[c]{@{}c@{}}\cite{Survey12} \\ (2020)\end{tabular}& \begin{tabular}[c]{@{}c@{}}Communication efficient \\ edge AI\end{tabular} & \cmark& \cmark& \xmark& \cmark& \cmark& \cmark& \cmark& \xmark& \xmark& \cmark\\ \hline
\ {}\begin{tabular}[c]{@{}c@{}}\cite{Survey29} \\ (2019)\end{tabular}& \begin{tabular}[c]{@{}c@{}}AI on mobile \\ and wireless networks\end{tabular} & \cmark& \begin{tabular}[c]{@{}c@{}}\cmark\\ 5G,\\ wireless \\ networks\end{tabular}& \xmark& \cmark& \cmark& \xmark& \xmark& \xmark& \cmark& \cmark\\ \hline
\ {}\begin{tabular}[c]{@{}c@{}}\cite{surveyV11,surveyV12} \\ (2020)\end{tabular}& \begin{tabular}[c]{@{}c@{}}Distributed Training \\of DNN\end{tabular} & \cmark& \begin{tabular}[c]{@{}c@{}}\xmark \end{tabular}& \xmark& \xmark& \xmark& \xmark& \xmark& \xmark& \xmark& \cmark\\ \hline
\ {}\begin{tabular}[c]{@{}c@{}}\cite{surveyV13} \\ (2021)\end{tabular}& \begin{tabular}[c]{@{}c@{}}Federated learning \\for IoT applications\end{tabular} & \xmark& \begin{tabular}[c]{@{}c@{}}\cmark \end{tabular}& \cmark& \cmark& \cmark& \xmark& \cmark& \xmark& \xmark& \xmark\\ \hline
\ {}\begin{tabular}[c]{@{}c@{}}\cite{Survey15} \\ (2020)\end{tabular}& \begin{tabular}[c]{@{}c@{}}Enabling protocols, \\ technologies\\ for federated learning\end{tabular} & \cmark& \cmark& \cmark& \cmark& \cmark& \xmark& \cmark& \xmark& \cmark& \xmark\\ \hline
\ {}\begin{tabular}[c]{@{}c@{}}\cite{Survey14,Survey17,Survey9} \\ (2020)\end{tabular}& \begin{tabular}[c]{@{}c@{}}Architecture, design and\\  applications of centralized,\\ distributed and federated \\learning\end{tabular} & \cmark& \cmark& \cmark& \cmark& \cmark& \xmark& \cmark& \xmark& \cmark& \cmark\\ \hline
\ {}\begin{tabular}[c]{@{}c@{}}\cite{Survey16} \\ (2020)\end{tabular}& \begin{tabular}[c]{@{}c@{}}Enabling protocols, \\technologies\\ for federated learning\end{tabular} & \begin{tabular}[c]{@{}c@{}}\cmark\\ Vehicular\\ IoT\end{tabular} & \xmark& \xmark& \xmark& \cmark& \xmark& \cmark& \xmark& \cmark& \xmark\\ \hline
\rowcolor[HTML]{ECF4FF} 
Our paper & Pervasive AI & \cmark& \xmark& \cmark& \cmark& \cmark& \cmark& \cmark& \cmark& \xmark& \cmark\\ \hline
\end{tabular}
\end{table*}

The intersection of pervasive computing and AI is still in its early stage, which attracts the researchers to review the existing works and provide innovative insights, as illustrated in Table \ref{tab:Related_works}.  First, many efforts discussed the applications of artificial intelligence that support edge networks, 
in order to meet the networking requirements. Multiple edge contexts are explored such as healthcare, smart cities, and grid energy. As an example, two recent surveys \cite{Survey6,Survey13} provided an in-depth discussion of the usage of AI in wireless and 5G networks to empower caching and offloading, resource scheduling and sharing, and network privacy. These surveys touched upon the pervasive AI, particularly federated learning and distributed inference. However, the distribution was discussed briefly as one of the techniques that further enables AI in the edge. In our survey, the applications of AI for pervasive networks is not the main topic. Instead, the deployment of AI on pervasive devices is the scope of this paper. 

The surveys in \cite{Survey1,Survey2,Survey3,Survey4,Survey28,Survey7} conducted a comprehensive review on the systems, architectures, frameworks, software, technologies, and algorithms that enable AI computation on edge networks and discussed the advantages of edge computing to support the AI deployment compared to cloud approaches. However, even though they dedicated a short part for distributed AI, these papers did not discuss the resource and communication challenges of pervasive computing nor the partitioning techniques of AI (e.g., splitting strategies of the trained DNN models or the training data.). Moreover, they did not consider the cloud computing as  indispensable part of the distributed system. Therefore, unlike the previous surveys \cite{Survey1,Survey2,Survey3,Survey4,Survey28,Survey7}, we present an in-depth review that covers the resources, communication and computation challenges of distributed AI among ubiquitous devices. More specifically, applying the same classical communication and computation techniques adopted in centralized approaches for pervasive AI is not trivial. As an alternative, both pervasive computing systems and distributed AI techniques are tailored to take into consideration the heterogeneous resources of participants, the AI model, and the requirements of the system.
These customized strategies for pervasive AI are the main focus of our survey.

The multi-agent reinforcement learning has not been reviewed by any of previous papers. Other papers surveyed the single agent and multi-agents RL, such as \cite{Survey22,Survey23,Survey27,Survey11,Survey24}. In these tutorials, the authors conducted comprehensive studies to show that the single-agent RL is not sufficient anymore to meet the requirements of emerging networks in terms of efficiency, latency, and reliability. In this context, they highlighted the importance of cooperative MARL to develop decentralized and scalable systems. They also surveyed the existing decision making models including  game theory and Markov decision process and they presented an overview of the evolution of cooperative and competitive MARL, in terms of rewards optimization, policy convergence, and performance improvement. Finally, the applications of MARL for networking problems are also reported. However, despite this recent popularity of MARL, the designed algorithms to achieve  efficient communication between agents and minimum computation are not surveyed yet. To the best of our knowledge, we are the first to survey the computation and communication challenges faced to achieve a consensus on the distributed RL policy. In other words, our focus is not the performance of the RL policy. Instead, we survey the computational load, communication schemes and architectures experienced by cooperative agents during learning and execution.

Unlike aforementioned papers, authors of \cite{surveyV14,DML3,DML2,DML} focused only on distributed machine learning. In these papers, they covered the training phase and data partitioning. The survey in \cite{DML3} discussed the issues of learning from a data characterised by its large volume, different types of samples, uncertainty, incompleteness, and low value density. Solutions to minimize the learning complexity and divide the data are introduced in \cite{DML2}, where authors reviewed the algorithms and decision rules to fragment large scale data into distributed datasets. The paper in \cite{surveyV14} described the architectures and topologies of nodes participating in the distributed training by presenting existing frameworks and communication patterns that can be employed to exchange states. Authors in \cite{DML} presented a contemporary and comprehensive survey of distributed ML techniques, which includes the applicability of such concept to wireless communication networks, and the computation and communication efficiency. However, this survey along with the previous works focus only on the training phase. Also, the authors do not provide a comprehensive and deep summary of the complexity, computation and communication efficiency witnessed by different decentralized architectures and the amount of data shared by participants, particularly for the multi-agent reinforcement learning. Our survey aims to bridge the gap by providing a comprehensive review of distributed AI, including both training and inference phases. More specifically, we thoroughly study the architectures of federated learning, active learning and reinforcement learning, and the partitioning strategies of DNN trained models. Then, for each approach, we show the impact on the communication and computation complexity and the algorithms scheduling the collaboration between devices. 

Finally, the authors in \cite{Survey5,Survey12,Survey29} provided a deep review of communication challenges of AI-based applications on edge networks. Specifically, the survey in \cite{Survey29} provided insights about allocating mobile and wireless networks resources for AI learning tasks. However, the distribution of AI techniques was not targeted in this latter paper. The surveys in \cite{Survey5,Survey12} are considered the closest ones to our topic as they explored the communication-efficient AI distribution. However, they mainly focused on the training phase, i.e., federated learning, whereas the pervasive inference and MARL were not studied. The inference distribution is briefly discussed in  \cite{Survey5} from a communication angle, without considering other constraints such as the memory and computation nor presenting the partitioning strategies (i.e., splitting of the trained model), which highly impact the distribution process, the parallelization technique, and participants orchestration. Our paper represents a holistic survey that covers all AI tasks that require cooperation between pervasive devices motivated either by the application requirements or by the system design and the AI model.

Our search of related papers has been conducted through different databases and engines, including  IEEE Xplorer, ScienceDirect, and ArXiv; and the papers have been chosen from a time frame set between 2017 and 2021, in addition to some well-established research works. More specifically, we selected all surveys with high citation rates that cover AI, pervasive computing, federated learning, reinforcement learning, bandit learning, deep learning applications in IoT systems, and AI deployment on edge networks. In the rest of our survey, we review the research conferences and journal papers with solid results that provide comprehensive studies on resource-efficient distributed inference and training.
 \section{Pervasive training}\label{Pervasive_training}
\vspace{-0.1cm}
In this section, we discuss the pervasive training, where the fitting of the model or the learning policy is accomplished within the distributed devices. Particularly, we present the resource management for federated learning, multi-agent reinforcement learning, and active learning.  These aforementioned techniques are distributed by design, which means their objective is to train the learning model within pervasive devices. More specifically the concept of federated learning and active learning is based on guaranteeing the privacy of the pervasive data by training each set locally at its source. Similarly, multi-agent reinforcement learning is designed to be implemented in a system comprising multiple independent/related entities that interact with the same environment. However, the  distribution concepts of these techniques are different. In fact, in federated learning, the data is distributed and each participant creates a local model. Then,  the global model is obtained by aggregating these pervasive models. Meanwhile, in active learning, the participants collaborate to label the data and each one can benefit from on-the-fly labeled samples incoming from the others. Finally, agents in MARL collaborate to converge to the policy that ensures the selection of the best actions.

\subsection{Federated Learning}\label{FL}
Despite the great potential of deep learning in different applications, it still has major challenges that need to be addressed. These challenges are mainly due to the massive amount of data needed for training deep learning models, which imposes severe communication overheads in both network design and end-users. Moreover, the conventional way of transferring the acquired data to a central server for training comes with many privacy concerns that several applications may not tolerate. In this context, the need for intelligent and on-device DL training has emerged. More specifically, instead of moving the data from the users to a centralized data center, pervasive data-sources engage the server to broadcast a pre-trained model to all of them. Then, each participant deploys and personalizes this generic model by training it on its own data locally. In this way, privacy is guaranteed as the data is processed within the host. The on-device training has been widely used in many applications\cite{Survey14}, such as the medical field, assistance services, and smart education. However, this no-round-trip training technique precludes the end-devices to benefit from others' experiences, which limits the performance of the local models. To this end, Federated Learning (FL) has been advanced, where end-users can fine-tune their learning models while preserving privacy and local data processing. Then, these local models (i.e., model updates) are aggregated and synchronized (averaged) at a centralized server, before being sent back to the end-users. This process is repeated several times (i.e., communication rounds) until reaching converge. Accordingly, each participant builds a model from its local data and benefits from other experiences, without violating privacy constraints. FL is proposed by Google researchers in 2016 \cite{synchFL}, and since then, it has witnessed unprecedented growth in both industry and academia. 

We present in what follows an overview for this emerging pervasive learning technique, i.e., Federated Learning. In particular, we introduce the computation and communication models of the FL techniques. Then, we present a brief summary of the related works in the literature, while highlighting a use case that considers the application of FL within UAV swarms. It is worth mentioning that the FL can be used for both online and offline learning (i.e., the training can be performed on static datasets at once, or continuously training on new data received by different participants). 

\subsubsection{Profiling computation and communication models \label{sec:Fundamentals}}
Generally, the FL system is composed of two main entities, which are the data-sources (i.e., owners of data or pervasive participants) and the centralized server (i.e., model owner). Let $N$ denote the number of data-sources. Each one of these devices has its own dataset $D_i$. This private data is used to train the local model $m_i$, and then the local parameters are sent to the centralized server. Next, the local models are collected and aggregated onto a global model $m_G=\bigcup_{i=1}^{N} m_i$. The FL is different from training in the remote server, where the distributed data are collected and aggregated first, i.e., $D_G=\bigcup_{i=1}^{N} D_i$, and then one model $m$ is trained centrally. We assume that  data-sources are honest and submit their real data or their true local models to the centralized server. Otherwise,
control and incentive techniques are used to guarantee the reliability of FL, including \cite{FL_incentive}.

Typically, the life cycle of FL is composed of multiple communication rounds that are completed when the centralized model reaches a satisfactory accuracy. Each round includes the following steps:

\begin{itemize}
    \item \textit{Initialization of FL:} The centralized server fixes the training task, the data shape, the initial model parameters, and the learning process (e.g., learning rate). This initial model $m_G^0$ is broadcasted to the selected participants.
    \item \textit{Training and updating the local models:} Based on the current global model $m_G^t$, each data-source $i$ utilizes its own data $D_i$ to update the local model $m_i^t$. We note that $t$ presents the current round index. Hence, at each step $t$, the goal of each participant is to  find the optimal parameters minimizing the loss function $L(m_i^t)$ defined as:
    \begin{equation}
    m_i^{t*}= argmin_{m_i^{t}} L(m_i^t).
    \end{equation}
    Subsequently, the updated parameters of the local models are offloaded to the server by all selected participants.
    \item \textit{Global model aggregation:} The received parameters are aggregated into one global model $m_G^{t+1}$, which will be sent back in its turn to the data owners. This process is repeated continuously, until reaching convergence. The server goal is to minimize the global loss function presented as follows: 
        \begin{equation}
      L(m^t_G)= \frac{1}{N} \sum\limits_{i=1}^{N}L(m^t_i).
    \end{equation}
    The aggregation of the global model is the most important phase of FL. A classical and straightforward aggregation technique, namely FedAvg, is proposed by Google reference paper \cite{synchFL}. In this technique, the centralized server tries to minimize the global loss function by averaging the aggregation following the equation below:
    \begin{equation}
    m_G^{t+1}=\sum\limits_{i=1}^{N}\frac{|D_i|}{\sum\limits_{j=1}^{N}|D_j|}m_i^{t+1},
    \end{equation}
    where $D_i$ is the local dataset. The FL system is iterated continuously until the convergence of the global loss function or  reaching a desirable accuracy.
\end{itemize}

A major challenge in FL is the large communication and energy overhead related to exchanging the models updates between different end-users, and the centralized server \cite{SigProc_M, ChinaComm}. Such overheads depend on multiple parameters, including the models' updates size, the number of participating users, the number of epochs per user, and the number of communication rounds required to maintain the convergence. Particularly, the energy consumed by an FL participant $i$ characterized by a frequency $f$, a local dataset $D_i$, and a number of local epochs $E$, is given by \cite{FL_energy,FL_energy2}:
\begin{equation}
\label{FL_energy}
e^c_i= E \times (\phi\gamma |D_i| f^2),
\end{equation}
where $\phi$ is the number of CPU cycles required to compute one input instance, and $\gamma$ is a constant related to the CPU. The latency required to compute the local model can be expressed as: 
\begin{equation}
\label{FL_latency}
t^c_i= E \times (\frac{\phi|D_i|}{f}).
\end{equation}
From the equations (\ref{FL_energy}) and (\ref{FL_latency}), we can see that a trade-off exists between the local training latency and the consumed energy. More specifically, for a fixed accuracy determined by the number of local epochs and a fixed frequency, the latency is accelerated depending on the size of the private data. If the data size and the accuracy are fixed, increasing the CPU frequency can help to minimize the local model computation. However, minimizing the latency comes at the expense of energy consumption that increases to the square of the operating frequency. 

\begin{figure*}[!h]
\centering
	\includegraphics[scale=0.4]{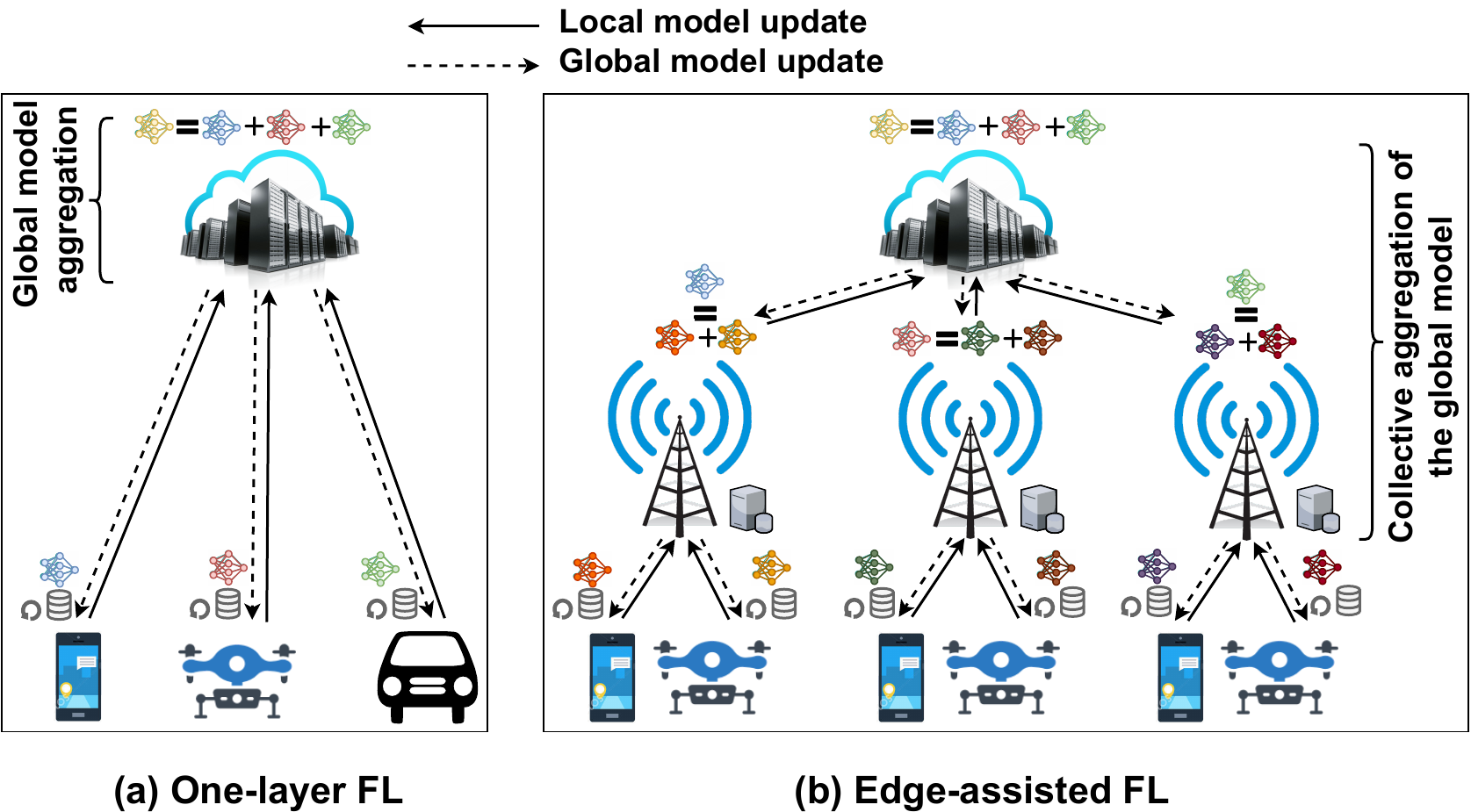}
	\caption{The FL architectures considered in the literature: (a) one-layer FL, (b) edge-assisted FL.}
	\label{fig: FL_arch}
\end{figure*}
The transmission time to share the model updates between the centralized servers and different FL participants mainly depends on the channel quality, the number of devices and the number of global rounds, illustrated as follows:
\begin{equation}
\label{FL_transmission}
t^T= T \times \sum\limits_{i=1}^{N}\frac{K}{\rho_i},
\end{equation}
where $K$ is the models' parameters size shared with the server and $\rho$ is the data rate of the participant $i$. On the other hand, the total energy consumed during the federated learning process using the local transmit powers $P_i$ is equal to:
\begin{equation}
e^T= T \times \sum\limits_{i=1}^{N}\frac{KP_i}{\rho_i},
\end{equation}
From the above equations, we can see that the local iterations $E$ and the global communication rounds $T$ are very important to optimize the energy, computation, and communication costs. Particularly, for a relative local accuracy $\theta_l$, $E$ can be expressed as follows \cite{FL_local}: 
\begin{equation}
\label{FL_E}
E= \alpha \times log(\frac{1}{\theta_l}),
\end{equation}
where $\alpha$ is a parameter that depends on the dataset size and local sub-problems. The global upper bound on the number of iterations to reach the targeted accuracy $\theta_G$ can be presented as \cite{FL_local}:
\begin{equation}
\label{FL_T}
E^g= \frac{\zeta log(\frac{1}{\theta_G})}{1-\theta_l}.
\end{equation}
We note that $\zeta log(\frac{1}{\theta_G})$ is used instead of $O(log(\frac{1}{\theta_G}))$, where $\zeta$ is a positive constant. The computation cost depending on the local iterations $E$ and the communication cost depending on the global rounds $T$ are contradictory. It means, minimizing $E$ implies maximizing $T$ to update the local parameters frequently, which results in increasing the convergence latency. 

To summarize, FL pervasiveness aspects that are being tackled by different studies, to reduce communication and energy overheads, may include: 
\begin{enumerate}
	\item reducing communication frequency, i.e., number of communication rounds;
	\item reducing the number of local iterations;
	\item selecting minimum number of participating users in the training process;  
	\item optimizing local devices operating frequencies;
	\item minimizing the entropy of the models updates by using lossy compression schemes;  
	\item using efficient encoding schemes in communicating models updates.   
\end{enumerate}


In what follows, we categorize different presented FL schemes in the literature, based on the system architecture, namely one-layer FL and edge-assisted FL. The former refers to user-cloud architecture, where different users share their learning models with a cloud or centralized server for aggregation, while the latter refers to user-edge-cloud architecture, where edge nodes are leveraged to reduce communication overheads and accelerate FL convergence (see Figure \ref{fig: FL_arch}). 

\subsubsection{Resource management for Federated learning}
\paragraph{One-layer FL \label{sec:Single}}  
The efficiency of FL concept has been proved by different experiments on various datasets.  
In particular, the proposed model in \cite{synchFL} presented a one-layer FL, where the available users/devices could exchange their local models with a centralized server that collects the local models and forms a global model.  
Afterward, several extensions have been proposed to the original FL. The investigated problems/approaches in FL, considering one-layer architecture, can be categorized into:
\begin{itemize}
	\item studying the convergence behaviour of the proposed FL schemes from a theoretical perspective, while optimizing the learning process given limited computational and communication resources \cite{zhao_federated_2018, Convergence_FedAvg, Wang2019, flOptimizationModel}; 
	\item considering partial user participation for the FL aggregation process in a resource-constrained environment while balancing between the model accuracy and communication cost \cite{Nishio2019ClientSF, wang_optimizing_2020, 9145182,  userSelection_Straggler2020}; 
	\item presenting communication-efficient schemes that aim at reducing the FL communications cost by adopting distinct sparsification and compression techniques  \cite{Sparse_Communication, Felix2020, FL_IOT}.  
\end{itemize} 
The effect of non-Independent and Identically Distributed (non-IID) data on the performance of FL has been investigated in \cite{zhao_federated_2018}. This work illustrated, theoretically and empirically, that highly skewed non-IID data (i.e., the local data at different users are not identically distributed) can substantially decrease the accuracy of the obtained trained model by up to $55\%$. 
To solve this issue, the authors suggested to share a small subset of data between all participants. By integrating these data from the neighboring participants with the local data at each participant, the local dataset will be less skewed.  However, sharing data among the available participants is not always feasible, given strict privacy constraints and communication cost of sharing such data.   
The convergence analysis of FedAvg scheme using non-IID data has been investigated  in \cite{Convergence_FedAvg} for strongly convex problems.  
In \cite{Wang2019}, the authors started first by studying the convergence behaviour of gradient-descent based FL scheme on non-IID data from a theoretical point of view.  
After that, the obtained convergence bound is used to develop a control mechanism, for resource-limited systems, by adjusting the frequency of the global model aggregation in real-time while minimizing the learning loss. 
A new FL algorithm, named FEDL, is presented in \cite{flOptimizationModel}.  This algorithm used a local surrogate function that enables each user to solve its local problem approximately up to a certain accuracy level. The authors presented the linear convergence rate of FEDL as a function of the local accuracy and hyper-learning rate. Then, a resource allocation problem over wireless networks was formulated, using FEDL, to capture the trade-off between the training time of FEDL and user's energy consumption.     

In \cite{Convergence_FedAvg}, the effect of considering the participation of all users in FL algorithm has been studied. Indeed, it is shown that increasing the number of participant users may lead to increasing the learning time since the central server have to wait for \textit{stragglers},  i.e., participants with bad wireless channels or large computational delay.  
To overcome the impact of \textit{stragglers}, different schemes have been proposed to select the best subset of users that can participate in the FL aggregation \cite{Nishio2019ClientSF, wang_optimizing_2020}.  
For instance, the authors in \cite{wang_optimizing_2020} presented a control algorithm, leveraging reinforcement learning, in order to accelerate the FL convergence by obtaining the subset of users that can participate in each communication round of FL, while accounting for the effect of non-IID data distribution.  
{
To maintain the balance between computational and communication costs, and global model accuracy, the authors in  \cite{9145182} presented a joint optimization model for data and users selection. } 
In \cite{userSelection_Straggler2020}, the problem of users selection to minimize the FL training time was investigated for Cell-Free massive Multiple-Input Multiple-Output (CFmMIMO) networks.   

Alternatively, sparsification and compression techniques are used to decrease the entropy of the exchanged models in FL process.  
In particular, instead of communicating  dense models' updates, the authors in \cite{Sparse_Communication} presented a framework that aims at accelerating the distributed stochastic gradient descent process by exchanging sparse updates (i.e., forwarding the fraction of entries with the biggest magnitude for each gradient).  
In \cite{Felix2020}, a sparse ternary compression technique was presented to compress both the upstream and downstream communications of FL, using sparsification, error accumulation, ternarization, and optimal Golomb encoding. 

\paragraph{Edge-assisted FL \label{sec:Hierarchical}}  
Some studies have considered edge-assisted FL architecture to tackle the problem of non-IID data. For example, the authors in \cite{Client-Edge-Cloud} extended the work in \cite{Wang2019} in order to  analytically prove the convergence of the edge-assisted FedAvg algorithm. Then, this work was further extended in \cite{wu_accelerating_2020} to mitigate the effect of \textit{stragglers} by proposing probabilistic users selection scheme. 
The authors in \cite{duan_self-balancing_2021} presented two strategies to prevent the bias of training caused by non-IID data. The first strategy was applied before training the global model by performing data augmentation to tackle the challenge of non-IID data. The second strategy utilized mediators, i.e., edge nodes, to reschedule the training of the participants based on the distribution distance between the mediators. 
In \cite{Naram2020}, the impact of non-IID data in edge-assisted FL architecture was investigated and compared to the centralized FL architecture. This study defined the main parameters that affect the learning process of edge-assisted FL. 
Table \ref{tab:Fl} presents the taxonomy of the federated learning techniques described in this section.
\begin{table*}[]
\centering
\footnotesize
\tabcolsep=0.09cm
\caption{Taxonomy of federated learning techniques.}
\label{tab:Fl}
\begin{tabular}{|l|l|l|l|l|l|l|l|}
\hline
\textbf{Refs} & \textbf{Year} & \textbf{FL devices} & \textbf{Architecture} & \textbf{Trained model} & \textbf{Aggregation algorithm} & \textbf{Dataset} & \textbf{Targeted metrics} \\ \hline
\cite{synchFL} & 2017 & \begin{tabular}[c]{@{}l@{}}Mobile\\ devices\end{tabular} & One-layer & \begin{tabular}[c]{@{}l@{}}- 2NN\\ - CNN \\ - LSTM\end{tabular} & FedAvg & \begin{tabular}[c]{@{}l@{}}- CIFAR-10 \cite{cifar}\\ - MNIST \cite{mnist} \end{tabular} & - Accuracy vs rounds \\ \hline
\cite{zhao_federated_2018} & 2018 & \begin{tabular}[c]{@{}l@{}}Mobile and\\ IoT devices\end{tabular} & One-layer & - CNN & Enhanced FedAvg & \begin{tabular}[c]{@{}l@{}}- CIFAR-10\\ - MNIST\\ - KWS \cite{KWS} \end{tabular} & \begin{tabular}[c]{@{}l@{}}- Accuracy vs rounds\\ - Shared data\\ - Weight divergence\end{tabular} \\ \hline
\cite{Convergence_FedAvg} & 2019 & End-users & One-layer & - Logistic regression & FedAvg & - MNIST & \begin{tabular}[c]{@{}l@{}}- Global loss vs rounds\\ - Rounds vs local epochs\end{tabular} \\ \hline
\cite{Wang2019} & 2019 & Edge nodes & One-layer & \begin{tabular}[c]{@{}l@{}}- Squared-SVM\\ - Linear regression,\\ - K-means\\ - CNN\end{tabular} & FedAvg & \begin{tabular}[c]{@{}l@{}}- MNIST \\ - Energy \cite{energy_data} \\ - User Knowledge \\ Modeling \cite{UKM}  \\ - CIFAR-10\end{tabular} & \begin{tabular}[c]{@{}l@{}}- Loss vs nodes\\ - Accuracy vs nodes\end{tabular} \\ \hline
\cite{flOptimizationModel} & 2019 & End-users & One-layer & \xmark & \begin{tabular}[c]{@{}l@{}}Non-weighted\\ averaging\end{tabular} & \xmark & \begin{tabular}[c]{@{}l@{}}- Communication vs \\   computation time\\ - Learning time vs energy\end{tabular} \\ \hline
\cite{Nishio2019ClientSF} & 2019 & End-users & One-layer & - CNN & Averaging & \begin{tabular}[c]{@{}l@{}}- CIFAR-10\\ - Fashion-MNIST \cite{fashionmnist} \end{tabular} & \begin{tabular}[c]{@{}l@{}}- Accuracy vs time\\ - Number of participants\end{tabular} \\ \hline
\cite{wang_optimizing_2020} & 2020 & \begin{tabular}[c]{@{}l@{}}Mobile\\ devices\end{tabular} & One-layer & - CNN & \begin{tabular}[c]{@{}l@{}}FedAvg with users seclection\\ Favor\end{tabular} & \begin{tabular}[c]{@{}l@{}}- MNIST\\ - Fashion-MNIST\\ - CIFAR-10\end{tabular} & -Accuracy vs rounds \\ \hline
\cite{9145182} & 2020 & End-users & One-layer & \begin{tabular}[c]{@{}l@{}}- MLP\\ - CNN\end{tabular} & \begin{tabular}[c]{@{}l@{}}FedAvg\end{tabular} & - MNIST & - Accuracy vs rounds \\ \hline
\cite{userSelection_Straggler2020} & 2020 & \begin{tabular}[c]{@{}l@{}}Mobile\\ devices\end{tabular} & One-layer & \xmark & \xmark & \xmark & \begin{tabular}[c]{@{}l@{}}- Transmission time\\ - Loss\end{tabular} \\ \hline
\cite{Felix2020} & 2020 & \begin{tabular}[c]{@{}l@{}}Mobile \\ devices\end{tabular} & One-layer & \begin{tabular}[c]{@{}l@{}}- VGG11\\ - CNN\\ - LSTM \\ - Logistic regression\end{tabular} & \begin{tabular}[c]{@{}l@{}}Weighted averaging with \\ Top-k sparsified  communication\end{tabular} & \begin{tabular}[c]{@{}l@{}}- CIFAR\\ - KWS\\ - MNIST\\ - Fashion-MNIST\end{tabular} & \begin{tabular}[c]{@{}l@{}}- Communication delay\\ - Accuracy\end{tabular} \\ \hline
\cite{FL_IOT} & 2020 & \begin{tabular}[c]{@{}l@{}}IoT\\ devices\end{tabular} & One-layer & \begin{tabular}[c]{@{}l@{}}- 2NN\\ - CNN\end{tabular} & \begin{tabular}[c]{@{}l@{}}communication-efficient\\  FedAvg (CE-FedAvg)\end{tabular} & \begin{tabular}[c]{@{}l@{}}- CIFAR-10\\ - MNIST\end{tabular} & \begin{tabular}[c]{@{}l@{}}- Uploaded data\\ - communication rounds\\ - convergence time\end{tabular} \\ \hline
\cite{9148776} & 2020 & UAVs & One-layer & \xmark & FedAvg & \xmark & - Rounds vs bandwidth \\ \hline
\cite{9143577} & 2020 & UAVs & One-layer & - FCN & FedAvg & CRAWDAD \cite{CRAWD} & \begin{tabular}[c]{@{}l@{}}- Accuracy vs rounds\\ - local learning time\end{tabular} \\ \hline
\cite{9159929} & 2020 & UAVs & One-layer & - CNN & FedAvg & - MNIST & - Utility of participants \\ \hline
\cite{9184079} & 2020 & UAVs & One-layer & \begin{tabular}[c]{@{}l@{}}- LSTM\\ - GRU\\ - AQNet \cite{aqnet} \end{tabular} & FedAvg & \begin{tabular}[c]{@{}l@{}}- Ground and aerial \\Sensing Data collected \\ by authors\end{tabular} & - Energy consumption \\ \hline
\cite{Client-Edge-Cloud} & 2020 & End-users & Edge-assisted & - CNN & Hierarchical FedAvg & \begin{tabular}[c]{@{}l@{}}- CIFAR-10\\ - MNIST\end{tabular} & \begin{tabular}[c]{@{}l@{}}- Accuracy vs epochs\\ - Training time\\ - Energy consumption\end{tabular} \\ \hline
\cite{wu_accelerating_2020} & 2020 & End-users & Edge-assisted & \begin{tabular}[c]{@{}l@{}}- FCN \\ - LeNet-5\end{tabular} & \begin{tabular}[c]{@{}l@{}}Weighted averaging with\\ Effective Data Coverage (EDC)\end{tabular} & \begin{tabular}[c]{@{}l@{}}- MNIST\\ - Aerofoil \cite{Airfoil}\end{tabular} & \begin{tabular}[c]{@{}l@{}}-  Accuracy vs rounds\\ - Training time\\ - Energy consumption\end{tabular} \\ \hline
\cite{duan_self-balancing_2021} & 2021 & \begin{tabular}[c]{@{}l@{}}Mobile\\ devices\end{tabular} & Edge-assisted & - CNN & FedAvg & \begin{tabular}[c]{@{}l@{}}- EMNIST \cite{EMNIST}\\ - CINIC-10 \cite{CINIC}\\ - CIFAR-10\end{tabular} & \begin{tabular}[c]{@{}l@{}}- Accuracy vs rounds\\ - Accuracy vs  epochs\\ - Storage requirement\end{tabular} \\ \hline
\cite{Naram2020} & 2021 & End-users & Edge-assisted & \begin{tabular}[c]{@{}l@{}}- FCN\\ - CNN\end{tabular} & FedAvg & \begin{tabular}[c]{@{}l@{}}- MNIST\\ - Fashion-MNIST\\ - CIFAR-10\end{tabular} & \begin{tabular}[c]{@{}l@{}}- Accuracy vs rounds\\ - Accuracy vs edge \\   distance distribution\\ - Speed\end{tabular} \\ \hline
\end{tabular}
\end{table*}

\subsubsection{Use case: Learning in the sky}\label{sec:Cases}

\begin{figure}[h!]
	\centering
		\scalebox{1.4}{\frame{\includegraphics[width=0.53\columnwidth]{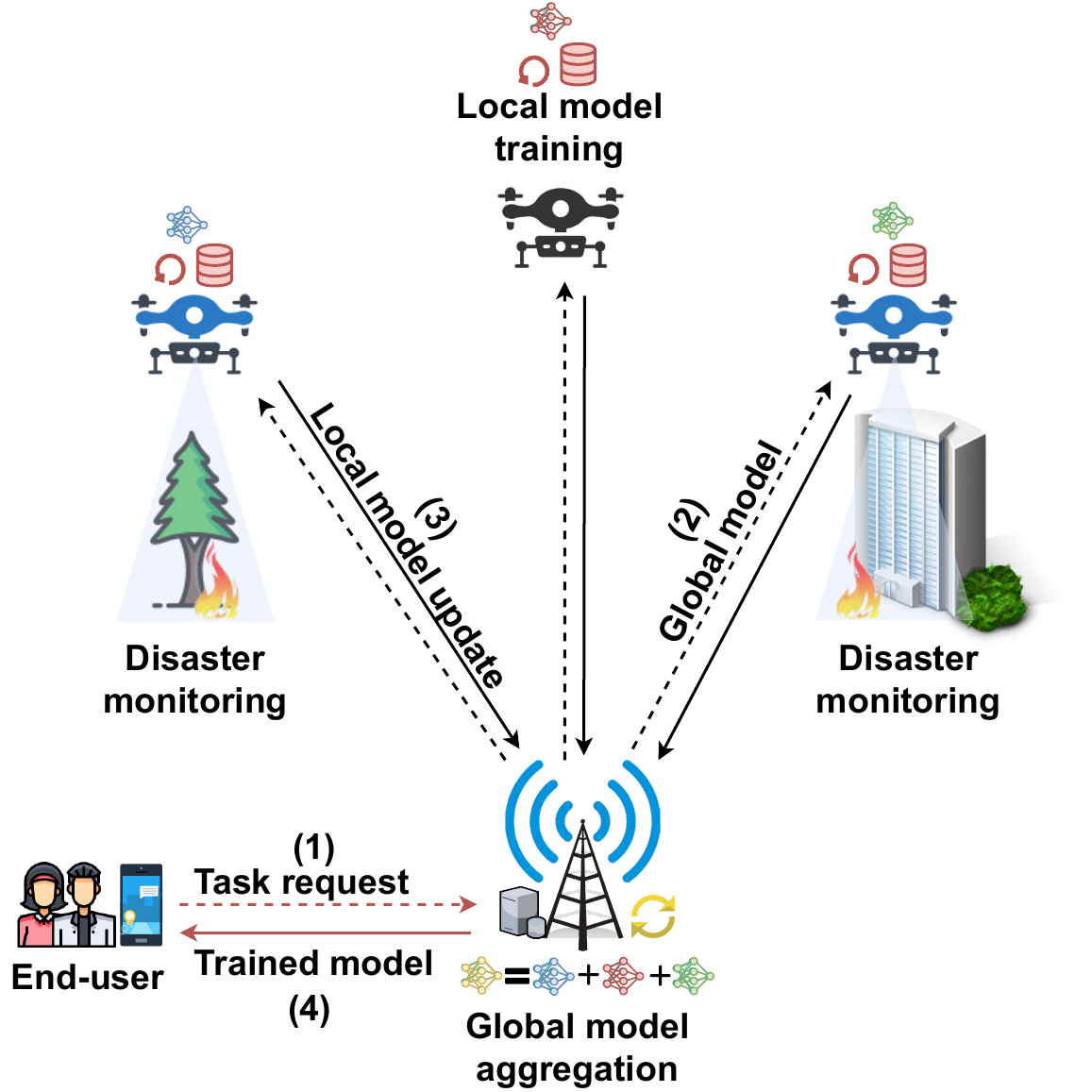}}}
	\caption{An example of FL applications in UAV-assisted environment.}
	\label{fig:FL_CS}
\end{figure}

Nowadays, deep learning has been widely used in Flying Ad-hoc Network (FANET). Different tasks can be executed using DL techniques at UAV swarms, such as coordinated trajectory planning \cite{9148776} and jamming attack defense \cite{9143577}. However, due to the related massive network communication overheads, forwarding the generated large amount of data from the UAV swarm to a centralized entity, e.g., ground base stations, makes implementing centralized DL challenging. 
As a promising solution, FL was introduced within a UAV swarm in several studies \cite{9148776, 9143577, 9159929, 9184079} to avoid transferring raw data, while forwarding only local trained models' updates to the centralized entity that generates the global model and send it to the end-user and all participants over the intra-swarm network (see Fig. \ref{fig:FL_CS}). 

In \cite{9148776}, the authors present a FL framework for the swarm of wirelessly connected UAVs flying at the same altitude. The considered swarm includes a leader UAV and a set of followers UAVs. It is assumed that each follower collects data while flying and implements FL for executing inference tasks such as trajectory planning and cooperative target recognition. Hence, each follower exploits its gathered data to train its own learning model, then forwarding its model's updates to the leading UAV. All received models are then aggregated at the leading UAV to generate a global FL model, that will be used by the following UAVs in the next iteration. 
Interestingly, \cite{9148776} investigates the impact of wireless factors (such as fading, transmission delay, and UAV antenna angle deviations) on the performance of FL within the UAV swarms. The authors present the convergence analysis of FL while highlighting the communication rounds needed to obtain FL convergence. Using this analysis, a joint power allocation and scheduling optimization problem is then formulated and solved for the UAV swarm network in order to minimize the FL convergence time. The proposed problem considers the resource limitations of UAVs in terms of: (1) the strict energy limitations due to the energy consumed by learning, communications, and ﬂying during FL convergence; and (2) delay constraints imposed by the control system that guarantees the stability of the swarm.   


\subsubsection{Lessons learned} 

Despite the prompt development of diverse DL techniques in different areas, they still impose a major challenge, which is: How can we efficiently leverage the massive amount of data generated from pervasive IoT devices for training DL models if these data cannot be shared/transferred to a centralized server? 
\begin{itemize}
\item FL has emerged as a promising privacy-preserving collaborative learning scheme to tackle this issue by enabling multiple collaborators to jointly train their deep learning models, using their local-acquired data, without the need of revealing their data to a centralized server \cite{Felix2020}.  
\item  The model aggregation mechanisms are the most discussed in the FL literature, which are applied to address the communication efficiency, system and model performance, reliability issues, statistical heterogeneity, data security, and scalability. More specifically, one-layer FL approaches are the most studied by previous works, even if researchers are recently investigating decentralized strategies.

\item A major dilemma in FL is the large communication overhead associated with transferring  the models' updates. Typically, by following the main steps of FL protocol, every node or collaborator has to send a full model update in every communication round. Such update follows the same size of the trained model, which can be in the range of gigabytes for densely-connected DL models \cite{9123563}. Given that large number of communication rounds may be needed to reach the FL convergence on big datasets, the overall communication cost of FL can become unproductive or even unfeasible. Thus, minimizing the communication overheads associated with the FL process is still an open research area.

\item  We also remark that despite the considerable presented studies that have provided significant insights about different FL scenarios and user selection schemes, optimizing the performance and wireless resources usage for edge-assisted FL is still missing. Most of the existing schemes for FL suffer from slow convergence. Also, considering FL schemes in highly dynamic networks, such as vehicular networks, or resource-constraint environments, such as healthcare systems, is still challenging.
\end{itemize}
\subsection{Multi-agent reinforcement learning}
In reinforcement learning \cite{sutton2018reinforcement}, the agent learns how to map the environment's states to actions in order to maximize a reward signal. This agent is not told which actions to choose but instead should discover which ones lead to the best reward by trying them. In the most general case, actions affect not only the immediate reward but also the next environment's states and potentially all subsequent rewards, which is called \textit{Markov decision process reinforcement learning}. In the simplified and special case setting of RL with a single state, the agent needs only to detect the best action that  maximizes the current reward (bandit objective) without accounting for the transition to any other state. This non-associative setting, namely \textit{multi-arm bandit learning}, has less modeling power of real systems but avoids much of the complexity of the full reinforcement learning problem.

In this section, we discuss the multi-agent Reinforcement Learning (MARL) with its both forms. The "Multi-agent" prefix indicates the existence of multiple collaborative agents \footnote{Note that while competitive settings can also be modeled, the focus of this section is on systems that aim to jointly optimize an objective function with minimum resource utilization (pervasive AI systems). Thus, competitive and zero-sum games will not be deeply surveyed.} that are aiming to optimize a specific criterion through learning from past experience (i.e., past interactions). We note that MARL was originally proposed to model a single agent interacting with the environment and aiming to maximize its reward. However, in pervasive computing systems, where there are numerous but resource-limited agents (i.e., devices), collaboration becomes essential to leverage the potential of the collective experience of these devices.

Motivated by the prevalence of collaboration in pervasive systems, we review in this section distributed MAB and MDP MARL algorithms from a resource utilization perspective. As has been the case throughout the paper, we are interested in the performance/resource-management trade-offs. Specifically, we propose a taxonomy based on the obtained performance with specific resource budgets (e.g., communication rounds).
\subsubsection{Multi-agent multi-arm bandit learning}
In this section, we first provide technical definitions of the single-agent stochastic bandit problem and then explain its multi-agent extension. 
\paragraph{Overview}
\begin{figure}[!h]
\centering
	\includegraphics[scale=0.3]{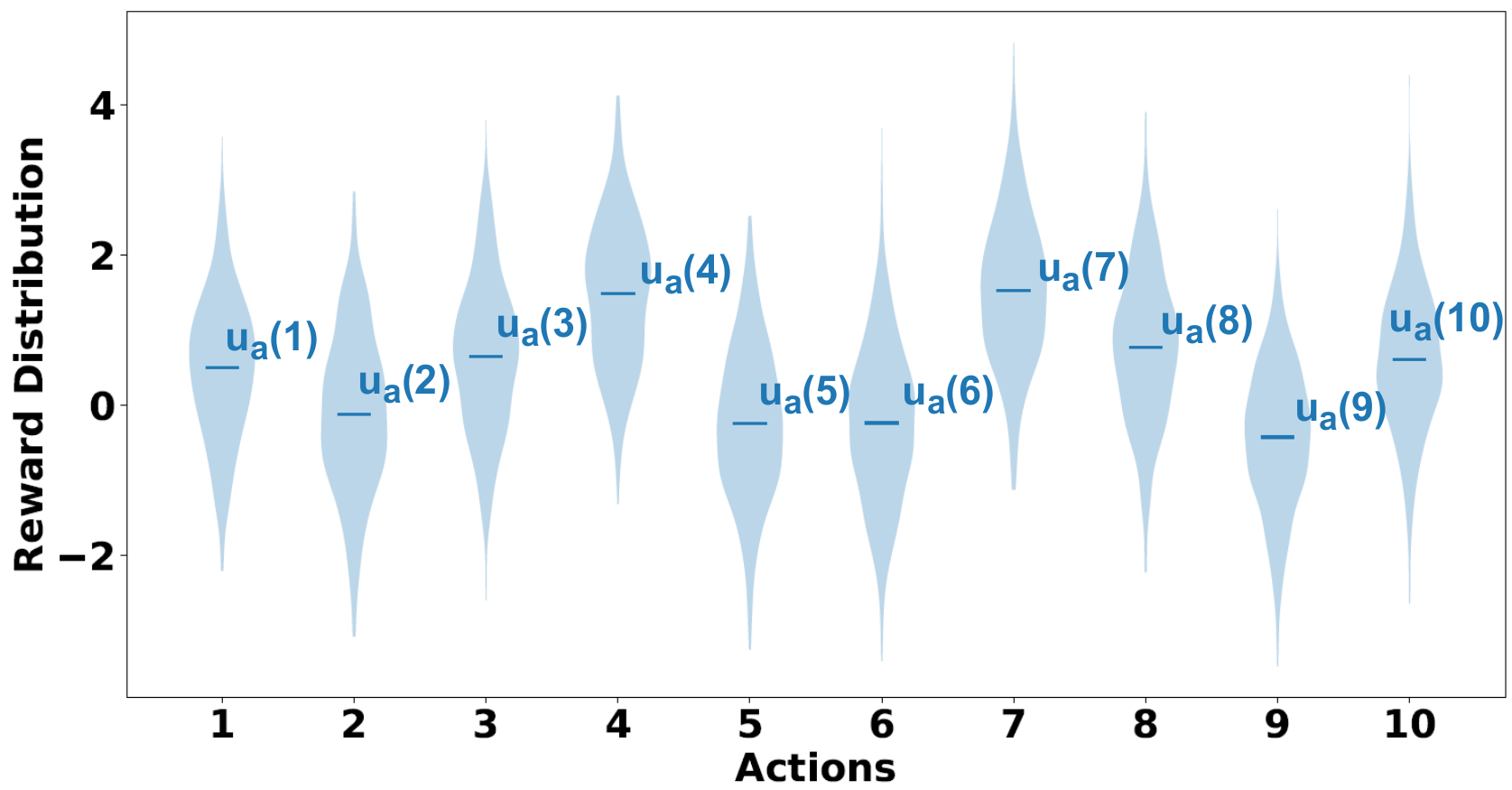}
	\caption{The basic bandit problem: a set of actions corresponding to different reward distributions.}
	\label{multi-bandits}
\end{figure}
The Bandit problem, introduced earlier in section \ref{AI}, is given in Algorithm \ref{alg:bp}, and visually illustrated in Fig. \ref{multi-bandits}. Fundamentally, there exists a set of actions $\mathcal{K}$ (10 actions in the figure), where each action $a$ results in a reward sampled from a distribution $\mathcal{D}_a$ (Gaussians in the example illustrated in Fig. \ref{multi-bandits}). 
\begin{algorithm}
\caption{Basic bandit problem}
\label{alg:bp}
\begin{algorithmic}[1]
 \renewcommand{\algorithmicrequire}{\textbf{Input:}}
 \renewcommand{\algorithmicensure}{\textbf{Output:}}
\Require The set of $K$ actions (arms) $\mathcal{K}$, 

\State \textbf{for} each round $t\in[T]$:
\State \hskip1em Algorithm picks an action $a_t$
\State \hskip1em Environment returns a reward $r_t \sim \mathcal{D}_{a_t}$
\end{algorithmic}
\end{algorithm}

The problem instance is fully specified by the time horizon $T$ and mean reward vector (the vector of the expected reward for each action/arm) $\boldsymbol{\mu}=u_a, a\in \mathcal{K}$, where $u_a = \mathbb{E}[\mathcal{D}_a]$. The optimal policy is simply choosing the action whose expected value is the highest, i.e., $a_* = \argmax_a u_a$. However, as this action is not known a priori ($\mathcal{D}_a$ is not known), it has to be estimated online from samples. Thus, it is inevitable that some sub-optimal actions will be picked, while building certainty on the optimal one. A reasonable performance measure is the \emph{Regret}, which is defined as the difference between the optimal policy's cumulative rewards, and the cumulative rewards achieved by a solution algorithm.
\begin{equation}
    R_T = \underbrace{u_* \times T}_{\text{\shortstack{Optimal policy's\\cumulative rewards}}} - \underbrace{\sum_{t=1}^T u_{at}}_{\text{\shortstack{An algorithm's cumulative\\ rewards}}}
\end{equation}

In other words, the regret $R_T$ is the sum of \emph{per-step regrets}. A per-step regret at time step $t$ is simply the difference between the best action's expected reward $u_*$ and the expected reward of the action chosen by an algorithm $u_{a,t}$ (i.e., $a$ is selected by the algorithm we are following). Thus, it represents how much rewards are missed because the best action is not known and has to be estimated from samples. Solution algorithms typically prove sub-linear regret growth (i.e., this difference goes to zero as time progresses. In this way, learning is achieved). The best achievable regret bound for the described bandit problem was proven to be $O(log T)$ \cite{lattimore2020bandit}.

\begin{figure}[!h]
\centering
	\includegraphics[scale=0.32]{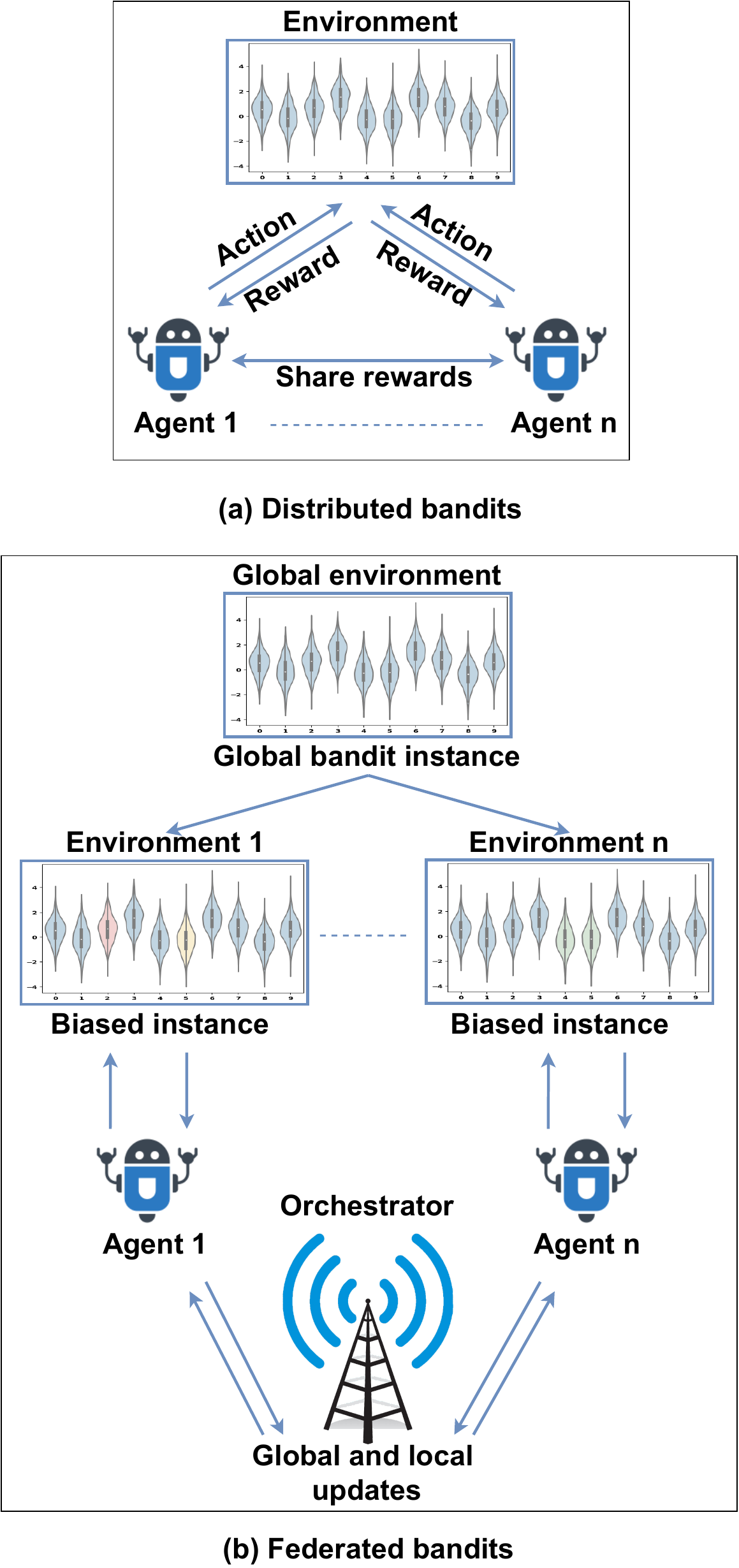}
	\caption{Multi-agent bandits formulations: (a) Distributed Bandits: each agent collaborates with others to identify the best action in the same environment (b) Federated bandits: each agent collaborates with others to identify the best global action using biased local samples. In this example, the local environments were generated (e.g., sampled) from a global one.}
	\label{fed-bandits}
\end{figure}

Several solution algorithms with optimal performance guarantees have been proposed in the literature \cite{lattimore2020bandit}, which fall generally into two categories, explore-then-commit and optimism-based algorithms. Explore-then-commit class, such as successive elimination algorithm, acts in phases and eliminates arms using increasingly sensitive hypothesis tests. On the other hand, the optimism algorithm, such as Upper Confidence Bound (UCB) algorithm, builds confidence for the reward of each action and selects the action with the highest upper bound. The asymptotic performance of both classes is similar. Note that performance guarantees are also classified into instance-dependent bound that depends on the problem information such as the difference between the best and second-best arms, and instance-independent regret (i.e., worst-case regret). These algorithms are recently being extended to model pervasive computing though two main MAB formulations: \emph{distributed} and \emph{federated} bandits, as shown in Fig. \ref{fed-bandits}. 

In distributed bandits, agents aim to solve the same bandit instance (i.e., quickly discover the best action), represented by the action set and their generating distributions. Meanwhile, in the federated bandit settings, agents handle different bandits instances and utilize each others' experiences to solve them. While the terms used to describe the exact problem is sometimes ambiguous in the literature (i.e., distributed, federated, and decentralized were sometimes used interchangeably), in this work, we adopt the recent convention on reserving the term federated for the case where each agent faces a different (but related to others) problem instance, while keeping the term distributed for the case where the instance is the same for all agents but the decision making is distributed across other agents. 

\paragraph{Distributed Bandits Formulations}
In many bandit problem instances, it is appealing to employ more agents to learn collaboratively and concurrently to speed up the learning process. In the distributed bandit problem, there exists a set of agents $[M]$ collaborating to solve the \emph{same} bandit instance (the $K$ arms are the same). These agents communicate according to some networking topologies. In many contexts, the sequential decision-making problem at hand is distributed by nature. For example, we can consider a recommender system deployed over multiple servers in different locations. While every server aims to always recommend the best item, it is intuitive to reuse each other's experiences and cut the time needed to learn individually. Furthermore, since their communication may violate the latency constraints, it is desirable that this collaboration and reuse of experience achieve minimum communication overhead. 

While the classical single-agent bandit algorithm has been proposed since the $~2002$, its multi-agent counterpart is much more recent, with new state-of-the-art algorithms being currently proposed. The work in \cite{kanade2012distributed} initiated the interest in the communication-regret trade-off. The authors established a non-trivial bound on the regret, given an explicitly stated bound on the number of exchanged messages. However, they focused on the full-information setting, assuming that the agents observe the rewards of all actions at each round, and not only the one picked, which is the case in bandit settings. Nonetheless, this work initiated the interest in studying the same trade-off under the bandit settings. The authors of \cite{hillel2013distributed} considered the partial feedback (i.e., bandit settings) and presented an optimal trade-off between performance and communication. This work did not consider regret as the performance criterion but rather assumed the less common ``best arm identification" setup, where the goal is to purely explore in order to eventually identify the best arm with high probability after some number of rounds. The authors in \cite{pmlr-v28-szorenyi13} studied the regret of distributed bandits with a gossiping-based P2P communication specialized to their setup, where at every step, each agent communicates only with two other agents randomly selected. \cite{kolla2018collaborative} studied the regret under the assumption that the reward obtained by each agent is observable by all its neighbors. \cite{landgren2016distributed} proposed a collaborative UCB algorithm on a graph-network of agents and studied the effect of the communication graph topology on the regret bound. \cite{martinez2019decentralized} improved this line of work as the approach requires less global information about the communication graph by removing the graph dependent factor multiplying the time horizon in the regret bound. 

Other works go beyond merely studying the effect of the network topology on the regret bound and explicitly accounting for the communication resources to use. The authors in \cite{wang2020optimal} deduced an upper bound on the number of needed communicated bits, proving the ability to achieve the regret bound in \cite{martinez2019decentralized} with a finite number of communication bits. However, the interesting question, particularly from the perspective of pervasive computing design, is whether the use of communication resources can also be bounded, i.e., can the order of optimal regret bound be guaranteed with a maximum number of communicated bits / communicated messages.

The work in \cite{Sankararaman2019Social} established the first logarithmic upper bound on the number of communication rounds needed for an optimal regret bound. The authors considered a complete graph network topology, wherein a set of agents are initialized with a disjoint set of arms. As time progresses, a gossiping protocol is used to spread the best performing arm with agents. The authors showed that, with high probability, all agents will be aware of the best arm while progressively communicating at less (halving) periods. The authors generalized this work with a sequel formulation \cite{chawla2020gossiping}, which relaxes the assumption of a complete graph, and introduces the option for agents to pull information. However, this approach is still using the same gossiping style of communication. According to \cite{agarwal2021multi}, this dependence on pair-wise gossiping communication results in a sub-optimal instance-independent regret bound. The authors in \cite{wang2019distributed} focused on the regret-communication trade-off in the distributed bandit problem. The networking model utilizes a central node that all agents communicate with. Initially, agents work independently to eliminate bad arms. Then, they start communicating with the central node at the end of each epoch, where epochs' duration grows exponentially, leading to a logarithmic bound on the number of needed messages. 

 \begin{table*}[!h]
\centering
\footnotesize
\tabcolsep=0.09cm
\caption{ Multi-agent stochastic bandit learning literature.}
\label{table:MAB_previous_work}
\begin{tabular}{|c|c|c|c|c|c|}
\hline
\textbf{Refs}  & \shortstack{\textbf{Problem} \\ \textbf{Formulation}} &	\shortstack{\textbf{Communication} \\ \textbf{Model}} &    \shortstack{\textbf{Communication}\\\textbf{Guarantee}}	& \textbf{Regret Guarantee} & \textbf{Method}\\  \hline

P2P-$\epsilon$-Greedy \cite{pmlr-v28-szorenyi13} & DB & Two Neighbors on a graph & $O(T)$& $O(T)$ & Gossiping arms estimates. \\ \hline 

coop-UCB2 \cite{landgren2016distributed} & DB & Neighbors on a graph & $O(T)$& $O(logT)$ & \shortstack{A running Consensus on \\the estimates of arms total rewards.}  \\ \hline

UCB-Network \cite{kolla2018collaborative} & DB & \shortstack{Multiple \\ (Graph and centralized)} & $O(T)$& $O(logT)$ & \shortstack{Identifying and utilizing \\dominating sets in the network.} \\ \hline

DDUCB \cite{martinez2019decentralized} & DB & Neighbors on a graph & $O(T)$& \shortstack{$O(logT)$ \\ (with improved constants)}& \shortstack{A running Consensus on \\the estimates of arms total rewards.} \\ \hline

\cite{Sankararaman2019Social}  & DB & \shortstack{Individual neighbors \\ on a complete graph} & $O(constant)$& $O(logT)$ & \shortstack{Gossiping among different \\local Poison clocks.} \\ \hline 

GosInE \cite{chawla2020gossiping}  & DB & \shortstack{Neighbors on \\ a complete graph}& $\Omega(T)$& $O(logT)+C_G$ & \shortstack{Gossiping and information \\ pulling.}\\ \hline 

DPE2 \cite{wang2020optimal}  & DB & Neighbors on a graph & $O(constant)$& $O(logT)$ & \shortstack{Leader-election to handle \\exploration (exploration is centralized).} \\ \hline

DEMAB \cite{wang2019distributed}  & DB & Centralized coordinator & $O(logT)$& $O(logT)$ &\shortstack{ Utilizing public randomness \\to divide arms among clients. }\\ \hline 

LCC-UCB \cite{agarwal2021multi}  & DB & \shortstack{Multiple \\(Graph and centralized)} & $O(logT)$& $O(logT)$ &\shortstack{ Communicating estimates \\after epochs of doubling lengths. }\\ \hline 


\cite{shahrampour2017multi}  & FB & Neighbors on a graph & $O(logT)$ & $O(logT+C_G)$ & \shortstack{Selecting the best arm \\according to voting.}\\ \hline 

GossipUCB \cite{zhu2021federated}  & FB & Neighbors on a graph &$O(T)$ & \shortstack{$O(max\{logT,log_{C_G}N\})$} & \shortstack{Maintaining local belief \\that is updated through gossiping.} \\ \hline 

\cite{shi2021federated}  & FB & Centralized coordinator & $O(logT)$ & $O(logT)$ & \shortstack{Aggregating estimates through \\the controller until a fixed point of time.} \\ \hline

\cite{shi_federated_2021}  & FB & Centralized coordinator & $O(logT)$ & $O(logT)$ & \shortstack{Mixed target learning objective\\ based on local and global objectives.}  \\ \hline

\cite{wang2020stochastic}  & FB & Neighbors on a graph& $O(T)$& $O(logT)$& \shortstack{Agent use estimates of their\\ neighbors weighted by a similarity metric.}\\ \hline
\end{tabular}
\end{table*}
The work in \cite{agarwal2021multi} presents a state of the art distributed bandit learning algorithm. The authors proposed algorithms for both fully connected and partially connected graphs (i.e., assuming that every agent can broadcast to everyone and assuming that agents can communicate with a subset of the others). Similar to elimination-based algorithms, the proposed algorithm proceeds with epochs of doubling lengths, only communicating at the end of an epoch, thus guaranteeing a logarithmic need for communication resources. The communicated messages are only the ID of the action played most often. Furthermore, the regret is proved to be optimal even in instance independent problems, for reasonable values of the time horizon (i.e., $log(T)>2^{14}$). During each epoch, agents maintain a set of arms that are recommended by other agents at the end of previous epochs and implement a UCB algorithm among them.

\paragraph{Federated Bandits Formulations}
The federated bandit formulation, shown in Fig. \ref{fed-bandits} (b), is a recently emerging framework akin to the federated learning framework discussed earlier. In this formulation, there exists a set of agents, each one is facing a \emph{different} instance of the bandit (but the instances are related to each other). This is different from the distributed bandit formulation discussed in the previous sub-section, where a set of agents collaborate to solve the \emph{same} instance of the multi-arms bandits. Recall that a bandit instance is determined by the mean reward vector $\boldsymbol{\mu}$. By ``related" instances in the federated bandit settings, we mean that each local bandit instance is a \emph{noisy} and potentially \emph{biased} observation of the mean vector. In light of this, collaboration is necessary, as even perfect local learning algorithms might not perform adequately due to their biased observations. 

The setting of federated bandits is first proposed by \cite{shahrampour2017multi} (although not under the same term). The authors proposed an algorithm, where agents agree on the best global arm, and they all play it at the beginning of each round. In this way, communication is needed at the beginning of each round. Recently, \cite{zhu2021federated} studied this federated setting, where the global arm mean vector is the average of the local ones. Although the authors did not propose a bound on the number of messages needed to be exchanged, the communication model considered a partially connected graph, where each agent communicates only with neighbors but with a focus on constrained communication resources. The algorithm contains two main steps: First, each agent shares a set of local information with other neighbors (the number of times an arm was sampled and its sampled mean). Second, a gossip update is performed, where each agent incorporates information received from neighbors in updating its estimate of each arm's mean.  

\cite{shi_federated_2021} presented a more general formulation, where the global mean vector is not necessarily the average of the local ones. Instead, the local means are themselves \emph{samples} from a distribution whose mean is unknown. The local observation for each agent is, in turn, samples from the local distributions. The communication model is similar to supervised federated learning, where agents communicate periodically with an orchestrator that updates the estimates of arms payoffs and instructs the agents on which arms to keep and which to eliminate. Although the communication is periodic, the total number of communication rounds is bounded (logarithmic with respect to the horizon $T$). This is because the number of agents incorporated in the learning process decays exponentially with time. Such an approach works since the average of clients' local means concentrates exponentially fast around that global mean (a known result from probability concentration analyses). 

A setting that is slightly different from the federated bandits was studied in\cite{wang2020stochastic}. The difference is that although agents have similar yet not identical local models, the reward for each agent is actually sampled from its local distribution. Thus, each agent is trying to identify the best arm in its local instance through using information from other ones on arms that are similar. This work is different from other aforementioned approaches where the agents' rewards are sampled from the \emph{global} distribution that they are collaboratively trying to estimate from biased local observations. 

Table \ref{table:MAB_previous_work} summarizes the works in MAB problems. It lists the problem formulation: distributed Bandits (DB) or federated Bandits (FB), the communication model (i.e., the network topology), the communication guarantee (i.e., number of messages needed to achieve the performance), the regret guarantee (i.e., the growth of the regret with respect to the time horizon),  and the main characteristics of the method, which describes how the rewards' estimates are communicated among the agents (Recall that the agents aim to collectively learn an accurate estimates of the rewards distributions). $C_G$ denotes a constant related to the communication graph or gossiping matrix and $N$ is the number of agents.
\paragraph{Use case: MAB for recommender systems}
Online learning systems are fundamental tools in recommender systems, which are, in turn, a cornerstone in the development of current intelligent user applications, from social media application feeds to content caching across networks. Due to the recent growth in data generation, local geo-distributed servers are often used to support applications that utilize recommender systems. Furthermore, privacy concerns sometimes limit the ability of these local servers to share data with other servers. The work in \cite{shi_federated_2021} studies the case of a set of servers that run a recommender system for their prospective clients. The goal of each one is to recommend the most popular content across all servers. However, due to latency constraints, communication at every decision-making step is infeasible. Besides, sharing individual samples of rewards violates privacy, as all servers will learn about a particular user's choice and preference. Due to these reasons, the authors proposed and utilized a federated bandits algorithm (Fed-UCB) which only communicates $logT$ times in a $T$ horizon to minimize recommendation latency. At each round, only the sample \emph{means} are communicated, preserving a certain level of privacy (additional improvements are also discussed). Finally, the performance of the system is shown to be near-optimal; thus, achieving the goal of recommending the best item across all servers while meeting the privacy and communication constraints.
\paragraph{Lessons Learned}
\begin{itemize}
    \item  Distributed bandits formulations are the most popular in the literature compared to the recent federated formulation. Specifically, we note that distributed bandits with gossip-style communication, like the one introduced in \cite{landgren2016distributed}, are a prevalent choice despite their sub-optimal communication resource utilization. This is attributed to the balance between complexity and the robustness resultant from the lack of a central controller.
    \item Communication-cognizant Multi-agent Bandit formulations: Online-learning systems need to account for the communication resources. Thus, recent works do not only analyze regret but explicitly optimize the communication resources. This is manifested through two main observations. First, the derived regret guarantees are always affected by the networking topology (e.g., parameters representing the connectivity of a communication graph, number of involved agents, or number of exchanged messages). Second, accompanied by every regret guarantee, an upper bound on communication resource usage is also provided (e.g., the maximum number of exchanged messages or exchanged bits).
    \item Towards the federation of the bandit framework: When the bandit instances faced by each agent are local biased instances, the federated bandits framework arises. In such a situation, agents need to learn with the help of a logically centralized controller, similar to supervised federated learning, in order to estimate the true global instance and the true best action\cite{shi_federated_2021}. However, if agents are not interested in solving a hidden global instance but rather only their own, they may reuse their peers' experience and an instance-similarity metric to help them solve their own instances \cite{wang2020stochastic}.
\end{itemize}

\subsubsection{Multi-agent Markov decision process learning}
This section presents an overview of Multi-agent MDP from a pervasive computing perspective. We specifically focus on the \emph{communication-performance} trade-off and classify previous works according to their approach to handle this trade-off. We note that our perspective is different from previous surveys (i.g., \cite{hernandez2019survey}, \cite{9043893}), which studied the technical merits and demerits of the learning algorithms. Instead, we are interested in the \emph{systems} aspects of the considered works. That is, what are the communication topology and protocol used between agents and how do these choices affect the performance (rewards obtained by all agents). 
\paragraph{Overview}
\begin{figure}[!h]
\centering
	\includegraphics[scale=0.54]{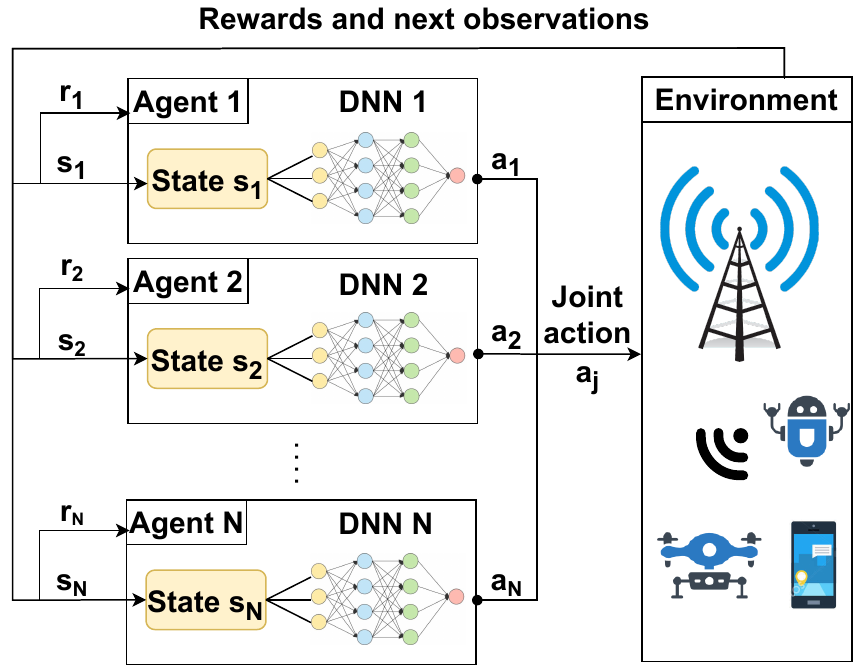}
	\caption{MARL Framework: multiple autonomous agents interact with an environment, observe (parts of) its state, and receive potentially different reward signals.}
	\label{fig:MARL}
\end{figure}

Unlike MAB formulations, in MDP, we have a \emph{state space}, which is a set of all possible states the environment might be in, along with a  \emph{transition operator} which describes the distribution over the next states given the current state and performed actions. Therefore, agents need not only to detect the best actions, which maximize the reward (bandits objective) but also to account for the possible next state, as it might be arbitrarily bad/good regardless of the current one. Hence, in MARL, the collaborative agents aim to maximize the current and \emph{future} expected sum of rewards, where the expectation is with respect to both the randomness of the state transitions and the action selection.

MARL problems, visualized in Fig. \ref{fig:MARL}, are often modeled as a Partially Observable Markov Game (POMG)\cite{POMG}, which is a multi-agent extension for Partially Observable Markov Decision Process (POMDP). POMGs are represented by the tuple $(\mathcal{N},\mathcal{S},\mathcal{A},\mathcal{O},\mathcal{T},\mathcal{R}, \gamma)$, where:
\begin{itemize}
    \item $\mathcal{N}$ is the set of all agents.
    \item  $s_t \in \mathcal{S}$ is a possible configuration of all the agents at time $t$.
    \item $a_t \in \mathcal{A}$ is a possible action vector for the agents, where $\mathcal{A} = \mathcal{A}_1 \times \mathcal{A}_2 \times ....\times \mathcal{A}_N$.
    \item $o_t \in \mathcal{O}$ is a possible observation of the agents, where $\mathcal{O} = \mathcal{O}_1 \times \mathcal{O}_2 \times .... \times \mathcal{O}_N$.
    \item $\mathcal{T}: \mathcal{O} \times \mathcal{A} \mapsto \mathcal{O}$ is the state transition probability.
    \item $\mathcal{R}$ is the set of rewards for all the agents $r_i: \mathcal{O} \times \mathcal{A} \mapsto \mathbb{R}$.
    \item $\gamma$ is the reward discount factor.
\end{itemize}
Each agent aims to find a policy $\pi_n$ that maximizes its own reward. If the rewards are not the same for all agents, the framework is referred to as mixed Decentralized-POMDP. When the rewards are similar for all agents (i.e., $r_n=r \quad \forall n \in \mathcal{N}$), the POMG is collaborative, and each agent's policy aims at maximizing the total reward. In the following, we discuss algorithms that might work on one or both settings. The main focus will be, on the communication aspects (i.e., topology and cost) of MARL algorithms. 

There exist results on the hardness of solving the POMG under several settings. We can cite the case of a tabular representation of the spaces and the cases where function approximation is used (linear or nonlinear). The main solution approaches are policy gradient and value-based methods \cite{sutton2018reinforcement}.
Policy gradient methods parametrize agents' policies within a class of functions and utilize gradient descent to optimize an objective function (i.e., reward obtained by the policy). Value-based methods aim to generalize the famous Q-learning algorithm to the multi-agent settings, either through making each agent learn its own Q-function and treating others as a part of a non-stationary environment, or through learning a global Q-function.

The optimization in policy gradient methods is done on the objective function: $J(\mathbf{\theta}\doteq v_{\pi_{\mathbb{\theta}}}(s_0))$, which is the cost of starting from the initial state $s_0$, and following the parametrized policy $\pi_{\mathbf{\theta}}$ thereafter. The gradient of this function can be written as:
\begin{equation}
    (G_t - b(S_t)) \frac{\nabla \pi (A_t|S_t, \mathbb{\theta}_t)} {\pi (A_t|S_t, \mathbb{\theta}_t)}.
\end{equation}
As shown in the policy gradient algorithm \cite{sutton2018reinforcement}, $b$ is any function of the state, referred to as the baseline. If it is the zero function, the resulting equation represents the ``reinforce'' algorithm.  Another popular option is the value function of the state. If this state value function is updated through bootstrapping, the resulting method is called Actor-critic. Thus, Actor-critic methods are policy gradients that use the state value function as a baseline ($b(s)=V(s)$) and update this function through bootstrapping.
Readers may refer to \cite{8103164} for more details and comparison between these approaches. As will be clarified next, each work tunes different parts of these main solution approaches according to the application. In the sequel, we present a classification of MARL literature from pervasive computing perspective. 


\paragraph{Centralized training and Decentralized Execution (CTDE)}
The Centralized Training and Decentralized Execution (CTDE) approach is first proposed in \cite{oliehoek2008optimal}. This approach leverages the assumption that, in most application scenarios, the initial training is done on centralized simulators, where agents can communicate with each other with no communication cost. This phase is denoted as centralized training. Then, at deployment, agents are assumed not to communicate at all, or conduct limited communication with each other, and they rely on their ``experience" at the training phase to actually execute their collaborative policies.

\textbullet{ Communication only at training:}
The advantage of such an approach is that it does not require communication between agents upon deployment and thus incurs no communication cost. However, this comes at the cost of losing adaptability, which is the major motivation behind online learning. Such loss might occur in case of a major shift in the environment model between the training and deployment, where the learned coordinated policies are no longer performant, and new coordination is needed. The main workaround is to monitor the agents' performance and re-initiating the centralized training phase to learn new coordinated policies whenever needed.

This approach has been popularized by recent methods such as VDN \cite{sunehag2018value}, QMIX \cite{rashid2018qmix}, and QTRAN \cite{son2019qtran}. These works adopt \emph{value function factorization} technique, where factorizations of the global value function in terms of individual (i.e., depending only on local observations) value function are learned during centralized training. Then, the global function (i.e., neural network) can be discarded at execution time, and each individual agent utilizes only the local function. When each agent acts greedily according to its local network, the global optimality can still be guaranteed since, at the training phase, these local networks were trained according to gradient signals with respect to the global reward.

Another approach to solving POMG is actor-critic. The CTDE version of actor-critic approaches is represented by learning a centralized critic, which evaluates the global action, and decentralized policy network, which takes an action based only on local observation. During training, the actor-critic networks are jointly learned, and hence the global critic ``guides" the training of the actors. Then, at execution, the global critic may be discarded, and only actors can be used. The works in \cite{lowe2017multi} present a deep deterministic policy gradient method that follows the described approach, where each agent learns a centralized critic and decentralized actor. Similarly, \cite{foerster2018counterfactual} follows the same approach, but all agents learn the same critic. Multiple other variations are done on the same DDPG algorithm aiming to either enhance performance \cite{iqbal2019actor} through incorporating an attention-mechanism, or reducing the use of communication resources (limited budget on the number of messages, or designing the message as (a part of) an agent's state). \cite{wang2020r}.

\begin{table*}
\centering
\footnotesize
\tabcolsep=0.09cm
\caption{ Communication-Cognizent Multi-Agent Reinforcement Learning literature.}
\label{table:rl_previous_work}
\begin{tabular}{|c|c|c|c|c|c|}
\hline
\textbf{Refs}  & \textbf{Framework} &	\shortstack{\textbf{Learning} \\ \textbf{algorithm}} &    \shortstack{\textbf{Communication}\\\textbf{scheme}}	& \shortstack{\textbf{Communication}\\\textbf{decision}}  \\  \hline

\shortstack{VDN \cite{sunehag2018value}, QMIX \cite{rashid2018qmix},\\ QTRAN \cite{son2019qtran}} & CTDE & Value-based &NA& \shortstack{Always during training,\\ None at execution.} \\  \hline

\shortstack{MADDPG \cite{lowe2017multi}, \\ COMA\cite{foerster2018counterfactual}} & CTDE & Actor-critic-based &NA& \shortstack{Always during training, \\ None at execution.}  \\  \hline

\cite{wang_learning_2020} IMAC & CTDE with learned  comm. & Policy gradient & \shortstack{Learned source \\ and destination} &\shortstack{At every step (limited size)}  \\  \hline

\cite{jiang2018learning} ATOC & CTDE with learned  comm. & Actor-critic based & \shortstack{Gated communication \\with neighbors} &\shortstack{When network topology \\changes.}  \\  \hline

\cite{singh2018learning} IC3Net  & CTDE with learned  comm. & Policy gradient & \shortstack{Gated communication \\with neighbors} &\shortstack{ Communicate when necessary, \\possibly many messages per round. } \\  \hline

\cite{mao_learning_2020} ACML &CTDE with learned  comm. & Actor-critic based  &  \shortstack{Gated communication \\with neighbors} & \shortstack{Communicate when necessary, \\respecting a limited bandwidth.}   \\  \hline

\cite{omidshafiei2017deep} & Fully decentralized & Value-based & Indirect & No message passing.  \\  \hline

\cite{zhang_fully_2018}& Fully decentralized & actor-critic-based & With neighbors & At every step.   \\  \hline

\cite{9029257}& Fully decentralized & actor-critic-based & With neighbors &  At every step (limited size).  \\  \hline

\cite{chen2018communication}& Fully decentralized & Policy gradient & \shortstack{Broadcast to all through \\central controller} & At every step.   \\  \hline

\end{tabular}
\end{table*}
\textbullet{ Learned communication:}
An important line of work within the MARL community is the study of learned communication between agents. In these settings, agents are allowed to send arbitrary bits through a communication channel to their peers in order to convey useful information for collaboration. These agents need to learn \emph{what} to send, and \emph{how} to interpret the received messages so that they inform each other of action selection. Thus, the agents are effectively learning communication protocols, which is a difficult task \cite{foerster2016learning}.

While the learned communication can be trained centrally and executed in a decentralized fashion, agents can still communicate at the execution phase through a limited bandwidth channel. Hence, we distinguish this setting from the works discussed in the previous subsection. Yet, similar approaches can be followed. For example, discarding the critic in execution (sometime used interchangeably with the term CTDE) but still maintaining the learned communication \cite{mao_learning_2020} and parameter sharing and gradient pushing in \cite{foerster2016learning}, where in execution, these messages are discretized.

Within the learned communication line of work, the authors in \cite{kim2019learning} aimed to learn to schedule communication between agents in a wireless environment and focused only on collision avoidance mechanism in the wireless environment. In \cite{wang_learning_2020}, information theoretic approach was used to compress the content of the message. In addition, source and destination are also learned through a scheduler. On the other hand, a popular line of work targeted the design of the so-called \emph{gating mechanism} techniques in order to accomplish the efficiency of the learned communication protocols. In this line of work, agents train a gating network, which generates a binary action to specify whether the agent should communicate with others or not at a given time step, limiting the number of communicated bits/messages needed to realize a certain desirable behavior. \cite{jiang2018learning} investigates the adaptability of these communication protocols and demonstrates the importance of communicating only with selected groups. Specifically, agents cannot distinguish messages that are particularly important to them (i.e., have implications on their actions) from the messages of all other agents. Thus, they introduce an attention scheme within each agent where an attention unit receives encoded local observation and action intention of the agent to decide whether a communication with others in its observable field is needed. The communication group dynamically changes and persists only when necessary.


The authors in \cite{singh2018learning} looked at \textit{communication at scale} and proposed an Individualized Controlled Continuous Communication Model (IC3Net), where agents are trained according to their own rewards (hence the approach can work for competitive scenarios also). Then, they demonstrated that their designed gating mechanism allows agents to block their communication, which is useful in competitive scenarios and reduces communication in cooperative scenarios by opting out from sending unnecessary messages. However, the effect of the gate on communication efficiency was not thoroughly studied, and the focus was instead on the emerging behavior. The work in \cite{mao_learning_2020} presents the state-of-the-art on efficient learned communication. The authors introduced Actor-Critic Message Learner (ACML), wherein the gate adaptively prunes less beneficial messages. To quantify the benefit of an action, Gated-ACML adopts a global Q-value difference as well as a specially designed threshold. Then, it applies the gating value to prune the messages, which do not hold values. The authors showed that surprisingly, not only the communication-efficiency significantly increases, but in specific scenarios, even the performance improves as a result of well-tuned communication. The reason behind this is that, since the communication protocol is learned, it is probable to hold redundant information that agents do not decode successfully. The proposed gating mechanism can also be integrated with several other learned communication methods.

\paragraph{Fully Decentralized Agents}
In fully decentralized reinforcement learning, there is no distinction between training and testing environments. Thus, the communication model stays the same throughout the agents' interaction with the environment. Under these settings, we recognize two extreme cases. First, agents do not communicate with each other, and learn to coordinate solely through the obtained rewards. In the case of no communication, the major challenge faced by the agents is the non-stationarity of the environment. A non-stationary environment from the perspective of the agents is when the distribution of the next states varies for the same current state and action pairs. The fully decentralized DRL was recently popularized by \cite{omidshafiei2017deep}. In \cite{omidshafiei2017deep}, the authors proposed a 3-dimensional reply buffer whose axes are the episode index, timestep index, and agent index. It was illustrated that conditioning on data from that buffer helps agents to minimize the effect of the perceived non-stationarity of the environment.

On the other extreme, agents can be modeled to be able to communicate at every step. Specifically, the problem of graph networked agents is investigated in \cite{zhang_fully_2018}. In this paper, agents are connected via a time-varying and possibly sparse communication graph. The policy of each agent takes actions that are based on the local observation and the neighbors' messages to maximize the globally averaged return. The authors fully decentralized actor-critic algorithms and provided convergence guarantees when the value functions are approximated by linear functions. However, a possible disadvantage of this algorithm is that the full parameter vector of the value function is required to be transmitted at each step. This has been addressed in \cite{9029257}, where also graph-networked agents are assumed, but each agent broadcasts only one (scaled) entry of its estimate of parameters. This significantly reduces communication cost (given that it occurs at every iteration). The paper also does not assume a bidirectional communication matrix and deals with only unidirectional ones, which is a more general formulation that appeals to more applications. The decentralized actor-critic-based algorithm also solves the distributed reinforcement learning problem for strongly connected graphs with linear value function approximation.

\cite{chen2018communication} considered the communication efficiency in fully decentralized agents, but with the assumption of a centralized controller. The paper utilizes policy gradient solution methods, where the controller aggregates the gradients of the agents to update the policy parameters. This process is akin to federated learning clients selection. The authors propose a process to determine which clients should communicate to the controller based on the amount of progress in their local optimization. They also propose a methodology to quantify the importance of local gradient (i.e., the local optimization progress) and then only involve agents who are above a certain threshold. Following this approach, the authors showed that the performance (i.e., cumulative reward) is similar to the case where all clients are participating, with considerable communication round savings.

Table \ref{table:rl_previous_work} summarises the works discussed above according to their communication model and the approach in handling the communication-performance tradeoff. We first identify the framework (CTDE, CTDE with learned communication, or fully decentralized) as well as the learning framework (value, policy gradient, or actor-critic). Note that these MARL algorithmic frameworks, which are based on a single-agent variant of the problem, involve learning the state space transition operator as it plays a major role in estimating the future expected sum of rewards. 
Then, we list two important configurations. First, the communication scheme, which states \emph{how} agents communicate with each other. In CTDE, the training is done in simulation. Thus, agents are logically centralized and do not communicate. If no messages are passed between agents and their collaboration is solely learned through rewards, then the communication scheme is \textit{indirect}. Otherwise, it is either \textit{gated with neighbors} directly or through a \textit{central controller}. Lastly, the communication decision states \emph {when} the communication is made, which can be at every step (with optimized message size or not), or according to other conditions as detailed in the discussion.

\paragraph{MDP MARL for UAV-assisted networks}
\begin{figure}[h!]
	\centering
		\scalebox{1.8}{\includegraphics[width=0.46\columnwidth]{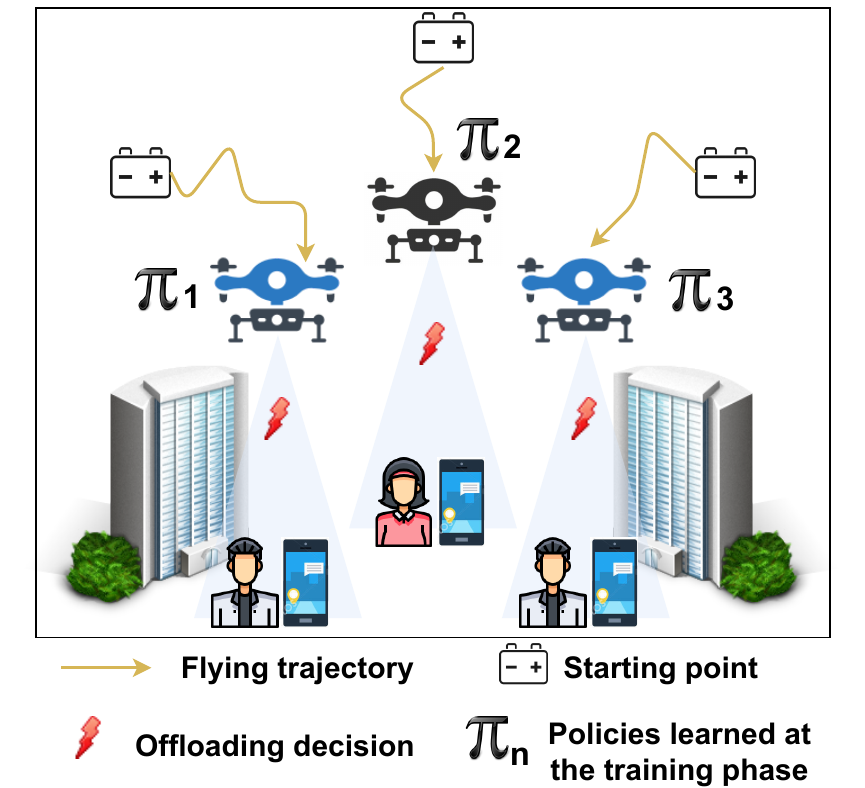}}
	\caption{UAV-assisted networks: UAV agents are trained to deduce a collaborative policies for providing compute/communication resources for on ground equipment.}
	\label{fig:uav_MARL}
\end{figure}
UAVs have provided new potentials for extending communication coverage and even compute capabilities of devices in areas where full networking infrastructure is not present. This is done through wireless communication between UAVs and on-ground equipment, enabling those equipment to extend their connectivity and potentially offload tasks to a broader network relayed by the flying devices.
The work in \cite{9209079} aims at utilizing UAVs to provide intermittent compute/communication resource support for user equipments (UEs) on the ground. The benefits of such UAV-assisted scenarios are numerous, including creating dynamic networking graphs without the need for full networking infrastructure, which can be of extreme value in catastrophes response for example. Nonetheless, the UAVs need to optimize their trajectory paths so that they cover the widest areas with minimum movement (i.e., energy consumption) and maximum utility (i.e., providing the resources to the UE that needs it the most). However, such optimization is shown to be intractable. Thus, the authors opted for learning-assisted (i.e., data-driven) methods. Since a centralized training was possible in their tackled scenario, they used a CTDE algorithm, specifically Multi-Agent DDPG (MADDPG). In MADDPG, agents aim to jointly optimize the UE-load of each UAV while minimizing the overall energy consumption experienced by UEs, that depends on the UAV's trajectory and offloading decisions. Following the MADDPG algorithm, the UAVs observations were \textit{communicated} among them to deduce the collaborative policy at training. At execution, no message sharing was needed. This resulted in a satisfactory performance due to the accurate simulator. However, as discussed earlier, environments that are expected to change require the use of other algorithms that maintain periodic, learned, or persistent communication after deployment.

We note that the application of reinforcement learning in resource-constrained environments (e,g., IoT devices), requiring the design of communication-aware techniques, is still scarce. Most testing for these methods is done on artificial testing environments like Open AI's Particle environments \cite{openaimultiagent-particle-envs_2021}, or video games like StarCraft II \cite{vinyals2019grandmaster}, which is a typical practice in the RL community since success in these environments is often indicative of broader applicability.
\paragraph{Lessons learned}
\begin{itemize}
    \item  Most of the MDP MARL works focused on performance gains and benchmarking, with little attention to resource utilization. This is because MARL is applied for games and other areas (e.g., robotics) where the priority is for performance. IoT applications, where resource utilization plays major role, is yet to make full use of state-of-the-art MARL algorithms.
    \item \textit{CTDE-a practical middle ground}: We note that CTDE is the most adopted in pervasive/IoT scenarios. We attribute this to the simplicity in the way agents communicate in this framework. Specifically, CTDE algorithm leverages the fact that training is often done in simulators, where there is no communication cost, and agents may share experience tuples, network parameters, and observations freely, in order to train policies that can be executed later on, based on only local observations. This approach seems to model most of the pervasive computing applications where agents do not need to start training while being decentralized. In this framework, the actor-critic-based algorithms are more popular, where a centralized critic network that uses the observations of all agents guides the training of a decentralized policy network that uses only the local observations. The critic network can be discarded at execution time, enabling decentralized execution. The framework is emerging as a possible alternative to the fully decentralized extremes, where agents communicate at every step or do not communicate at all and try to indirectly and independently learn collaborative policies \cite{lowe2017multi,wang2020r}. 
    
    
    \item \textit{Scheduling for efficient learned communication}: In learned communication, agents learn to encode and decode useful messages. In this area, gating mechanisms are the main tools towards efficient communication \cite{jiang2018learning,singh2018learning,mao_learning_2020}. In gate design, agents learn when to send and refrain from sending a messages by quantifying the benefit (i.e., reward) of actions following this communication. More general \emph{schedulers} modules investigate the design of communication module that learn to minimize the content of the messages as well (i.e., compressing the communication messages) \cite{wang_learning_2020}. Overall, scheduling mechanisms are being increasingly used in MARL settings with learned communication, in order to face the limited bandwidth problems often encountered in practical scenarios.
    
\end{itemize}


{ 
\subsection{Active Learning (AL)} \label{AL}

As far as pervasive training schemes have been tackled, AL has emerged as a promising and effective concept. Herein, we first present a brief overview for the concept of AL, then we discuss some recent applications of AL presented in the literature.  

\paragraph{Overview} \label{AL_Overview}

The main idea behind AL is that an active learner is allowed to actively select over time  the most informative data to be added to its training dataset in order to enhance its learning goals \cite{AL2017}, \cite{AL2014}. 
Hence, in AL framework, the training dataset is not static, which means that the training dataset and learning model are progressively updated in order to continuously promote the learning quality  \cite{ASPAL2018}.   
Specifically, the main steps of AL are: (1) acquiring new data from the contiguous nodes; (2) picking out the most informative data to append to the training dataset; (3) retraining the learning model using newly-acquired data. Hence, the communication overheads associated with different AL schemes will depend on:
\begin{itemize}
	\item Type and amount of exchanged data between the contiguous nodes. We remark here that contiguous nodes can exchange labels, features, or samples. Hence, based on the type and amount of changed data there will be always a tradeoff between enhancing the performance and decreasing communication overheads. 
	\item Number of selected nodes that will be considered in the AL process.  
\end{itemize}

It is worth mentioning also that FL allows multiple nodes to cooperatively train a global model without sharing their local data, which differs from AL in many ways. In particular, FL seeks for obtaining a synchronization between different cooperative nodes, in addition to the presence of a centralized node (or server) to generate the global model. Thus, AL and FL are addressing orthogonal problems – the former leverages the newly-acquired data from the contiguous nodes to retrain its model, while the latter trains its model in a distributed manner by sharing the model's updates with the contiguous nodes \cite{infocom2020}.  

\paragraph{Applications of AL} \label{AL_Applications}

Traditionally, AL algorithms depend on the presence of an accurate classifier that generates the ground-truth labels for unlabeled data. However, this  assumption becomes hard to maintain in several real-time applications, such as crowdsourcing applications and  automated vehicles. Specifically, in crowdsourcing, many sources   are typically weak labelers, which may generate noisy data, i.e., data that may be affected by errors due to low resolution and age of information problems.  
However, most of the existing studies on AL investigate the noisy data  (or imperfect labels) effect on the  binary classification problems \cite{niosyBinaryClass2016,niosyBinaryClass2015}, while few works consider the general problem of multi-class or multi-labeled data  \cite{DeepLearningFromCrowds,Imbalanced_Label2015,Subset_selection2019}.  

One of the main problems in crowdsourcing is how to collect large amount of labeled data with high quality, given that the labeling can be done by volunteers or non-expert labelers. Hence, the process of acquiring large amount of labeled data turned to be challenging, computationally demanding, resource hungry, and often redundant. Moreover, crowdsourced data with cheap labels comes with its own problems. Despite being labels cheap, it is still expensive to handle the problem of noisy labels. 
Thus, when data/labelers are not selected carefully, the acquired data may be very noisy \cite{Crowds2018, WMV2017}, due to many reasons such as varying degrees of competence, labelers biases, and disingenuous behavior, which significantly affects the performance of supervised learning. 
Such challenges have encouraged the researcher to design innovative schemes that can enhance the quality of the acquired data from different labelers. 
For instance, \cite{DeepLearningFromCrowds}  tackles the problem of demanding deep learning techniques to large datasets by presenting an AL-based solution that leverages multiple freely accessible crowdsourced geographic data to increase datasets' size. However, in order to effectively deal with the noisy labels extracted from these data and avoid performance degradation, the authors have proposed a customized loss function that integrates multiple datasets by assigning different weights to the acquired data based on  the estimated noise.  
\cite{Subset_selection2019} enhances the performance of supervised learning with noisy labels in crowdsourcing systems by introducing a simple quality metric and selecting the $\epsilon$-optimal labeled data samples. The authors investigate the data subset selection problem based on the Probably Approximately Correct (PAC) learning model. Then, they consider the majority voting label integration method and propose two data selection algorithms that optimally select a subset of $k$ samples with high labelling quality. 
In \cite{Imbalanced_Label2015}, the authors investigate the problem of   imbalanced noisy data, where the acquired labeled data are not uniformly distributed across different classes. 
The authors therein aim to label training data given received noisy labels from diverse sources. Then, they used their learning model to predict the labels for new unlabeled data,  and update their learning model until some conditions are met (e.g., the performance of the learned model meets a predefined requirement, or it cannot be improved any more). Specifically, for labeled data, they implemented a label integration and data selection scheme that considers data uncertainty and class imbalance level, while  classifying the unlabeled data using the trained model before adding them to the training dataset. Hence, the proposed framework presents two core procedures: label integration and sample selection. In the label integration procedure, a Positive LAbel Threshold (PLAT) algorithm is used to infer the correct label from the received noisy labels of each sample in the training set. After that, three sample selection schemes are proposed to enhance the learning performance. These schemes are respectively based on the uncertainty derived from the received-noisy labels, the uncertainty derived from the learned model, and the combination method.  
%

A different application of AL is investigated in \cite{ASPAL2018}, where AL is exploited for  incremental face identification. Conventional incremental face recognition approaches,  such as incremental subspace approaches, have limited performance on complex and large-scale environment. Typically, the performance may drastically drop when the training data of face images is either noisy or insufficient. Moreover, most of existing incremental methods suffer from noisy data or outliers when updating the learning model. Hence, the authors in \cite{ASPAL2018} present an active self-paced learning framework, which combines: active learning and Self-Paced Learning (SPL). The latter refers to a recently developed learning approach that mimics the learning process of humans by gradually adding to the training set the easy to more complex data, where easy data is the one with high classification confidence. 
In particular, this study aims to solve the incremental face identification problem by building a classifier that progressively selects and labels the most informative samples in an active self-paced way, then adds them to the training set. 

AL has been also considered in various applications of intelligent transportation systems. For instance, the authors in \cite{VehicleRecognition2019} investigate the vehicle type recognition problem, in which labeling a sufficient amount of data in surveillance images is very time consuming. To tackle this problem, this work leveraged fully labeled web data to decrease the required labeling time of surveillance images using deep transfer learning. Then, the  unlabeled images with  high uncertainty are selected to be queried in order to be added later to the training set. Indeed, the cross-domain similarity metric is linearly combined with the entropy in the objective function of the query criteria to actively select the best samples. 
Ultimately, we highlight that most of the presented studies so far consider in their AL framework specific classifiers (or learning models), which cannot be easily used in other learning models \cite{multiclass2018}. Accordingly, obtaining an optimal label integration and data selection strategy that can be used with a generic  multi-class classification techniques is still worth further investigation.}



{

\paragraph{Use case: AL for Connected Vehicles}
\begin{figure}[h!]
	\centering
		\scalebox{1.9}{\includegraphics[width=0.49\columnwidth]{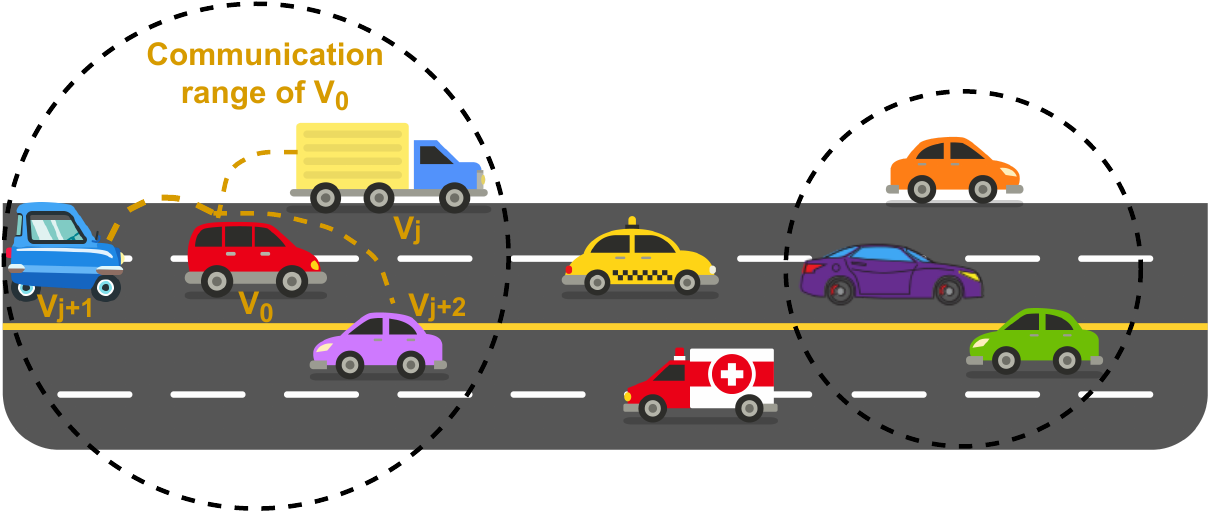}}
	\caption{AL for time-varying vehicular network.}
	\label{fig:CS}
\end{figure}
Traditional machine learning models require massive, accurately labeled datasets for training in order to ensure high classification accuracy for new data as it arrives \cite{dlcomp2018}. This assumption cannot be guaranteed in many real-time applications, such as connected and autonomous vehicles. Indeed, vehicles are typically weak labelers (i.e., data sources that generate label with low classification confidence). Hence, they may acquire/generate noisy data, e.g., data generated by cameras in the presence of fog or rain. Also, in a highly dynamic environment like vehicular network, not only the generated data by the vehicles' classifiers can have low classification accuracy, but also the data received from neighboring vehicles may be prone to noise and communication errors. Hence, the authors in \cite{TVT2021} have tackled these challenges by proposing a cooperative AL framework for connected vehicles. The main goal of this work is two-fold: (1) selecting the optimal set of labelers, those considered to be the most reliable ones; (2) selecting a maximally diverse subset of high quality data that are locally-generated and/or received from neighboring vehicles to be used for updating the learning model at each vehicle. 

In \cite{TVT2021}, the time-varying vehicular network shown in Figure \ref{fig:CS} is considered. 
It is assumed that each vehicle can communicate and exchange information only with the neighboring vehicles that are located within its communication range. For instance, the set $\mathcal{N}_{v_0}(t) = \left\{v_{j}, v_{j+1}, v_{j+2}\right\}$ at time $t$ means that there are only three vehicles staying in the communication range of vehicle $v_0$.   
Furthermore, this framework considers two types of data:  multiple-labeled online dataset and offline/historical labeled dataset.   
The online dataset is considered as sequences of samples that arrive from neighboring  vehicles or generated at vehicle $v_0$ within time $T$ (i.e., refers to the period of time during which a vehicle $v_0$ is exposed to a certain road view). At time $T$, vehicle $v_0$ receives a sequence of training samples/labels that contains input features and associates with multiple noisy labels generated from the vehicles sending data to $v_0$. 
The presented framework in \cite{TVT2021} includes five main stages, as described below: 
\begin{enumerate}
	\item \textbf{Offline Learning:} Initially, each vehicle with its own offline/historical training data generates an initial learning model with a certain accuracy level. 
	\item \textbf{Online labeling:} The vehicle starts to collect new information  through its local sensors or from neighboring vehicles. These information can be labels, features, or samples, depending on the adopted operational mode. 
	\item \textbf{Label integration:} After acquiring the new information, each vehicle obtains an aggregated label for the received data using different proposed label integration strategies. 
	\item \textbf{Labeler selection:} After monitoring the behavior of the neighboring vehicles, each vehicle selects a subset of high-quality labelers, based on their reputation values that are estimated from the past interactions using subjective logic model.  
	\item \textbf{Data selection and models update:} Finally, each vehicle selects the maximally diverse collection of high-quality samples to update its learning model.    
\end{enumerate}

The proposed AL framework in \cite{TVT2021} depicts its efficiency for connected automated vehicles, as follows: 
(1) it allows to increase the amount of acquired data at different vehicles during the training phase; 
(2) it accounts for the labelers' accuracy, data freshness, and  data diversity while selecting the optimal subset of labelers and data to be included in the training set; 
(3) Using different real-world datasets, it could provide $5-10\%$ increase in the classification accuracy compared to the state-of-the-art approaches that consider random data selection, while enhancing the classification accuracy by $6\%$ compared to random labelers selection approaches. 
}

 \section{Pervasive Inference}\label{PI}
\begin{figure}[!h]
\centering
	\includegraphics[scale=0.37]{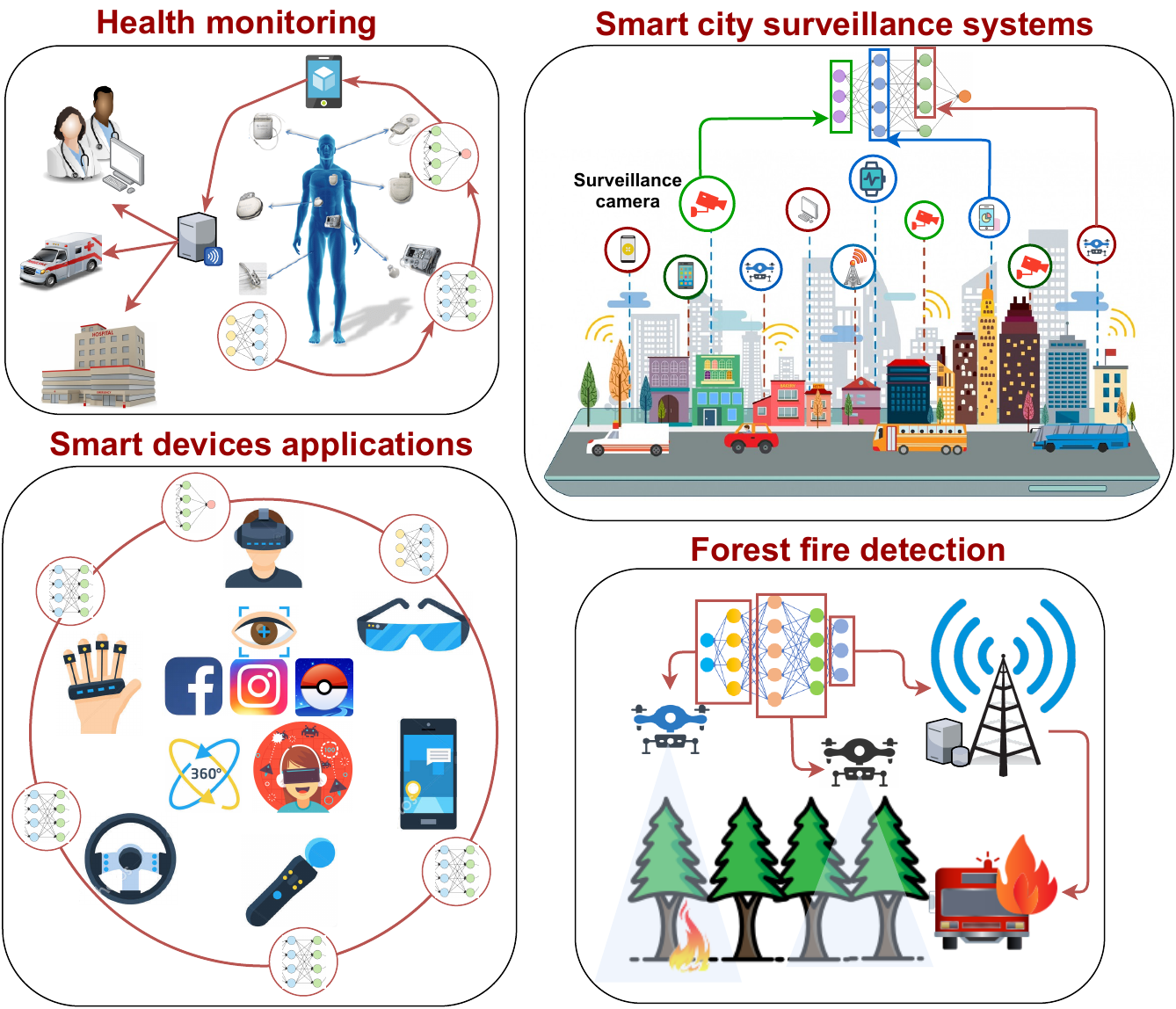}
	\caption{Pervasive inference system in multiple scenarios. }
	\label{pervasive_AI}
\end{figure}

In this section, we discuss the pervasive inference, where the trained model is partitioned and different segments are distributed among ubiquitous devices. It is worth mentioning that the training method of the distributed model is out of the scope of this section. Fig. \ref{pervasive_AI} illustrates different scenarios, where the distribution can solve the challenges presented by the centralized approaches. In the following subsections, the communication and computation components of the pervasive inference are introduced. Then, the resource management approaches for the distribution are reviewed and a use cases is described.
\subsection{Profiling computation and communication models}\label{profiling}
The computation and communication models present the mechanisms to formulate different operations and functions into an optimization problem in order to facilitate the theoretical analysis of DNN distribution. More specifically, we discuss the computational requirements of different DNN tasks, the wireless communication latency between different pervasive participants and their energy consumption.
\subsubsection{Computation models}
Various parameters play a critical role to model the computational tasks of different segments of the DNN network including latency, generality, scalability and context awareness. In this subsection, we describe the computation models of two popular splitting strategies adopted in the literature, which are the per-layer and per-segment splitting. These models are presented after introducing some definitions.
\paragraph{Overview and definitions}\mbox{}\\
\textbf{\indent Binary offloading}:  Relatively simple or highly complex tasks that cannot be divided into sub-tasks and have to be computed as a whole, either locally at the source devices or sent to the remote servers because of resource constraints, are called binary offloading tasks. These tasks can be denoted by the three-tuple notation $T(K,\tau, c)$. This commonly used notation illustrates the size of the data to be classified presented by $K$ and the constraint $\tau$ (e.g., completion deadline, the maximum energy, or the required accuracy). The computational load to execute the input data of the DNN task is modeled as a variable $c$, typically defined as the number of multiplications per second \cite{b3}.
Although binary offloading has been widely studied in the literature, we note that it is out of the scope of this survey covering the pervasivity and distribution of AI tasks.
\newline
\textbf{ \indent Partial offloading}: In practice, DNN classification is composed of multiple subtasks (e.g., layers execution, multiplication tasks, and feature maps creation), which allows to implement fine-grained (partial) computations. More specifically, the AI task can be split into two or more segments, where the first one can be computed at the source device and the others are offloaded to pervasive participants (either remote servers or neighboring devices). 
\begin{figure}[h!]
\centering
	\includegraphics[scale=0.6]{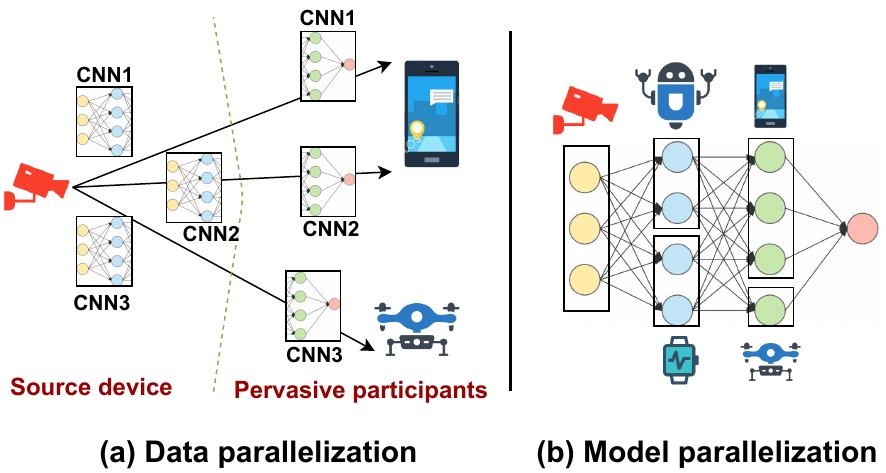}
	\caption{Inference parallelization: data and model parallelization}
	\label{parallelization}
\end{figure}

\textbf{Data parallelization:} The most manageable task of partial offloading is the data parallelization, where duplicated offloaded segments are independent and can be arbitrarily divided into different groups and executed by different participants of the pervasive computing system, e.g., segments from different classification requests (as shown in Fig. \ref{parallelization} (a)). We highlight that the input data to  parallel segments are independent and can be different or akin.

\textbf{Model parallelization:} A more sophisticated partial offloading pattern is the model parallelization, where the execution of one task is split across multiple pervasive devices. Accordingly, the input data is also split and fed to different parallel segments. Then, their outputs are merged again. In this offloading pattern, the dependency between different tasks cannot be ignored as it affects the execution of the inference. Particularly, the computation order of different tasks (e.g., layers) cannot be determined arbitrarily because the outputs of some segments serve as the inputs of others (as shown in Fig. \ref{parallelization} (b)). In this context, the inter-dependency between different computational parts of the DNN model needs to be defined. It is worth mentioning that many definitions of data and model parallelism are presented in the literature, which are slightly different. In our paper, we opted for the definitions presented in \cite{robots}.

\begin{figure}[!h]
\centering
	\includegraphics[scale=0.6]{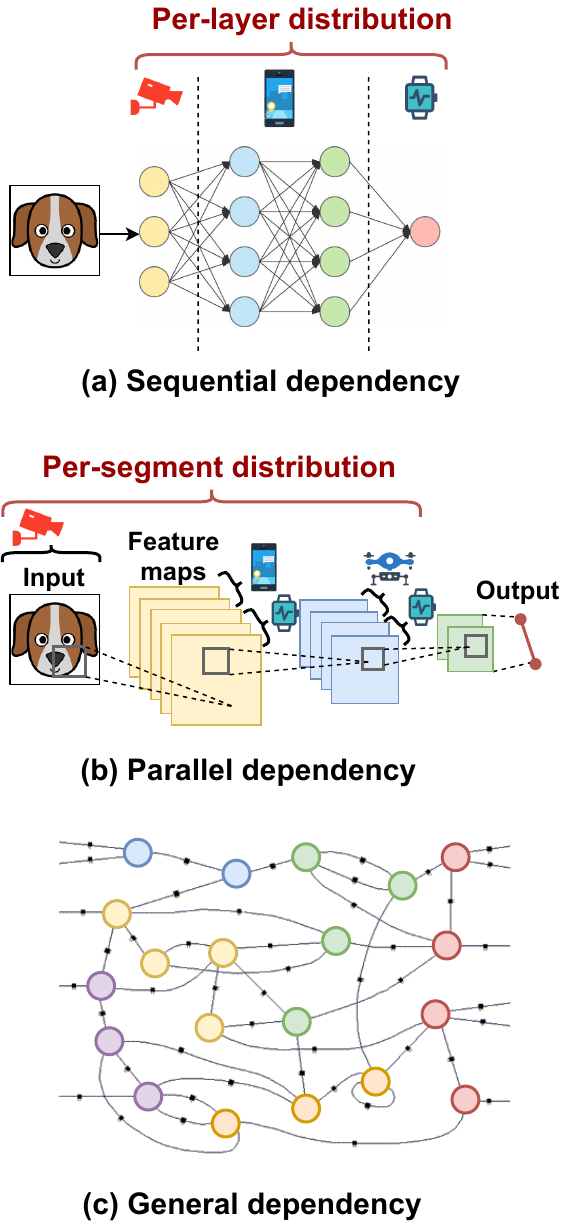}
	\caption{Typical typologies of DNNs and partitioning strategies.}
	\label{partitioning}
\end{figure}

\textbf{Typical dependencies:} Different DNN networks can be abstracted as task-call graphs.  These graphs are generally presented by Directed Acyclic Graphs (DAGs), which have a finite directed structure with no cycles. Each DNN graph is defined as $G(V,E)$, where the set of vertices $V$ presents different segments of the network, while the set of edges $E$ denotes their relations and dependencies. Typically, three types of dependencies contribute to determining the partition strategies, namely the sequential dependency which occurs in the conventional CNN networks with sequential layers and without any residual block (e.g., VGG \cite{VGG}), the parallel dependency which depicts the relation between different tasks in the same layer (e.g., different feature maps transformations), and the general dependency existing in general DNN models (e.g., randomly wired CNN \cite{Randomly_wired}). Different dependencies are depicted in Fig. \ref{partitioning}. The required computation workload and memory are specified for each vertex $V$ and the amount of the input and output data can be defined on the edges.

Based on the presented dependencies, two partition strategies can be introduced, namely per-layer and per-segment partitioning (see Fig. \ref{partitioning}). Per-layer partitioning defines dividing the model into layers and allocating each set of layers within a pervasive participant (e.g., IoT device or remote servers). On the other hand, per-segment partitioning denotes segmenting the DNN model into smaller tasks such as feature maps transformations, multiplication tasks and even per-neuron segmentation.

\textbf{Computation latency:}
The primary and most common engine of the pervasive devices to perform local computation is the CPU. The performance of the CPU is assessed by cycle frequency/ clock speed $f$ \cite{Survey33} or the multiplication speed $e$ \cite{CNNDist}. In the literature, authors adopt the multiplication speed to control the performance of the devices executing the deep inference. In practice, $e$ is bounded by a maximum value $e_{max}$ reflecting the limitation of the device computation capacity. Based on the model introduced for binary offloading, the computation latency of the inference task $T(K,\tau,c)$ is calculated as follows \cite{CNNDist}:
\begin{equation}
    \begin{aligned}
               t^c=\frac{c}{e}.
    \end{aligned}
    \label{eq:1}
\end{equation}
Importantly, a higher computational capacity $e_{max}$ is desirable to minimize the computation latency at the cost of energy consumption. As end-devices are energy constrained, the energy consumption of the local computation is considered as a key measure for evaluating the inference efficiency. More specifically, a high amount of energy consumed by AI applications is not desirable by end-devices due to their incurred cost. Similarly, significant energy consumption of edge nodes (e.g., access points or MEC servers.) increases the cost envisaged by the service providers.
\begin{table*}[]
\footnotesize
\tabcolsep=0.09cm
\caption{Characteristics of different splitting strategies:\\
{\footnotesize A: After, B: Before, $N_{Fc}$: number of fully connected layers, $n$: number of input neurons (Fc), $m$: number of output neurons (Fc), $H_1, W_1, D_1$: dimensions of the input data (Conv), $H_2, W_2, D_2$: dimensions of the output data (Conv), $H_f, W_f, D_1$: dimensions of the filter (Conv), $k/D_2$: number of filters, $d_x,d_y$: dimensions of the spatial splitting (Conv), $N$: Number of participants, $k'_i$: Number of segments per participant.}}
\label{tab:splitting}
\begin{tabular}{|c|c|c|c|c|c|c|c|c|}
\hline
\begin{tabular}[c]{@{}c@{}}\textbf{Partitioning}\\ \textbf{strategy}\end{tabular} & \begin{tabular}[c]{@{}c@{}}\textbf{$N^{o}$ of smallest} \\\textbf{segments}\end{tabular} & \begin{tabular}[c]{@{}c@{}}\textbf{Activation}\\ \textbf{task}\end{tabular} & \begin{tabular}[c]{@{}c@{}}\textbf{Inputs}\\ \textbf{per segment}\end{tabular} & \begin{tabular}[c]{@{}c@{}}\textbf{Filters weights}\\ \textbf{per device}\end{tabular} & \begin{tabular}[c]{@{}c@{}}\textbf{Outputs}\\ \textbf{per segment}\end{tabular} & \begin{tabular}[c]{@{}c@{}}\textbf{Computation}\\ \textbf{per segment}\end{tabular} &\begin{tabular}[c]{@{}c@{}}\textbf{Transmitted data}\\ \textbf{per layer}\end{tabular} & \begin{tabular}[c]{@{}c@{}}\textbf{Merging}\\ \textbf{strategy}\end{tabular} \\ \hline
\begin{tabular}[c]{@{}c@{}}Per-layer:\\ Fully-connected (Fc)\end{tabular} & $N_{Fc}$ & A & $n$ & \xmark & $m$& $n \times m$ &$n+m$ & Seq\\ \hline

\begin{tabular}[c]{@{}c@{}}Per-segment: Output\\ splitting for Fc layers\end{tabular} & $\sum\limits^{N_{Fc}}_{i=1} m_i$ & B/A & $n$ &  \xmark& 1 & $n$  & $n \times N +m$& Concat  \\ \hline

\begin{tabular}[c]{@{}c@{}}Per-segment: Input \\ splitting for Fc layers\end{tabular} & $\sum\limits_{i=1}^{N_{Fc}} n_i$& A & 1 & \xmark & $m$ & $m$  & $N \times m +n$& Sum \\ \hline

\begin{tabular}[c]{@{}c@{}}Per-layer:\\ Convolution (Conv)\end{tabular} & $N_{Conv}$ & A & $H_1 \times W_1 \times D_1$ & \begin{tabular}[c]{@{}c@{}}$k \times D_1 \times$ \\ $(H_f \times W_f)$\end{tabular}  & $H_2 \times W_2 \times k$ &  \begin{tabular}[c]{@{}c@{}} $cp= D_1 \times $\\ $(W_f \times H_f) \times$  \\ $ k \times (W_2 \times H_2)$  \end{tabular} & \begin{tabular}[c]{@{}c@{}} $H_1 \times W_1 \times D_1 +$\\ $H_2 \times W_2 \times k$\end{tabular} & Seq \\ \hline

\begin{tabular}[c]{@{}c@{}}Per-segment: channel\\ splitting for Conv\end{tabular} & \begin{tabular}[c]{@{}c@{}} $ \sum\limits_{i=1}^{N_{Conv}}k_i$ \end{tabular} & B/A & $H_1 \times W_1 \times D_1$ & \begin{tabular}[c]{@{}c@{}}$k'_i \times D_1 \times$ \\ $(H_f \times W_f)$\end{tabular} &  $H_2 \times W_2$ & $\frac{cp}{k}$ &\begin{tabular}[c]{@{}c@{}} $(N \times H_1 \times W_1 \times D_1)$ \\$+ (k \times H_2 \times W_2)$\end{tabular}  &Concat \\ \hline

\begin{tabular}[c]{@{}c@{}}Per-segment: spatial\\ splitting for Conv\end{tabular} & \begin{tabular}[c]{@{}c@{}}$\sum\limits_{i=1}^{ N_{Conv}}\frac{H_1^i\times W_1^i}{d^i_x \times d^i_y}$ \end{tabular}& B/A & \begin{tabular}[c]{@{}c@{}}$\frac{H_1 \times W_1 \times D_1}{d_x \times d_y} +$ \\ $padding$\end{tabular} & \begin{tabular}[c]{@{}c@{}}$k \times D_1 \times$ \\ $(H_f \times W_f)$\end{tabular} & $\frac{H_2 \times W_2 \times k}{d_x \times d_y}$ & $cp /\frac{H_1 \times W_1}{d_x\times d_y}$&\begin{tabular}[c]{@{}c@{}} $H_1 \times W_1 \times D_1 +$\\ $H_2 \times W_2 \times k+$ \\  $N\times padding$\end{tabular}& Concat \\ \hline

\begin{tabular}[c]{@{}c@{}}Per-segment: filter \\ splitting for Conv\end{tabular} & \begin{tabular}[c]{@{}c@{}} $\sum\limits_{i=1}^{N_{Conv}}D^i_1\times k_i$\end{tabular} & A & $H_1 \times W_1$ & \begin{tabular}[c]{@{}c@{}}$k'_i \times$ \\ $(H_f \times W_f)$\end{tabular} & $H_2 \times W_2$ &$ \frac{cp}{D_1 \times k}$ & \begin{tabular}[c]{@{}c@{}} $(D_1 \times H_1 \times W_1)+$ \\$ (N \times H_2 \times W_2 \times k)$\end{tabular}&  \begin{tabular}[c]{@{}c@{}}Sum+ \\concat \end{tabular}\\ \hline
\end{tabular}
\end{table*}

\textbf{Computation energy:}
If the inference is executed at the data source, the consumed energy is mainly associated to the task computation. In contrast, if the task is delegated to remote servers or to neighboring devices, the power consumption consists of the required energy to transfer the data between participants, the amount of energy consumed for the computation of different segments, and the energy required to await and receive the classification results. Suppose that the inference task/sub-task $T_i$ takes a time $t^c_i$ to be computed locally in the device participating in the pervasive inference and let $P_i$ denote the processing power to execute the task per second. The energy consumed to accomplish an inference task $T_i$ locally at the computing device is equal to \cite{energy-Aware-dist}:
\begin{equation}\label{eq:2}
    \begin{aligned}
              e^{local}_i= t^c_i \times P_i.
    \end{aligned}
\end{equation}

Next, we profile the DNN partitioning strategies presented in the literature, in terms of computation and memory requirements first and then in terms of communicated  data to offload the output of segments. The key idea of partitioning a DNN network is to evenly or unequally distributing the computational load and the data weights across pervasive devices intending to participate in the inference process, while minimizing the classification latency. A partitioning can be achieved by simply segmenting the model per-layer or set of layers (see Fig. \ref{parallelization} (a)) or by splitting the layers' tasks (see Fig. \ref{parallelization} (b)). Then, each part is mapped to a participant.
\paragraph{Per-layer splitting} As previously mentioned, the computational load of each layer is measured as the number of multiplications required to accomplish the layer's goal \cite{b1}.
\newline
\textbf{\indent Fully-connected layers:} The computation requirement of a fully-connected layer can be calculated as follows:
\begin{equation}\label{eq:4}
    \begin{aligned}
   c^{Fc}=n \times m,
    \end{aligned}
\end{equation}
where $n $ represents the number of the input neurons and $m$ is the number of the output neurons. 
\newline
\textbf{\indent Convolutional layers:} The computation load of a convolution layer can be formulated as follows \cite{b1}:
\begin{equation}\label{eq:5}
    \begin{aligned}
   c^{conv}=D_1 \times (W_f \times H_f) \times D_2 \times (W_2 \times H_2).
    \end{aligned}
\end{equation}
We remind that $D_1$ is the number of input channels of the convolutional layer which is equal to the number of feature maps generated by the previous layer, $(W_f \times H_f)$ denotes the spatial size of the layer’s filter, $D_2$ represents the number of filters and $(W_2 \times H_2)$ represents the spatial size of the output feature map (see Fig. \ref{Conv}).

\begin{figure}[h]
\centering
	\includegraphics[scale=0.27]{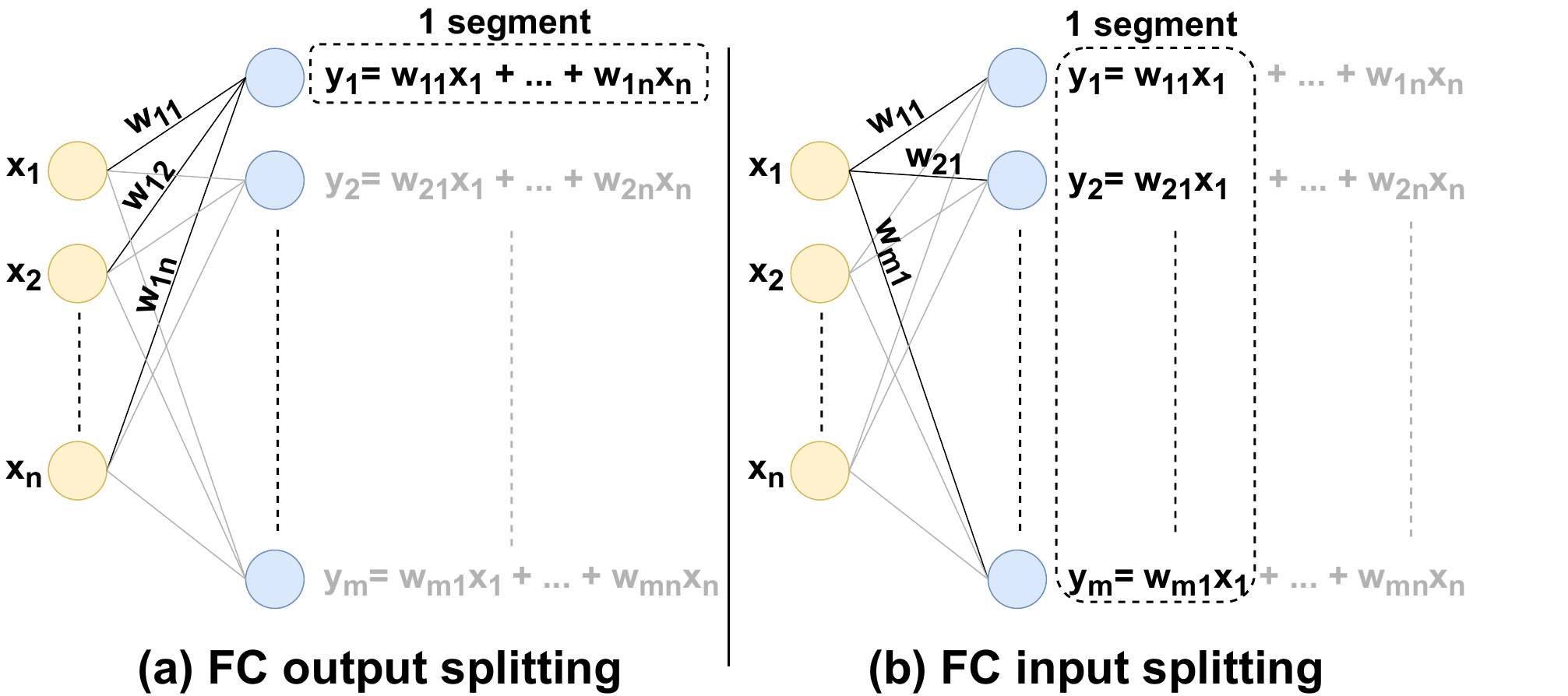}
	\caption{Partitioning of fully connected layers.}
	\label{FC}
\end{figure}

The computational load introduced by pooling and ReLU layers can be commonly neglected, as these layers do not require any multiplication task \cite{b1}. We highlight that the per-layer splitting is motivated by the sequential dependency between layers. This dependency does not permit the model parallelism nor the latency minimization. Instead, it allows the resource-constrained devices to participate in the AI inference.
\paragraph{Per-segment splitting} \mbox{}\\
\textbf{\indent Fully-connected layers:} We start by profiling the fully-connected layer partitioning. More specifically, the computations of different neurons $y_i$ in a fully-connected layer are independent. Hence, their executions can be distributed, and model parallelism can be applied to minimize the inference latency. Two methods are introduced in the literature (e.g., \cite{IoTInferencing,FullyDistribution}), which are the output and input partitioning as shown in Fig. \ref{FC}. 
\begin{itemize}
    \item \textit{Output splitting}: the computation of each neuron $y_i$ is performed in a single participant that receives all input data $\{x_1,x_2,...,x_n\}$, as highlighted in Fig. \ref{FC} (a). Later, when the computation of all neurons is done, results are merged by concatenating the output of all devices in the correct order. The activation function can be applied on each device or after the merging process. 
    \item \textit{Input splitting}: each participant computes a part of all output neurons. Fig. \ref{FC} (b) illustrates an example, where each device executes $\frac{1}{n}$ of the required multiplications. By adopting this partitioning method, only a part of the input, $x_i$, is fed to each participant. Subsequently, when all participants accomplish their tasks, summations are performed to build the output neurons. However, in contrast to the output-splitting method, the activation function can only be applied after the merging process. 
\end{itemize}
\begin{figure*}[!h]
\centering
	\includegraphics[scale=0.38]{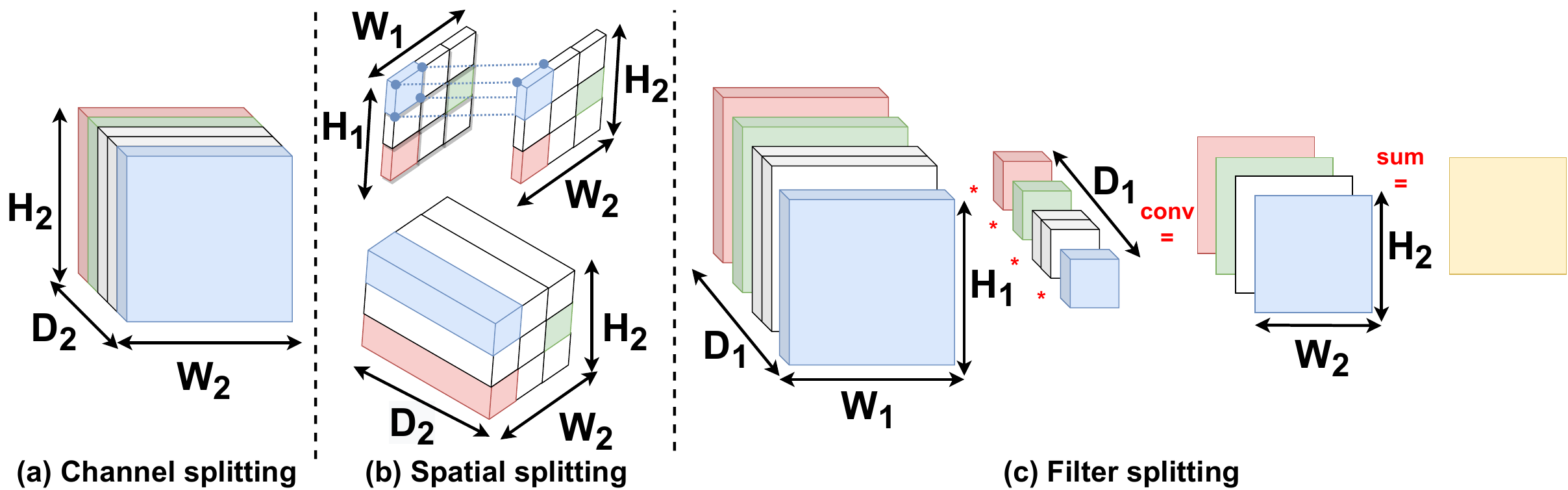}
	\caption{Partitioning of convolutional layer: (a) is an output splitting, and (b) and (c) are input splittings.}
	\label{splitting}
\end{figure*}
\textbf{\indent Convolutional layers:} 
Next, we illustrate different partitioning strategies of the convolutional layer. As described in the previous section \ref{CNN}, each filter is responsible to create one of the feature maps of the output data (Fig. \ref{Conv}). We remind that the dimensions of the input data are $H_1 \times W_1 \times D_1$, the dimensions of the $k$ filters are $H_f \times W_f \times D_f$, and the dimensions of the output feature maps are defined by $H_2 \times W_2 \times D_2$. We note that by definition $D_1$ is equal to $D_f$ and $k$ is equal to $D_2$. Furthermore, each filter contains $D_1 \times (H_f \times W_f)$ weights and performs $D_1 \times (H_f \times W_f)$ multiplications per output element. Similarly to the fully connected layers, two partitioning strategies characterize the convolutional layer, namely the input and output splitting. In this context, the output splitting includes the channel partitioning, meanwhile, the input splitting consists of the spatial and filter partitioning strategies (see Fig. \ref{splitting}). These splitting strategies are introduced and adopted by multiple recent works, including \cite{IoTInferencing,DeepThings,MoDNN}, for which we will thoroughly review the resource management techniques in the following section.
\begin{itemize}
    \item \textit{Channel splitting}: each participant computes one or multiple non-overlapping output feature maps, which serve as input channels for the next layer. This implies that each device $i$ possesses only $1 \leq k'_i \leq k$ filters responsible to generate $k'_i$ feature maps, where $\sum_i k_i'=k$.  In addition to the $k'_i$ filters, the entire input data is fed to each device to compute different outputs. In this way, filters’ weights are distributed across participants, ($k'_i \times D_1 \times H_f \times W_f$) each, and the total number of multiplications is equal to $D_1 \times (H_f \times W_f) \times k'_i \times (W_2 \times H_2)$ per device. The channel partitioning strategy allows model parallelization, and consequently inference acceleration. At the end, when all devices finish their tasks, different feature maps are concatenated depth-wise, with a complexity equal to $O(k)$. We emphasize that the activation function can be applied before merging at each device or once at the concatenation device. Fig. \ref{splitting} (a) shows an example of channel partitioning.
    \item \textit{Spatial splitting}: this fine-grained splitting divides the input spatially, in the x or y axis, in order to jointly assemble the output data, as shown in Fig. \ref{splitting} (b). Let $d_x$ and $d_y$ define the split dimensions on the x-axis and y-axis, respectively. Therefore, the input data are partitioned into segments of size ($d_x \times d_y$), and each group of segments can be transmitted to a device.  Furthermore, each part allocated to a participant needs to be extended with overlapping elements from the neighboring parts, so that the convolution can be performed on the borders. Compared to the latter splitting, in which all the input data should be copied to all participants with parts of the filters, the spatial splitting distributes only parts of the data with all the filters to each device. It means, in addition to the segment of the input data, an amount of ($k \times D_1 \times W_f \times H_f$) weights should be transmitted and stored at each device. Note that storing the filters is considered as a one-time memory cost, as they will be used for all subsequent inferences. Also, the total number of multiplications is reduced per-device and each one executes only $(\frac{d_x \times d_y}{H_1 \times W_1})$ of the computational load per segment. When all computations are done, the output data is concatenated spatially with a complexity of $O(\frac{H_2\times W_2}{d_x \times d_y})$, and the activation function can be applied before or after the merging process. Note that for simplicity, we presented, for spatial splitting, the case where filters do not apply any size reduction.
    \item  \textit{Filter splitting}: in this splitting strategy, both filters and input data are split channel wise on a size of $k'_i$ for each participant $i$. Figure \ref{splitting} (c) illustrates the convolution of the input data by one filter in order to produce one feature map. In this example, the input channels and one of the filters are divided into 4 devices, which implies that  each device stores only its assigned channels of the input data and the filter, so the memory footprint is also divided. The computational load is also reduced, in such a way each participant executes $k'_i \times (H_f \times W_f)  \times (W_2 \times H_2)$ multiplications. In the end, all final outputs are summed to create one feature map and the activation function can only be applied after the merging process. A concatenation task is performed, when all features are created. 
\end{itemize}
Table \ref{tab:splitting} summarizes the computation and memory characteristics of different splitting strategies. In this table, we present the number of smallest segments per model, the input, output and computation requirements for each small segment, the weights of filters assigned to each device owing $k'_i$ segments, and the transmitted data per layer when having $N$ participants.
\subsubsection{Communication models}
The latency is of paramount importance, in AI applications. Hence, minimizing the communication delay and the data transmission by designing an efficient DNN splitting is the main focus of pervasive inference. 
\paragraph{Overview}\mbox{}\\
\textbf{ \indent Communication latency}: In the literature, the communication channels between different pervasive devices are abstracted as bit-pipes with either constant rates or random rates with a defined distribution. However, this simplified bit-pipe model is insufficient to illustrate the fundamental properties of wireless propagation. More specifically, wireless channels are characterized by different key aspects, including: (1) the multi-path fading caused by the reflections from objects existing in the environment (e.g., walls, trees, and buildings); (2) the interference with other signals occupying the same spectrum due to the broadcast nature of the wireless transmissions, which reduces their Signal-to-Interference-plus-Noise-Ratios (SINRs) and increases the probability of errors; (3) bandwidth shortage,  motivating the research community to exploit new spectrum resources, design new spectrum sharing and aggregation, and propose new solutions (e.g., in-device caching and data compression). Based on these characteristics, the communication/upload latency between two devices, either resource-constrained devices or high-performant servers, can be expressed as follows:
\begin{equation}\label{eq:6}
    \begin{aligned}
  t^u=\frac{K}{\rho_{i,j}},
    \end{aligned}
\end{equation}
where $K$ is the size of the transmitted data and $\rho_{i,j}$ is the achievable data rate between two participants $i$ and $j$.

The total transmission latency $t^T$ of the entire inference is related to the type of dependency between different layers of the model. This latency is defined in eq. (\ref{eq:9}), if the dependency is sequential (e.g., layers) and in eq. (\ref{eq:10}) if the dependency is parallel (e.g., feature maps). In case the dependency is general (e.g., randomly wired networks), we formulate the total latency as the sum of sequential communication and the maximum of parallel transmissions.
\begin{equation}\label{eq:9}
    \begin{aligned}
t^T=\sum_{s=1}^{S} t^u_s.
    \end{aligned}
\end{equation}
\begin{equation}\label{eq:10}
    \begin{aligned}
t^T= max(t^u_s, \quad \forall s \in \{1...S\}).
    \end{aligned}
\end{equation}
\newline
\textbf{ \indent Communication energy}: The energy consumption to offload the inference sub-tasks to other participants consists of the amounts of energy consumed on outwards data transmissions and when receiving the classification results generated by the last segment of the task $T$. This energy is formulated as follows \cite{Survey33}\cite{energy-Aware-dist}:
\begin{equation}\label{eq:3}
    \begin{aligned}
              e^{ofd}_i= t^u_i. P_i+\sum_s\sum_k\sum_j \frac{K_s}{\rho_{k,j}}.P_s.X_{k,s}X_{j,s+1},
    \end{aligned}
\end{equation}
where $t^u_i$ is the upload delay to send the original data/task $i$ to the first participant, $K_s$ is the output of the segment $s$ (e.g., layers or feature maps), $\rho_{k,j}$ denotes the data rate of the communication, and $X_{k,s}$ is a binary variable indicating if the participant $k$ executes the segment $s$. 

Using only the onboard battery and resources, the source-generating device may not be able to accomplish the inference task within the required delays and the energy constraint. In such a case, partitioning the task among neighboring devices or offloading the whole inference to the remote servers are desirable solutions. 
\paragraph{Per-layer splitting}
Per-layer partitioning is characterized by a simple dependency between different segments and a higher data transmission per device. Indeed, the computation of one Fc layer per participant costs the system a total communication overhead equal to ($n+m$). Meanwhile, the allocation of a convolutional layer requires a transmission load equal to ($H_1 \times W_1 \times D_1) + (H_2 \times W_2 \times k$). 
\paragraph{Per-segment splitting}
The per-segment partitioning requires a higher total transmission load with less computation and memory footprint per device. Meaning, this type of partitioning trades communication with the memory. More details are illustrated in Table \ref{tab:splitting}, where the output and input splitting of the fully-connected layers have a total communication overhead of $n \times (N-1)$ and $m \times (N-1)$ respectively compared to the per-layer distribution. Hence, depending on the input and output sizes, namely $n$ and $m$, the optimal partition strategy can be selected. Regarding the convolutional layers, the channel splitting has an overhead of $(N-1) \times H_1 \times W_1 \times D1$ since a copy of the entire input data needs to be broadcasted to all devices, the spatial splitting pays an overhead of the padding equal to $N \times padding$, and the filter splitting has an overhead of $(N-1) \times H_2 \times W_2 \times k$ incurred in the merging process.
\begin{figure*}[!h]
\centering
	\includegraphics[scale=0.37]{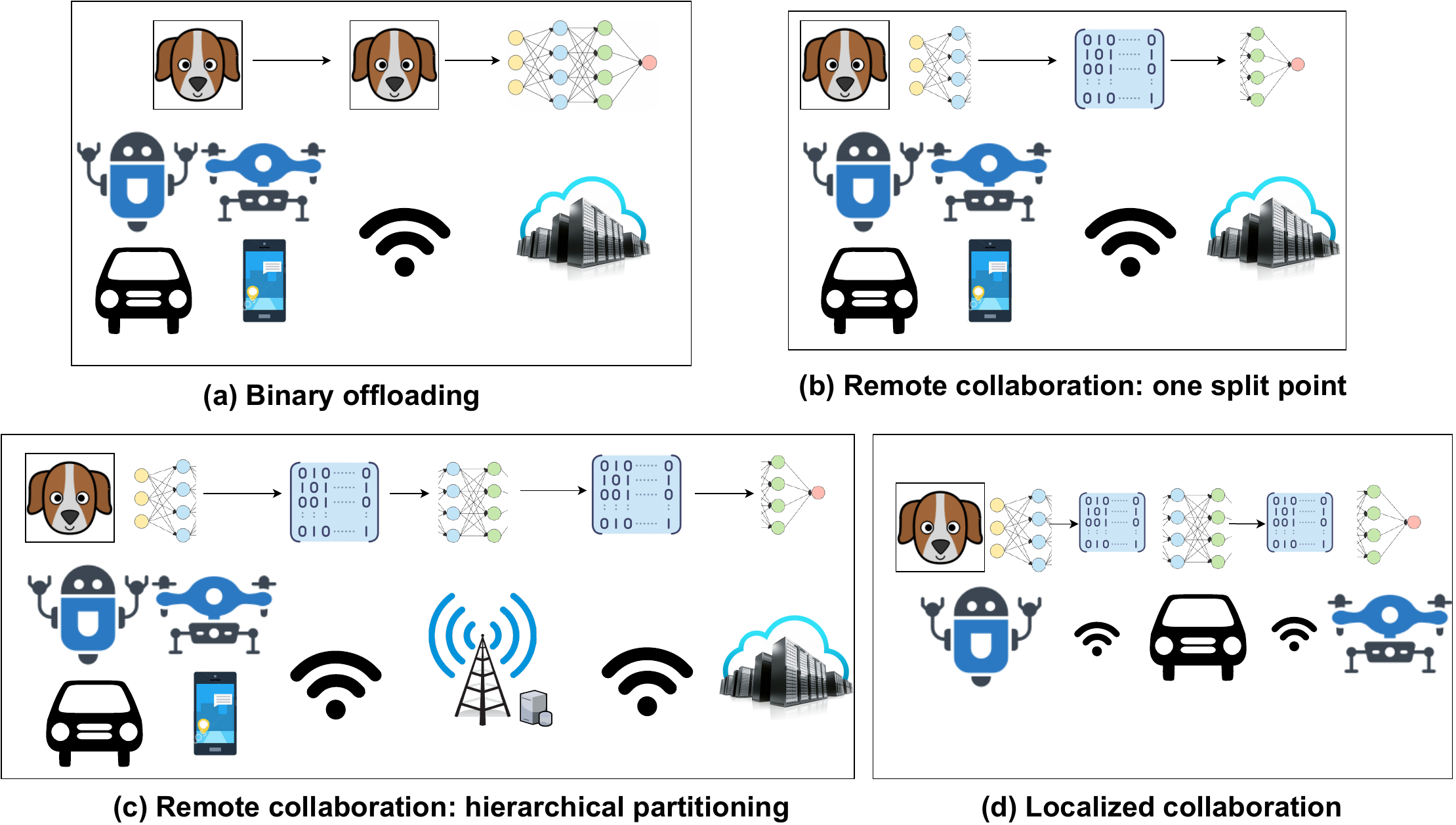}
	\caption{Resource management for distributed inference.}
	\label{distribution}
\end{figure*}
\subsubsection{Lessons learned}
The main lessons acquired from the review of the splitting strategies are:
\begin{itemize}
    \item The performance of model parallelism is always better than that of data parallelism in terms of latency minimization, as it allows computing multiple sub-tasks simultaneously. Meanwhile, the data parallelism pays the high costs of merging and transmitting the same inputs, either for fault-tolerance purposes or to handle multiple concurrent requests.
    \item The choice of the parallelism mode, highly depends on the partitioning strategy and the dependency between different segments. For example, in the per-layer splitting with a sequential dependency, the model parallelization cannot be applied to compute different fragments. On the other hand, the general and parallel dependencies pave the way to distribute concurrent segments.
    \item Data parallelism is highly important for AI applications with a high load of inference requests, such as 24/7 monitoring systems and VR/AR applications. In such scenarios, the classifications and feature learning are required every short interval of time. Generally, source devices do not have sufficient resources to compute this huge load of inferences. In this case, distributing the requests within neighboring devices and parallelizing their computations, contribute to minimizing the queuing time.
    \item Understanding the characteristics of the pervasive computing system is compulsory for selecting the partition strategy. More specifically, the per-layer distribution is more adequate for systems with a lower number of participants and higher pervasive capacities. For example, VGG19 has 19 layers and accordingly needs a maximum of 19 participants. More importantly, these devices are required to be able to accommodate the computation demand of convolutional layers. Meanwhile, opting for fine-grained partitioning results in small fragments that fit in resource-limited devices, such as sensors. However, a high number of sensors (e.g., $\sum_{i}^{N_{conv}}D_1^i \times k_i$ segments for filter splitting.) should be involved to accomplish the inference. 
    \item Choosing the most optimal partitioning for the per-segment strategy highly depends on the proprieties of the DNN network, including the channel sizes, the number of filters, the size of feature maps, and the number of neurons. Particularly, for Fc splitting, $m$ and $n$ are the decisive variables for choosing input or output partitioning. For convolutional layers, the size of the channels and filters, and the capacities of participants are the decisive parameters to select the strategy. In terms of memory requirements, the channel splitting requires copying the whole input channels to all devices along with a part of the filters. Meanwhile, the spatial splitting copies all the filters and a part of the data, whereas the filter splitting needs only a part of the channels and filters. In terms of transmission load, the spatial  splitting has less output data per segment compared to channel and filter strategies. Finally, the channel splitting has a higher computational load per device. Still, it incurs less dependency between segments.
\end{itemize}
\vspace{-0.2cm}
\subsection{Resource management for distributed inference}
The joint computational and transmission resource management plays a key role in achieving low inference latency and efficient energy consumption. In this section, we conduct a comprehensive review of the existing literature on resource management for deep inference distribution and segments allocation on pervasive systems. We start by discussing the remote collaboration, which consists of the cooperation between the data source and remote servers to achieve the DNN inference. In this part, we determine the key design methodologies and considerations (e.g., partitioning strategies and number of split points) in order to shorten the classification delays. Subsequently, more complex collaboration, namely localized collaboration, is examined, where multiple neighboring devices are coordinated to use the available computational and wireless resources and accomplish the inference tasks with optimized energy, delays, and data sharing. 
\subsubsection{Remote collaboration}
The remote collaboration encompasses two approaches, the binary and the partial offloading defined in the previous section. The binary offloading consists of delegating the DNN task from a single data-generating device to a single powerful remote entity (e.g., edge or cloud server), with an objective to optimize   the classification latency, accuracy, energy, and cost (see Fig. \ref{distribution} (a)). The decision will be whether to offload the entire DNN or not, depending on the hardware capability of the device, the size of the data, the network quality, and the DNN model, among other factors. Reference papers covering binary offloading of deep learning include DeepDecision \cite{DeepDecision,offloading} and MCDNN
\cite{MCDNN}. The authors of these papers based their studies on empirical measurements of trade-offs between different aforementioned parameters. The binary offloading has been thoroughly investigated in the literature for different contexts. However, the DNN offloading has a particular characteristic that distinguishes it from other networking tasks, namely the freedom to choose the type, the parameters, and the depth of the neural network, according to the available resources.

As the scope of this survey, is the pervasive AI, we focus on the partial offloading that covers the per-layer distribution applying one or multiple splitting points along with the per-segment distribution.
\paragraph{Per-layer distribution - one split point} The partial offloading leverages the unique structure of the deep model, particularly layers, to allow the collaborative inference between the source device and the remote servers. More specifically, in such an offloading approach, some layers are executed in the data-generating device whereas the rest are computed by the cloud or the edge servers, as shown in Fig. \ref{distribution} (b). In this way, latency is potentially reduced owing to the high computing cycles of the powerful remote entities. Furthermore, latency to communicate the intermediate data resultant from the DNN partitioning  should lead to an overall classification time benefit. The key idea behind the per-layer partitioning is that after the shallow layers, the size of the intermediate data is relatively small compared to the original raw data thanks to the sequential filters. This can speed up the transmission over the network, which motivates the partition at deep layers.  

Neurosurgeon \cite{Neurosurgeon} is one of the first works that investigated layer-wise partitioning, where the split point is decided intelligently depending on the network conditions. Particularly, the authors examined deeply the status quo of the cloud and in-device inference and confirmed that the wireless network is the bottleneck of the cloud approach and that the mobile device can outperform the cloud servers only when holding a GPU unit. As a next step, the authors investigated the DNN split performance in terms of computing and output data size of multiple state-of-the-art DNNs over multiple types of devices and wireless networks and concluded that layers have significantly different characteristics. Based on the computation and the latency to transmit the output data of the DNN layers, the optimal partition points that minimize the energy consumption and end-to-end latency are identified. Finally, after collecting these data, Neurosurgeon is trained to predict the power consumption and latency based on the layer type and network configuration and dynamically partition the model between the data source and the cloud server.

\begin{figure}[!h]
\centering
	\includegraphics[scale=0.5]{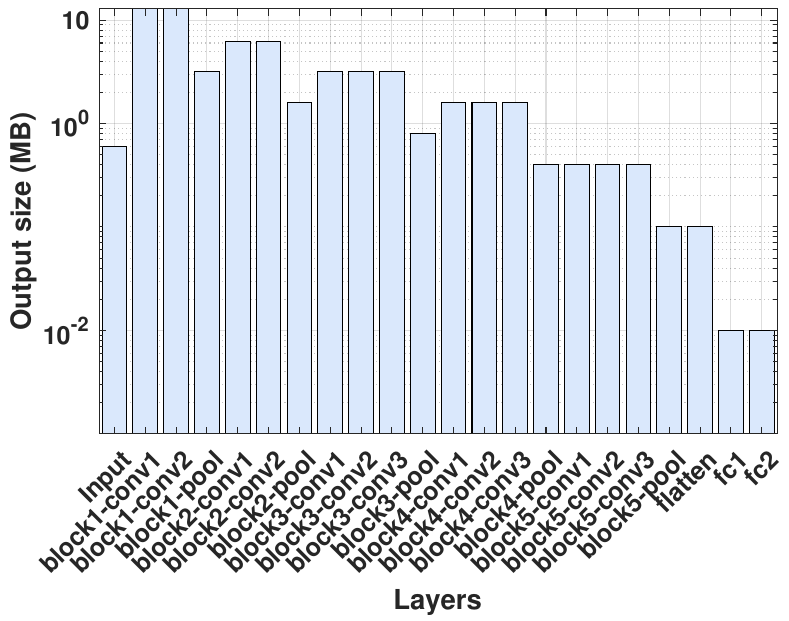}
	\caption{The transmitted data size between different layers of the VGG16 network.}
	\label{VGG16}
\end{figure}

However, while the DNN splitting significantly minimizes the inference latency by leveraging the computational resources of the remote server, this strategy is constrained by the characteristics of intermediate layers that can still generate high-sized data, which is the case of VGG 16 illustrated in Fig. \ref{VGG16}. 
\newline
To tackle the problem of sized intermediate data, the authors of \cite{li2018edge} proposed to combine the early-exit strategy, namely BranchyNet \cite{BranchyNet}, with their splitting approach. The objective is to execute only few layers and exit the model without resorting to the cloud, if the accuracy is satisfactory. In this way, the model inference is accelerated, while sacrificing the accuracy of the classification. We note that BranchyNet is a model trained to tailor the right size of the network with minimum latency and higher accuracy. Accordingly, both models cooperate to select the optimal exit and split points. 
The authors extended the work by replacing both trained models with a reinforcement learning strategy \cite{Boomerang}, namely Boomerang. This RL approach offers a more flexible and adaptive solution to real-time networks and presents less complex and more optimal split and exit points’ selection. The early-exit strategy is also proposed along with the layer-wise partitioning by the ADDA approach \cite{ADDA}, where authors implemented the first layers on the source device and encouraged the exit point before the split point to use only local computing and eliminate the transmission time.  Similarly, authors in \cite{energy-Aware-dist2}, formulated the problem of merging the exit point selection and the splitting strategy, while aiming to minimize the transmission energy, instead of focusing on latency. 

In addition to using the early-exit to accelerate the inference, other efforts adopted compression combined with the partitioning to reduce the shared data between collaborating entities. Authors in \cite{featureEncoding} introduced a distribution approach with feature space encoding, where the edge device computes up to an intermediate layer, compresses the output features (loss-less or lossy), and offloads the compressed data to the host device to compute the rest of the inference, which enhances the bandwidth utilization. To maintain  high accuracy, the authors proposed to re-train the DNN with the encoded features on the host side.  The works in \cite{JALAD, AutoTuning} also suggested compressing the intermediate data through quantization, aiming at reducing the transmission latency between edge and cloud entities. The authors examined the trade-off between the output data quantization and the model accuracy for different partitioning scenarios. Then, they designed accordingly a model to predict the edge and cloud latencies and the communication overhead. Finally, they formulated an optimization problem to find the optimal split layer constrained by the accuracy requirements. To make the solution adaptive to runtime, an RL-based channel-wise feature compression, namely JALAD, is introduced by the authors in \cite{JALAD}. Pruning is another compression technique proposed in \cite{2stepsPruning} to be joined with the partitioning strategy. The authors introduced a 2-step pruning framework, where the first step mainly focuses on the reduction of the computation workload and the second one handles the removal of non-important features transmitted between collaborative entities, which results in less computational and offloading latency. This can be done by pruning the input channels, as their height, length, and number impact directly the size of the output data and the computation requirements, which we illustrated in Table \ref{tab:splitting}.
\paragraph{Per-layer distribution - back and forth, and hierarchical distribution} Solely offloading the deep learning computation to the cloud can violate the latency constraints of the AI application requiring real-time and prompt intervention. Meanwhile, using only the edge nodes or IoT devices can deprive the system from powerful computing resources and potentially increase the processing time. Hence, a judicious selection of multiple cuts and distribution between different resources, i.e., IoT device – edge server – cloud, contribute to establishing a trade-off between minimizing the transmission time and exploiting the powerful servers. Additionally, the layers of the DNN model are not always stacked in a sequential dependency. More specifically, layers can be arranged in a general dependency as shown in Fig. \ref{partitioning} (c), where some of them can be executed in parallel or do not depend on the output of the previous ones. In this case, adopting an optimized \textit{back and forth} distribution strategy, where the end-device and the remote servers parallelize the computation of the layers and merge the output, can be beneficial for the inference latency. Authors in \cite{DNNSurgery} designed a Dynamic Adaptive DNN Surgery (DADS) scheme that optimally distributes complex structured deep models, presented by DAG graphs, under variable network conditions. In case the load of requests is light, the min-cut problem \cite{min-cut} is applied to minimize the overall delay of processing one frame of the DNN structure. When the load condition is heavy, scheduling the computation of multiple requests (data parallelization) is envisaged using the 3-approximation ratio algorithm \cite{3-appro} that maximizes the parallelization of the frames from different requests. Complex DNN structures were also the focus in \cite{joinDNN}, where the authors used the shortest path problem to formulate the allocation of different frames of the DNN \textit{back and forth} between the cloud and the end-device. The path, in this case, is defined as latency or energy of the end-to-end inference.

On the other hand, \textit{hierarchical architecture} for sequential structures is very popular as a one way distribution solution to establish a trade-off between transmission latency and computation delay (see Fig. \ref{distribution} (c)). The papers in \cite{HierarDis, HierarDistGlobecom, AR} proposed to divide the trained DNN over a hierarchical distribution, comprising “IoT-edge-cloud” resources. Furthermore, they adopted the state-of-the-art work BranchyNet \cite{BranchyNet} to early exit the inference if the system has a good accuracy. In this way, fast, private, and localized inference using only shallow layers becomes possible at the end and edge devices, and an offloading to the cloud is only performed when additional processing is required. Hierarchical distribution can also be combined with compressing strategies to reduce the size of the data to be transmitted and accordingly minimize the communication delay and the time of the entire inference, such as using the encoding techniques as done in \cite{IoTDNN}. Authors in \cite{HDDNN,FailureDis} also opted for hierarchical offloading, while focusing primarily on fault-tolerance of the shared data. Particularly, authors in \cite{HDDNN} considered two fault-tolerance methods, namely reassigning and monitoring, where the first one consists of assigning all layers tasks at least once, and then the unfinished tasks are reassigned to all participants regardless of their current state. This method, is generating a considerable communication and latency overhead related to allocating redundant tasks, particularly to devices with  limited-capacities. Hence, a second strategy is designed to monitor the availability of devices before the re-assignment. Finally, the work in \cite{FailureDis} proposed to add skip blocks \cite{ResNet} to the DNN model and include at least one block in each partition, to enhance the robustness of the system in case the previous layer connection fails.

\begin{table*}[]
\centering
\footnotesize
\tabcolsep=0.09cm
\begin{threeparttable}
\caption{Performance of distribution strategies compared to:
\protect \begin{tikzpicture}
\protect\filldraw[color=black!60, fill=white!5,  thick](-1,0) circle (0.15);
\protect\end{tikzpicture} cloud only;
\protect\begin{tikzpicture}
\protect\filldraw[color=black!60, fill={rgb,255:red,218; green,232; blue,252},  thick](-1,0) circle (0.15);
\protect\end{tikzpicture} on-device only;\\
\protect\begin{tikzpicture}
\protect\filldraw[color=black!60, fill={rgb,255:red,213;green,232;blue,212},  thick](-1,0) circle (0.15); 
\end{tikzpicture} edge-server only.
}
\label{tab:performance}
\begin{tabular}{|l|l|l|l|l|l|l|}
\hline
\begin{tabular}[c]{@{}l@{}}\textbf{Refs}\end{tabular} & \textbf{Latency} & \textbf{Bandwidth} & \textbf{Energy} & \textbf{computation/ memory} & \textbf{throughput} & \begin{tabular}[c]{@{}l@{}}\textbf{Inference}\\ \textbf{rate}\end{tabular} \\ \hline
Neurosurgeon \cite{Neurosurgeon} & 3.1 $\times$  $\rightarrow$ 40.7 $\times$ & \xmark & 59.5 \% $\rightarrow$ 94.7\% & \xmark & 1.5 $\times$  $\rightarrow$ 6.7 $\times$ & \xmark \\ \hline
\begin{tabular}[c]{@{}l@{}}Edgent  \cite{li2018edge}\\ Boomerang \cite{Boomerang}\end{tabular} & \cellcolor[HTML]{DAE8FC}2.3 $\times$ & \cellcolor[HTML]{DAE8FC}\xmark & \cellcolor[HTML]{DAE8FC}\xmark & \cellcolor[HTML]{DAE8FC}\xmark & \cellcolor[HTML]{DAE8FC}\xmark & \cellcolor[HTML]{DAE8FC}\xmark \\ \hline
 & 1.2 $\times$  $\rightarrow$ 2 $\times$ & \xmark & \xmark & \xmark & \xmark & \xmark \\ \cline{2-7} 
\multirow{-2}{*}{ADDA \cite{ADDA}} & \cellcolor[HTML]{DAE8FC}1.7 $\times$ $\rightarrow$ 3 $\times$ & \cellcolor[HTML]{DAE8FC}\xmark & \cellcolor[HTML]{DAE8FC}\xmark & \cellcolor[HTML]{DAE8FC}\xmark & \cellcolor[HTML]{DAE8FC}\xmark & \cellcolor[HTML]{DAE8FC}\xmark \\ \hline
 & \cellcolor[HTML]{DAE8FC}\xmark & \cellcolor[HTML]{DAE8FC}\xmark & \cellcolor[HTML]{DAE8FC}15.3 $\times$ & \cellcolor[HTML]{DAE8FC}\xmark & \cellcolor[HTML]{DAE8FC}16.5 $\times$ & \cellcolor[HTML]{DAE8FC}\xmark \\ \cline{2-7} 
\multirow{-2}{*}{\cite{featureEncoding}} & \cellcolor[HTML]{D5E8D4}\xmark & \cellcolor[HTML]{D5E8D4}\xmark & \cellcolor[HTML]{D5E8D4}2.3 $\times$ & \cellcolor[HTML]{D5E8D4}\xmark & \cellcolor[HTML]{D5E8D4}2.5 $\times$ & \cellcolor[HTML]{D5E8D4}\xmark \\ \hline
JALAD \cite{JALAD} & 1.1 $\times$  $\rightarrow$ 11.7 $\times$ & \xmark & \xmark & \xmark & \xmark & \xmark \\ \hline
JointDNN \cite{joinDNN} & 3 $\times$ & \xmark & 7 $\times$ & \xmark & \xmark & \xmark \\ \hline
 & 8.08 $\times$ & \xmark & \xmark & 14.01 $\times$ & \xmark & \xmark \\ \cline{2-7} 
\multirow{-2}{*}{DADS \cite{DNNSurgery}} & \cellcolor[HTML]{D5E8D4}6.45 $\times$ & \cellcolor[HTML]{D5E8D4}\xmark & \cellcolor[HTML]{D5E8D4}\xmark & \cellcolor[HTML]{D5E8D4}8.31 $\times$ & \cellcolor[HTML]{D5E8D4}\xmark & \cellcolor[HTML]{D5E8D4}\xmark \\ \hline
Auto tuning \cite{AutoTuning} & 1.13 $\times$ $\rightarrow$ 1.7 $\times$ & \xmark & \xmark & 85\% $\rightarrow$ 99\% & \xmark & \xmark \\ \hline
DDNN \cite{HierarDis} & \xmark & 20 $\times$ & \xmark & \xmark & \xmark & \xmark \\ \hline
 & 2 $\times$ & \xmark & \xmark & \xmark & \xmark & \xmark \\ \cline{2-7} 
\multirow{-2}{*}{COLT-OPE \cite{HierarDistGlobecom}} & \cellcolor[HTML]{DAE8FC}4 $\times$ & \cellcolor[HTML]{DAE8FC}\xmark & \cellcolor[HTML]{DAE8FC}\xmark & \cellcolor[HTML]{DAE8FC}\xmark & \cellcolor[HTML]{DAE8FC}\xmark & \cellcolor[HTML]{DAE8FC}\xmark \\ \hline
 & 48.11 \% & \xmark & \xmark & \xmark & \xmark & \xmark \\ \cline{2-7} 
\multirow{-2}{*}{\cite{AR}} & \cellcolor[HTML]{DAE8FC}39.75 \% & \cellcolor[HTML]{DAE8FC}\xmark & \cellcolor[HTML]{DAE8FC}70\% & \cellcolor[HTML]{DAE8FC}\xmark & \cellcolor[HTML]{DAE8FC}\xmark & \cellcolor[HTML]{DAE8FC}\xmark \\ \hline
DINA\cite{acceleration} & \cellcolor[HTML]{D5E8D4}2.6 $\times$ $\rightarrow$ 4.2 $\times$ & \cellcolor[HTML]{D5E8D4}\xmark & \cellcolor[HTML]{D5E8D4}\xmark & \cellcolor[HTML]{D5E8D4}\xmark & \cellcolor[HTML]{D5E8D4}\xmark & \cellcolor[HTML]{D5E8D4}\xmark \\ \hline
MoDNN\cite{MoDNN} & \cellcolor[HTML]{DAE8FC}2.17 $\times$ $\rightarrow$  4.28 $\times$ & \cellcolor[HTML]{DAE8FC}\xmark & \cellcolor[HTML]{DAE8FC}\xmark & \cellcolor[HTML]{DAE8FC}\xmark & \cellcolor[HTML]{DAE8FC}\xmark & \cellcolor[HTML]{DAE8FC}\xmark \\ \hline
AAIoT \cite{AAIoT} & \cellcolor[HTML]{D5E8D4}1 $\times$ $\rightarrow$  10 $\times$ & \cellcolor[HTML]{D5E8D4}\xmark & \cellcolor[HTML]{D5E8D4}\xmark & \cellcolor[HTML]{D5E8D4}\xmark & \cellcolor[HTML]{D5E8D4}\xmark & \cellcolor[HTML]{D5E8D4}\xmark \\ \hline
DeepWear \cite{DeepWear} & \cellcolor[HTML]{DAE8FC}5.08 $\times$ $\rightarrow$  23 $\times$ & \cellcolor[HTML]{DAE8FC}\xmark & \cellcolor[HTML]{DAE8FC}53.5\% $\rightarrow$  85.5\% & \cellcolor[HTML]{DAE8FC}\xmark & \cellcolor[HTML]{DAE8FC}\xmark & \cellcolor[HTML]{DAE8FC}\xmark \\ \hline
\cite{IoTInferencing} & \cellcolor[HTML]{DAE8FC}2 $\times$ $\rightarrow$  6 $\times$ & \cellcolor[HTML]{DAE8FC}\xmark & \cellcolor[HTML]{DAE8FC}\xmark & \cellcolor[HTML]{DAE8FC}\xmark & \cellcolor[HTML]{DAE8FC}\xmark & \cellcolor[HTML]{DAE8FC}\xmark \\ \hline
\cite{MDPIDist} & \cellcolor[HTML]{DAE8FC}\xmark & \cellcolor[HTML]{DAE8FC}\xmark & \cellcolor[HTML]{DAE8FC}\xmark & \cellcolor[HTML]{DAE8FC}\xmark & \cellcolor[HTML]{DAE8FC}\xmark & \cellcolor[HTML]{DAE8FC}1.7 $\times$ $\rightarrow$  4.69 $\times$ \\ \hline
DeepThings \cite{DeepThings} & \cellcolor[HTML]{DAE8FC}0.6 $\times$ $\rightarrow$  3 $\times$ & \cellcolor[HTML]{DAE8FC}\xmark & \cellcolor[HTML]{DAE8FC}\xmark & \cellcolor[HTML]{DAE8FC}68\% & \cellcolor[HTML]{DAE8FC}\xmark & \cellcolor[HTML]{DAE8FC}\xmark \\ \hline
\end{tabular}
\begin{tablenotes}
   \footnotesize
   \item - The results in the table present the enhancement of the proposed strategies compared to the baseline approaches.
   \item - $\times$ stands for the number of times the metric is improved, i.e., how many times the latency, bandwidth usage, energy, computation, and memory are reduced, and how many times the throughput and inference rate are increased compared to the baselines.
\end{tablenotes}
\end{threeparttable}
\end{table*}
\paragraph{Per-segment distribution}
The per-segment partitioning is generally more popular when distributing the inference among IoT devices with limited capacities, as some devices, such as sensors, cannot execute the entire layer of a deep network. Furthermore, per-segment partitioning creates a huge dependency between devices; and consequently, multiple communications with remote servers are required. That is why only few works adopted this strategy for inference collaboration between end devices and edge/fog servers, including \cite{acceleration}. Authors in \cite{acceleration} proposed a spatial splitting (see Fig. \ref{splitting} (b)) that minimizes the communication overhead per device. Then, a distribution solution is designed based on the matching theory \cite{matching_theory} and the swap matching problem \cite{swap_matching}, to jointly accomplish the DNN inference. The matching theory is a mathematical framework in economics that models interactions between two sets of selfish agents, each one is competing to match agents of the other set. The objective was to reduce the total computation time while increasing the utilization of the resources related to the two sets of IoT and fog devices. 
\subsubsection{Localized collaboration} 
Another line of work considers the distribution of DNN computation across multiple edge participants, as shown in Fig. \ref {distribution} (d). These participants present neighboring nodes that co-exist in the same vicinity, e.g., IoT devices or fog nodes. The model distribution over neighboring devices can be classified into two types: the per-layer distribution where each participant performs the computation of one layer or more and the per-segment allocation where smaller segments of the model are allocated to resource-limited devices.
\paragraph{Per-layer distribution}
The layer-wise partitioning can itself be classified under two categories, the one splitting point strategy where only two participants are involved and multiple splitting points where two or more devices are collaborating. For example, the DeepWear \cite{DeepWear} approach splits the DNN into two sub-models that are separately computed on a wearable and a mobile device. First, the authors conducted in-depth measurements on different devices and for multiple models to demystify the performance of the wearable-side DL and study the potential gain from the partial offloading. The derived conclusions are incorporated into a prediction-based online scheduling algorithm that judiciously determines how, and when to offload, in order to minimize latency and energy consumption of the inference. On the other hand, authors in \cite{CNNDist} proposed a methodology for optimal placement of CNN layers among multiple IoT devices, while being constrained by their computation and memory capacities. This methodology minimizes the latency of decision-making, which is measured as the total of processing times and transmissions between participants. Furthermore, this proposed technique can be applied both to CNNs in which the number of layers is fixed and CNNs with an early-exit. Similarly, authors in \cite{AAIoT} proposed a CNN multi-splitting approach to accelerate the inference process, namely AAIoT. Unlike the above-mentioned efforts, AAIoT deploys the layers of the neural network on multi-layer IoT architecture. More specifically, the lower-layer device presents the data source, and the higher-layer devices have more powerful capacities. Offloading the computation to higher participants implies sacrificing the transmission latency to reduce the computation time. However, delivering the computation to lower participants does not bring any benefit to the system. An optimal solution and an online algorithm that uses dynamic programming are designed to make the best architectural offloading strategy. Other than capacity-constrained IoT devices, the distribution of the inference process over cloudlets in a 5G-enabled MEC system is the focus of the work in \cite{energy-Aware-dist}, where authors proposed to minimize the energy consumption, while meeting stringent delay requirements of AI applications, using a RL technique.
\paragraph{Per-segment distribution}
The per-segment distribution is defined as allocating fine-grained partitioned DNN on lightweight devices such as Raspberry Pis. The partitioning strategy is based on the system configuration and the pervasive network characteristics, including the memory, computation, and communication capabilities of the IoT devices and their number. The segmentation of the DNN models varies from neurons partitioning to channels, spatial, and filters splitting, as discussed in section \ref{profiling}. For example, the work in \cite{MoDNN} opted for the spatial splitting (see Fig. \ref{splitting} (b)), where the input and the output feature maps are partitioned into a grid and distributed among lightweight devices. The authors proposed to allocate the cells along the longer edge of the input matrix (rows or columns) to each participant, in order to reduce the padding overhead produced by the spatial splitting. Different segments are distributed to IoT devices according to the load-balancing principles using the MapReduce model. The same rows/columns partitioning is proposed in \cite{CCNN}, namely the data-lookahead strategy. More specifically, each block contains data from other blocks within the same layer such that its connected blocks in subsequent layers can be executed independently without requesting intermediate/padding data from other participants. The spatial splitting is also adopted in \cite{DeepThings}, where authors proposed a Fused Tile Partitioning (FTP) method. This method fuses the layers and divides them into a grid. Then, cells connected across layers are assigned to one participant, which largely reduces the communication overhead and the memory footprint. 

The previous works introduced homogeneous partitioning, where segments are similar. Unlike these strategies, authors in \cite{MDPIDist, MDPIDist2} proposed a heterogeneous partitioning of the input data to be compatible with the IoT system containing devices with different capabilities ranging from small participants that fit only few cells to high capacity participants suitable for layer computation. For the same purpose, authors in \cite{EDDL} jointly conducted per-layer and per-segment partitioning, where the neurons and links of the network are modeled as a DAG. In this work, grouped convolutional techniques \cite{grouped_conv} are used to boost the model parallelization of different nodes of the graph. The papers in \cite{FullyDistribution, MusicChair,MusicChair2, IoTInferencing} studied different partitioning strategies of the convolutional layers (channel, spatial and filters splitting) and fully connected layers (output and input splitting). Next, they emphasized that an optimal splitting depends greatly on the parameters of the CNN network and that the inference speedup depends on the number of tasks to be parallelized, which is related to the adopted splitting method. Hence, one partitioning approach cannot bring benefits to all types of CNNs. Based on these conclusions, a dynamic heuristic is designed to select the most adequate splitting and model parallelism for different inference scenarios.

Table \ref{tab:performance} shows the performance of these techniques in terms of latency, bandwidth, energy, computation, memory, and throughput, whereas Table \ref{tab:my-table} presents a comparison between different distributed inference techniques introduced in this section.
\subsubsection{Lessons learned}
The lessons acquired from the literature review covering the DNN distribution can be summarized as follows:
\begin{itemize}
    \item  The per-layer strategy with remote collaboration is the most studied approach in the literature, owing to its simple splitting scheme and its assets in using high-performance servers while reducing the transmitted data. However, such strategies may not be efficient in terms of privacy or for networks with unstable transmission links, and hence may not be suitable for all applications.
    \item In per-layer strategies, selecting the split points depends on multiple parameters, which are the capacity of the end device that constrains the length of the first segment, the characteristics of the network (e.g., wi-fi, 4G, or LTE) that impact the transmission time, and the DNN topology that determines the intermediate data size.
    \item The deep neural networks with a small-reduction capacity of pooling layers or with fully-connected layers of similar sizes undergo small variations in the per-layer data size. In this case, remote collaboration is not beneficial for data transmission. Hence, compression (e.g., quantization, pruning, and encoding) can be a good solution to benefit from remote capacity with the minimum of communication overhead.
    \item  Recently, even if it is still not mature yet, multiple efforts have focused on the localized inference through per-segment distribution that allows to involve resource-limited devices and avoid the transmission to remote servers. This kind of works targeted the model parallelization and aimed to maximize the concurrent computation of different segments within the same request. However, fewer works covered data parallelization and real-time adaptability to the dynamics and number of requests. Particularly, the load of inferences highly impacts the distribution of segments to fit them to the capacity of participants.
    \item Adopting a mixed partitioning strategy is advantageous for heterogeneous systems composed of high and low-capacity devices and multiple DNNs, which allows to fully utilize the pervasive capacities while minimizing the dependency and data transmission between devices.  
\end{itemize}
\begin{table*}[]
\centering
\footnotesize
\tabcolsep=0.09cm
\caption{Comparison between Distributed Inference techniques.}
\label{tab:my-table}
\begin{tabular}{|c|c|c|c|c|c|c|c|c|c|c|c|}
\hline
\textbf{Refs} & \textbf{Year} & \textbf{End-Device} & \begin{tabular}[c]{@{}c@{}}$N^{o}$ \textbf{. of}\\  \textbf{end}\\ \textbf{Devices}\end{tabular} & \begin{tabular}[c]{@{}c@{}}\textbf{Localized} \\ \textbf{inference}\end{tabular} & \textbf{Context} & \begin{tabular}[c]{@{}c@{}}\textbf{Real-time}\\\textbf{processing}\end{tabular} & \begin{tabular}[c]{@{}c@{}}\textbf{Partitioning}\\ \textbf{mechanism}\end{tabular} & \begin{tabular}[c]{@{}c@{}}$N^{o}$\textbf{. of}\\ \textbf{partitions}\end{tabular} & \begin{tabular}[c]{@{}c@{}}\textbf{Model or data}\\ \textbf{parallelism}\end{tabular} & \begin{tabular}[c]{@{}c@{}}\textbf{other}\\  \textbf{techniques}\end{tabular} & \begin{tabular}[c]{@{}c@{}}\textbf{Runtime} \\ \textbf{adaptability}\end{tabular} \\ \hline
\begin{tabular}[c]{@{}c@{}}Neuroseurgeon\\ \cite{Neurosurgeon}\end{tabular} & 2017 & Tegra  TKI & 1 & \xmark & \xmark & \xmark & Per-layer & 1 & \xmark & \xmark & \xmark \\ \hline
DDNN \cite{HierarDis} & 2017 & \xmark & Many & \xmark & \xmark & \cmark & Per-layer & Many & Data & Early exit & \xmark \\ \hline
MoDNN \cite{MoDNN} & 2017 & LG Nexus 5 & 4 & \cmark & \xmark & \xmark & Per-segment & Many & Model & \xmark & \xmark \\ \hline
Edgent \cite{li2018edge} & 2018 & RaspBerry Pi 3 & 1 & \xmark & \xmark & \cmark & Per-layer & 1 & \xmark & Early exit & \xmark \\ \hline
\cite{featureEncoding} & 2018 & \xmark & 1 & \xmark & \xmark & \xmark & Per-layer & 1 & \xmark & Compression & \xmark \\ \hline
\begin{tabular}[c]{@{}c@{}}DeepThings \\ \cite{DeepThings}\end{tabular} & 2018 & RaspBerry Pi 3 & Many & \cmark & \xmark & \cmark & Per-segment & Many & Model & \xmark & \xmark \\ \hline
\begin{tabular}[c]{@{}c@{}}Collaborative\\ robots \cite{robots}\end{tabular} & 2018 & RaspBerry Pi & 12 & \cmark & \begin{tabular}[c]{@{}c@{}}Robots and\\ image \\recognition\end{tabular} & \cmark & Per-segment & Many & Both & \xmark & \cmark \\ \hline
\begin{tabular}[c]{@{}c@{}}Musical chair\\ \cite{MusicChair,MusicChair2}\end{tabular} & 2018 & RaspBerry Pi & Many & \cmark & \begin{tabular}[c]{@{}c@{}}object/action \\ recognition\end{tabular} & \cmark & Per-segment & Many & Both & \xmark & \cmark \\ \hline
HDDNN \cite{HDDNN} & 2018 & \xmark & Many & \xmark & \xmark & \cmark & Per-layer & Many & Data & Encryption & \xmark \\ \hline
\begin{tabular}[c]{@{}c@{}}Auto tuning\\ \cite{AutoTuning}\end{tabular} & 2018 & Jetson TX2 & Many & \xmark & \xmark & \xmark & Per-layer & Many & \xmark & Quantization & \xmark \\ \hline
JALAD \cite{JALAD} & 2018 & \begin{tabular}[c]{@{}c@{}}GPU \\Quadro k620 \end{tabular}& 1 & \xmark & \xmark & \xmark & Per-layer & 1 & \xmark & Quantization & \cmark \\ \hline
\begin{tabular}[c]{@{}c@{}} KLP \\ \cite{MDPIDist, MDPIDist2}\end{tabular} & \begin{tabular}[c]{@{}c@{}}2018\\ 2019\end{tabular} & STM32F469 & Many & \cmark & \xmark & \xmark & Per-segment & Many & Model & \xmark & \xmark \\ \hline
ADDA \cite{ADDA} & 2019 & RaspBerry Pi 3 & 1 & \xmark & \xmark & \xmark & Per-layer & 1 & \xmark & Early exit & \xmark \\ \hline
\begin{tabular}[c]{@{}c@{}}Boomerang\\ \cite{Boomerang}\end{tabular} & 2019 & RaspBerry Pi 3 & 1 & \xmark & \xmark & \cmark & Per-layer & 1 & \xmark & Early exit & \cmark \\ \hline
\cite{Byzantine} & 2019 & Krait CPU & 12 & \cmark & \begin{tabular}[c]{@{}c@{}}sensors\\  fault tolerance\end{tabular} & \cmark & \xmark & Many & Model & \xmark & \cmark \\ \hline
\cite{CNNDist} & 2019 & \begin{tabular}[c]{@{}c@{}}RaspBerry Pi\\ STM32H7\end{tabular} & Many & \cmark & \xmark & \xmark & Per-layer & Many & Data  & \xmark & \xmark \\ \hline
DADS \cite{DNNSurgery} & 2019 & \begin{tabular}[c]{@{}c@{}}RaspBerry Pi 3\\ model B\end{tabular} & 1 & \xmark & \xmark & \cmark & Per-layer & Many & \xmark & \xmark & \cmark \\ \hline
\begin{tabular}[c]{@{}c@{}}COLT-OPE \\ \cite{HierarDistGlobecom} \end{tabular} & 2019 & \xmark & 1 & \xmark &\xmark  & \xmark & Per-layer & Many & \xmark & Early exit & \cmark \\ \hline
EDDL \cite{EDDL} & 2019 & Fog nodes & Many & \cmark & \xmark & \xmark & \begin{tabular}[c]{@{}c@{}}Per-layer\\ Per-segment\end{tabular} & Many & Model & \begin{tabular}[c]{@{}c@{}}Sparsification\\ Early exit\end{tabular} & \xmark \\ \hline
\cite{energy-Aware-dist2} & 2019 & \begin{tabular}[c]{@{}c@{}}GPU \\ GTX1080 \end{tabular}& 1 & \xmark & \xmark & \cmark & Per-layer & 1 & \xmark & \xmark & \cmark \\ \hline
\cite{FullyDistribution} & 2019 & \xmark & 7 & \cmark & \xmark & \xmark & \begin{tabular}[c]{@{}c@{}}Per-layer\\ Per-segment\end{tabular} & Many & Model & \xmark & \xmark \\ \hline
\begin{tabular}[c]{@{}c@{}}deepFogGuard\\\cite{FailureDis} \end{tabular} & 2019 & \xmark & Many & \xmark & \xmark & \xmark & Per-layer & Many & \xmark & \xmark & \xmark \\ \hline
\begin{tabular}[c]{@{}c@{}}2steps-pruning\\ \cite{2stepsPruning} \end{tabular} & 2019 & \xmark & 2 & \xmark & \xmark & \xmark & Per-layer & 1 & \xmark & pruning & \xmark \\ \hline
\begin{tabular}[c]{@{}c@{}}JointDNN \\ \cite{joinDNN} \end{tabular}& 2019 & jetson tx2 & 1 & \xmark & \xmark & \xmark & Per-layer & 1 & \xmark & \xmark & \cmark \\ \hline
AAIoT \cite{AAIoT} & 2019 & \begin{tabular}[c]{@{}c@{}}Raspberry Pi,\\  Mobile PC, \\ Desktop PC, \\ Server\end{tabular} & Many & \cmark & \xmark & \xmark & Per-layer & Many & \xmark & \xmark & \xmark \\ \hline
MWWP \cite{AccelerateDI} & 2020 & \xmark & Many & \xmark & health care & \cmark & Per-layer & Many & Data & \xmark & \cmark \\ \hline
\cite{Commeff} & 2020 & Raspberry Pi & Many & \cmark & \begin{tabular}[c]{@{}c@{}}multi-view \\ object \\ detection\end{tabular} & \xmark & Per-segment & Many & Model  & Compression & \xmark \\ \hline
\begin{tabular}[c]{@{}c@{}}CONVENE \\ \cite{CCNN}\end{tabular} & 2020 & \xmark & 1 & \xmark & \begin{tabular}[c]{@{}c@{}}Parallel data \\ sharing on \\antennas\end{tabular} & \xmark & Per-segment & Many & Model  & \xmark & \cmark \\ \hline
DINA \cite{acceleration} & 2020 & \xmark & Many & \xmark & \xmark & \xmark & Per-segment & Many & Both & \xmark & \cmark \\ \hline
\cite{vehicles} & 2020 & \xmark & Many & \cmark & \begin{tabular}[c]{@{}c@{}}Intelligent \\Connected\\  Vehicles\end{tabular} & \cmark & \xmark & Many & \xmark & \xmark & \xmark \\ \hline
\cite{IoTDNN} & 2020 & \xmark & 1 & \xmark & \xmark & \xmark & Per- layer & 1 & \xmark & Compression & \xmark \\ \hline
\cite{AR} & 2020 & Huawei & 1 & \xmark & \begin{tabular}[c]{@{}c@{}}augmented \\reality\\  in 5G\end{tabular} & \cmark & Per-layer & 2 & Data & Early-exit & \cmark \\ \hline
\cite{IoTInferencing} & 2020 & Raspberry Pi 3 & Many & \cmark & \begin{tabular}[c]{@{}c@{}}Visual based \\ applications\end{tabular} & \xmark & Per-segment & Many & Both & \xmark & \xmark \\ \hline
\begin{tabular}[c]{@{}c@{}}Deep Wear \\ \cite{DeepWear} \end{tabular}& 2020 & Android wear & 2 & \cmark & \begin{tabular}[c]{@{}c@{}} Wearable\\ devices\end{tabular} & \cmark & Per-layer & 1 & \xmark & Compression & \cmark \\ \hline
\cite{energy-Aware-dist} & 2021 & Cloudlet & Many & \cmark & 5 G & \cmark & Per-layer & Many & Data & \xmark & \cmark \\ \hline
\begin{tabular}[c]{@{}c@{}}DistPrivacy \\ \cite{DistPrivacy}\end{tabular} & 2021 & \begin{tabular}[c]{@{}c@{}}Raspberry Pi\\ STM32H7\\ LG Nexus 5\end{tabular} & Many & \cmark & Data privacy & \cmark & Per-segment & Many & Both & \xmark & \cmark \\ \hline
\end{tabular}
\end{table*}
\subsection{Use case: Distribution on moving robots}
Currently, robotic systems have been progressively converging to computationally expensive AI networks for tasks like path planning and object detection. However, resource-limited robots, such as low power UAVs, have insufficient on-board power-battery or computational resources to scalably execute the highly accurate neural networks. 
\begin{figure}[!h]
\centering
	\frame{\includegraphics[scale=0.53]{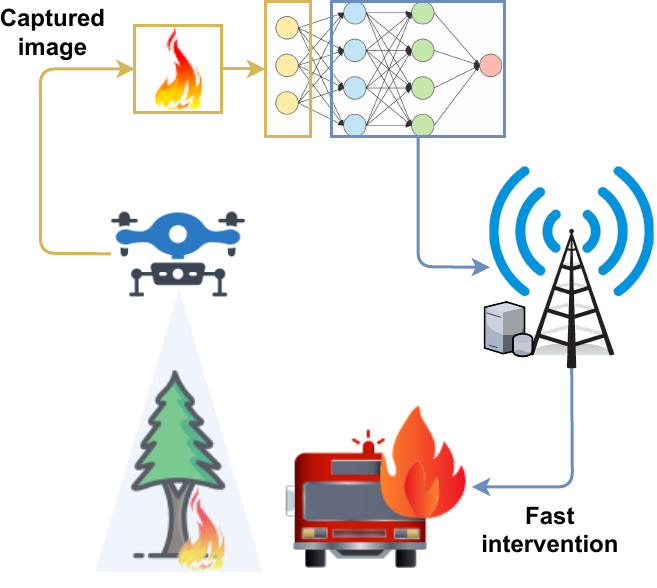}}
	\caption{A fire detection scenario with distributed DNN.}
	\label{UAVs_mobility}
\end{figure}

The work in \cite{uavs,uavs2} examined the case of per-layer distribution with one split point between one UAV and one MEC server (see Fig. \ref{UAVs_mobility}). More specifically, the authors proposed a framework for AI-based visual target tracking system, where low-level layers of the DNN classifier are deployed in the UAV device and high-level layers are assigned to the remote servers. The classification can be performed using only the low-level layers, if the image quality is good. Otherwise, the output of these layers should be further processed in the MEC server, for higher accuracy. In this context, the authors formulated a weighted-sum cost minimization problem for binary offloading and partial offloading, while taking into consideration the error rate/accuracy, the data quality, the communication bandwidth, and the computing capacity of the MEC and the UAV. The offloading probability is derived for the binary offloading and the offloading ratio (i.e., the segment of DNN to execute in the MEC) is obtained for the partial offloading scheme. In this model, the mobility of the UAVs (i.e., the distance between the UAV and the server) is involved through the transmission data rate between the device and the MEC. Additionally, the distance between the UAV and the target impacts the quality of the image and consequently impacts the offloading decisions. In the proposed framework, multiple trade-offs are experienced:
\begin{itemize}
    \item The accuracy is achieved at the expense of delay and transmitted data: if most of the images have bad quality, the system is not able to accomplish low average latency as on-board inference is not sufficient. For this reason, different inferences should be extended wisely using the segment allocated in the MEC, particularly if the environment is challenging such as bad weather or when the target is highly dynamic.
    \item A trade-off exists also between the accuracy and latency, and the position of the UAVs: when the device is close to the target, high resolution images can be taken, which allows obtaining good accuracy on-board and avoiding the data offloading. Being close to the targets is not always possible, particularly  in harsh environments or when the surveillance should be hidden.
    \item The battery life is increased at the expense of the inference latency: the battery can be saved, if the processing coefficient is decreased, which enlarges the computation time of the classification. 
    \item  The split point selection: if the intermediate data is smaller than the raw data, the offloading is encouraged to enhance the accuracy.
\end{itemize}
An online solution of this offloading trade-off using reinforcement learning is presented in \cite{cloudRobotics}.

The previous works adopted the per-layer wise with one split point and remote collaboration approach. This strategy is more adequate for flying devices that can enhance their link quality by approaching the MEC stations. However, for ground robots, offloading segments of the inference to remote servers costs the system a large transmission overhead and high energy consumption. Authors in \cite{robots_power} studied the distribution of the DNN network among ground robots and profiled the energy consumed for such tasks, when moving or being in idle mode. Several conclusions are stated:
\begin{itemize}
    \item When the robot is idle, the DNN computation and offloading increase the power consumption of the device by 50\%.
    \item If the device is moving, the DNN execution causes high spikes in power consumption, which may limit the device to attain a high performance as this variation incurs a frequent change of the power saving settings in the CPU.
    \item Distributing the inference contributes to reducing the energy consumed per device, even-though the total power consumption is higher. This is due to the reduced computation and memory operations per device and the idle time experienced after offloading the tasks.
\end{itemize}
Based on the energy study of moving robots, the authors proposed to distribute the DNN model into smaller segments among multiple low-power robots to achieve an equilibrium  of performance in terms of energy and number of executed tasks \cite{robots}. Still, the distribution of the model into small segments (e.g., filter splitting) requires the intervention of a large number of robots that are highly dependent, which is not realistic.
\section{Privacy of pervasive AI systems}\label{privacy_AI}
Even though the pervasive AI has presented unprecedented opportunities to empower IoT applications, it gave rise to novel security and privacy concerns. In fact, if servers and participants are not controlled or owned by one operator, they are considered malicious by nature. Particularly, sensitive information can be leaked while sharing intermediate data or updates between participants. Moreover, an untrusted participant can alter the local data or send wrong parameters to slow the learning or mislead the system. In this section, we overview the privacy (i.e., one of the devices revealing private information about others) and security (i.e., one of the devices injects false information to disrupt the collective behavior of the devices) challenges and we survey different approaches that address these issues.
\subsection{Privacy for pervasive training}
\subsubsection{Privacy and security challenges}
In some federated learning settings, participants can randomly join or leave the training process, which raises various vulnerabilities from different sources, including malicious servers, insider adversaries and outsider attackers. More specifically, aggregation servers can be honest but curious to inspect the models without introducing any changes. On the other hand, potential malicious servers \cite{FL_privacy1}, as well as untrusted participants can tamper the model during the learning rounds or dissimulating participation in order to obtain the final aggregated model without actually contributing with any data or only contributing with a small number of samples. This attack is called free-riding \cite{freerider}. Outsider eavesdroppers can also intercept the communication channels between trusted devices and the server to spoof the model or inject noisy data (data poisoning). Authors in \cite{FL_privacy2} proved that it is possible to extract sensitive information from a trained model, as it implies the correlation between the training samples. The research work in \cite{FL_privacy3} showed that confidence information returned by ML classifiers introduce new model inversion attacks that enable the adversary to reconstruct samples of training subjects with high accuracy. Inferring the sensitive information is also possible through querying the prediction model.

Bandit and MARL algorithms are also prone to attacks if one of the agents is compromised. In general, if any of the agents starts communicating false data (i.e., false data injection attacks), the regret-guarantees in bandits, and convergence behavior in MARL no longer hold. In addition to these expected effects, recent works in \cite{OL_privacy1,OL_privacy2} have demonstrated that a malicious agent may not only be disruptive but can also actively sway the policy into malicious objectives by driving other agents to reach a policy of its choice.

\subsubsection{Defense techniques and solutions}
\paragraph{Differential Privacy (DP)}
DP is a data perturbation technique that was first introduced  for ML in \cite{DP1}. In DP, a statistical noise is injected to the sensitive data to mask it and an algorithm is called differentially private, if its output cannot give any insight or reveal any information about its input. The DP has been widely used to preserve the privacy of the learning, although it is always criticized by its effect on the accuracy of the results due to noise growth. For this, a careful calibration between the privacy level and the model usability is needed. The use of differential privacy for distributed learning systems becomes a very active research area. The authors in \cite{DP2} proposed a differentially private stochastic gradient descent technique \cite{DP3} that adds random noise (e.g., Gaussian mechanism) to the trained parameters, before sending them to the aggregation server. Then, during the local training, each participant keeps calculating the probability that an attacker succeeds to exploit the shared data, until reaching a predefined threshold at which it stops the process. Moreover, in each round, the aggregation server chooses random participants. In this way, neither the local parameters can be used, nor the global model, as the attacker has no information about the devices participating in the current round. DP has been also used to ensure the agent's privacy in federated bandits \cite{OL_privacy3}, where the authors considered the federated bandit formulation with contextual information (in both, centralized aggregator and p2p communication style). The authors provided regret and privacy guarantees so that the peers, or the central aggregator, do not learn individual agent's samples.

While concealing the agents contribution during the  training, a trade-off between the privacy and
learning performance should be established. In this context, authors in \cite{DP33} tested the performance of FL applied to real-world healthcare datasets while securing the private data using differential privacy techniques. Results show that a significant performance loss is witnessed, even though the privacy level is increased. This encouraged the research community to propose alternative approaches to ensure privacy in federated learning.

\paragraph{Homomorphic Encryption (HE)}
HE is a form of encryption that consists of performing computational operations on cipher texts while having the same results that can be generated by the original data.
The approach in \cite{encr_FL} and \cite{HE11} ensures the integrity of DL training process against outsider attackers as well as  honest-but-curious server. The key idea is to encode and compress the parameters of the trained neural networks before sharing them with the server. Then, these aggregated updates are directly computed with decoder on the server. This guarantees their privacy during the communication and after decoding. Although the encryption technique can preclude the server from extracting information of local models, it costs the system more communication rounds and cannot prevent the collusion between the server and a malicious participant. To solve this problem, authors in \cite{HE12} proposed to adopt hybrid solution which integrates both lightweight homomorphic encryption and differential privacy. In this work, intentional noises are added to perturb the original parameters in case the curious server accomplices with one of the participants to get encryption parameters.

Even though the encryption is a robust approach to achieve privacy preservation for many applications, its adoption for deep learning is facing various challenges as it can only be deployed on tasks with certain degrees and complexities. In other words, the fully homomorphic schemes are still not efficient for practical use.
\paragraph{Blockchain-based solutions}
Blockchain is a recent distributed ledger system initially designed for cryptocurrency and later increasingly applied to the IoT systems, where a record of transactions is deployed distributively in a peer-to-peer network \cite{blockchain}. Authors in \cite{block2} proposed to use a blockchain-based communication scheme to exchange updates in a distributed ML system, with the aim of leveraging the blockchain's security features in the learning process.
In such practice, local models are shared and verified in the trusted blockchain network. Furthermore, this framework can prevent participants from free-riding as their updates are checked and they receive rewards proportional to the number of trained data samples. However, in contrast to vanilla FL, Block FL needs to take into consideration the extra delay incurred by the blockchain network. To address this, the Block FL is formulated by considering communication, computation, and the block generation rate, i.e., the proof of work difficulty. A possible drawback of this approach is its vulnerability against any latency increase.
Also, the use of blockchain implies the addition of a significant cost to implement and  maintain miners.

\paragraph{Secure Multi-Party Computation (SMC)}
SMC is a sub-field of cryptographic protocols that has as a goal to secure the data except the output when multiple participants jointly perform an arbitrarily function over their private input. A study in \cite{SMC} has used SMC to build FL systems. The proposed protocols consider secret sharing, which adds new round at the beginning of the process for the keys sharing, double-masking round that protects from potential malicious server, and server-mediated key agreement that minimizes trust.
\paragraph{Prevention against data poisoning}
Data poisoning is one of the attacks that is very destructive for ML, where an attacker injects poisoned data (e.g., mislabeled samples and wrong parameters) into the dataset, which can mislead the learning process. Authors in \cite{poisoning1} and \cite{poisoning2} propose secure decentralized techniques to protect the learning against data poisoning, as well as other system attacks. A zero-sum game is proposed to formulate the conflicting objectives between  honest participants that utilize Distributed Support Vector Machines (DSVMs) and a malicious attacker that can change data samples and labels. This game characterizes the contention between the honest learner and the attacker. Then, a fully distributed and iterative algorithm is developed based on Alternating Direction Method of Multipliers (ADMoM) \cite{ADMoM} to procure the instantaneous responses of different agents. Blockchain-based solutions can also be used to prevent the FL system from data poisoning attacks.
\paragraph{Other techniques}
Most of the aforementioned techniques protect the private data from outsider attackers while assuming that the server is trustful and participants are honest. However, one malicious insider can cause serious privacy threats. Motivated by this challenge, authors in \cite{ot} proposed a collaborative DL framework to solve the problem of internal attackers. The key idea is to select only a small number of gradients to share with the server and similarly receive only a part of the global parameters instead of uploading and updating the whole set of parameters. In this way, a malicious participant cannot have the whole information and hence cannot infer it. However, this approach suffers from accuracy loss. Furthermore, authors in \cite{ot1} presented a new attack based on Generative Adversarial Networks (GANs) that can infer sensitive information from a victim participant even with just a portion of shared parameters. In the same context, a defense approach based on GANs is designed by authors in \cite{ot2}, in which participants generate artificial data that can replace the real samples. In this way, the trained model is called federated generative model and the private data parameters are not exposed to external malicious devices. Still, this approach can lead to potential learning instability and performance reduction due to the fake data used in the training.
\subsection{Privacy for pervasive inference}
\subsubsection{Privacy and security challenges}
The data captured by end-devices and sent to remote servers (e.g., from cameras or sensors to cloud servers) may contain sensitive information such as camera images, GPS coordinates of critical targets, or vital signs of patients. Exposing these data has become a big security concern for the deep learning community. This issue is even more concerning when the data is collected from a small geographical area (e.g., edge computing) involving a set of limited and cooperating users. In fact, if an attacker reveals some data (even public or slightly sensitive), a DL classifier can be trained to automatically infer the private data of a known community. These attacks, posing severe privacy threats, are called inference attacks that analyze trivial or available data to illegitimately acquire knowledge about more robust information without accessing it, by only capturing their statistical correlations. An example of a popular inference attack is the Cambridge Analytica scandal in 2016, where public data of Facebook users were exploited to predict their private attributes (e.g., political view and location).  Some well-known inference attacks are summarized in Table \ref{inference_attacks}.
\begin{table}[h]
\centering
\caption{Examples of inference attacks.}
\label{inference_attacks}
\begin{tabular}{|c|c|c|}
\hline
\textbf{Inference attacks} & \textbf{Exposed data} & \textbf{Sensitive data} \\ \hline
\begin{tabular}[c]{@{}c@{}}Side-channel attacks\\ \cite{side_channel}\end{tabular} & \begin{tabular}[c]{@{}c@{}}Processing time,\\ power consumption.\end{tabular} & \begin{tabular}[c]{@{}c@{}}Cryptographic\\ keys\end{tabular} \\ \hline
\begin{tabular}[c]{@{}c@{}}Location inference\\ attacks \cite{location_attack}\end{tabular} & \begin{tabular}[c]{@{}c@{}}smartphones' sensor\\ data.\end{tabular} & Location \\ \hline
\begin{tabular}[c]{@{}c@{}}Feature inference\\ attacks \cite{feature_attack}\end{tabular} & \begin{tabular}[c]{@{}c@{}}Prediction results,\\ partial features of the\\ DNN model.\end{tabular} & DNN structure \\ \hline
\begin{tabular}[c]{@{}c@{}}Membership inference\\ attacks \cite{membership_attack}\end{tabular} & \begin{tabular}[c]{@{}c@{}}confidence level \\ of classes,\\ gradients.\end{tabular} & \begin{tabular}[c]{@{}c@{}}membership of \\ a sample to a\\ dataset.\end{tabular} \\ \hline
\begin{tabular}[c]{@{}c@{}}attribute inference \\ attacks \cite{attribute_attack}\end{tabular} & \begin{tabular}[c]{@{}c@{}}social data, likes, \\ friends.\end{tabular} & \begin{tabular}[c]{@{}c@{}}Gender, ages,\\ preferences.\end{tabular} \\ \hline
\end{tabular}
\end{table}

Edge computing naturally enhances privacy of the sensitive information by minimizing the data transfer to the cloud through the public internet. However, additional privacy techniques should be adopted to further protect the data from eavesdroppers. In this context, in addition to its ability to allow the pervasive deployment of neural networks, the DNN splitting was also used for privacy purposes. Meaning, by partitioning the model,  partially processed data is sent to the untrusted party instead of transmitting raw data. In fact,
in contrast to the training data that belongs to a specific dataset and generally follows a statistical distribution, the inference samples are random and harder to be reverted.
Furthermore, the model parameters are independent from the input data, which makes the inference process reveal less information about the sample \cite{white-box}. While preserving privacy, the inevitable challenge of DNN partitioning that remains valid, is selecting the splitting point that meets the latency requirements of the system.
\begin{table*}[!h]
\centering
\caption{Comparison between privacy-aware distribution strategies.\\
(H: High, M: Medium, L: Low).}
\label{tab:privacy}
\begin{tabular}{|c|c|c|c|c|c|c|c|}
\hline
\begin{tabular}[c]{@{}c@{}}\textbf{Privacy-aware}\\ \textbf{strategy}\end{tabular} & \begin{tabular}[c]{@{}c@{}}\textbf{Privacy}\\ \textbf{level}\end{tabular} & \begin{tabular}[c]{@{}c@{}}\textbf{Accuracy}\\ \textbf{preserving}\end{tabular} & \begin{tabular}[c]{@{}c@{}}\textbf{DNN}\\ \textbf{re-training}\end{tabular} & \begin{tabular}[c]{@{}c@{}}\textbf{Compatibility}\\ \textbf{with IoT and DNNs}\end{tabular} & \begin{tabular}[c]{@{}c@{}}\textbf{Partitioning}\\ \textbf{strategy}\end{tabular} & \begin{tabular}[c]{@{}c@{}}\textbf{Communication}\\ \textbf{overhead}\end{tabular} & \begin{tabular}[c]{@{}c@{}}\textbf{Computation}\\ \textbf{overhead on}\\ \textbf{source-device}\end{tabular} \\ \hline
\begin{tabular}[c]{@{}c@{}}Deep split \cite{white-box}\end{tabular} & H & \cmark & \xmark & \cmark & per-layer & L & H \\ \hline
\begin{tabular}[c]{@{}c@{}}Feature extraction \\ \cite{DistEncoder,hybrid_privacy}\end{tabular} & L & \xmark & \xmark & \cmark & per-layer & L & M \\ \hline
\begin{tabular}[c]{@{}c@{}}Noise addition\\  \cite{DistNoise,shredder_privacy,diffPrivacy}\end{tabular} & M & \xmark & \cmark & \xmark & per-layer & M & H \\ \hline
\begin{tabular}[c]{@{}c@{}}Cryptography  \cite{dist_privacy2}\end{tabular} & H & \xmark & \cmark & \xmark & per-layer & M & H \\ \hline
\begin{tabular}[c]{@{}c@{}}Privacy-aware\\  partitioning \cite{DistPrivacy}\end{tabular} & M & \cmark & \xmark & \cmark & \begin{tabular}[c]{@{}c@{}}Filter\\ splitting \end{tabular} & H & L \\ \hline
\end{tabular}
\end{table*}
\subsubsection{Defense techniques and solutions}
\paragraph{Features extraction}
Authors in \cite{DistEncoder} proposed to extract the features sufficient and necessary to conduct the classification from the original image or from one of the layers' outputs using an encoder and transmit these data to the centralized server for inference. This approach prevents the exposure of irrelevant information to the untrusted party that may use it for unwanted inferences.
The work in \cite{hybrid_privacy} also proposed feature extraction for data privacy, while achieving a trade-off between on-device computation, the size of transmitted data, and security constraints. In fact, selecting the split layer from where the data will be extracted intrinsically presents a security  compromise. Particularly, as we go deeper in the DNN network, the features become more task specific and the irrelevant data that can involve sensitive information are mitigated \cite{transformDNN}. Hence, if the split is performed in a deep layer, the privacy is more robust and the transmission overhead is lower. However, a higher processing load is imposed on the source device.  The latter work \cite{hybrid_privacy}, along with the work in \cite{white-box}, advised to perform deep partition in case the source device has enough computational capacity. If the source device is resource-constrained, the model should be partitioned in the shallow layers, although most of the output features are not related to the main task. Authors in \cite{hybrid_privacy} proposed a solution based on Siamese fine-tuning \cite{Siamase} and dimensionality reduction to manipulate the intermediate data and send only the primary measures without any irrelevant information. In addition to enhancing privacy, this mechanism contributes to reducing the communication overhead between the end-device and the remote server. 

However, to this end, the arms race between attacks and defenses for DNN models has come to a forefront, as the amount of extracted features can be sufficient for adversary approaches to recover the original image. Whereas, less shared features may also result in low classification accuracy. The works in \cite{white-box,adversial_attacks1,adversial_attacks2} proposed adversarial attacks to predict the inference input data (or the trained model), using only available features from shared outputs between participants. Authors in \cite{white-box} focused particularly on the privacy threats presented by the DNN distribution; and accordingly, designed a white-box attack assuming that the structure of the trained model is known and the intermediate data can be inverted through a regularized Maximum Likelihood Estimation (rMSE). Additionally, a black-box attack is also proposed, where  the malicious participant only has knowledge about his segment and attempts to design an inverse DNN network to map the received features to the targeted input and recover the original data. Authors demonstrated that reversing the original data is possible, when the neural system is distributed into layers.
\paragraph{Noise addition}
Adding noise to the intermediate data is adopted in \cite{DistNoise}. In this paper, the authors proposed to perform a simple data transformation in the source-device to extract relevant features and add noise. Next, these features extracted from shallow layers are sent to the cloud to complete the inference. To maintain a high classification accuracy, the neural network is re-trained with a dataset containing noisy samples. However, adding noise to the intermediate data costs the system  additional energy consumption and computational overhead. Therefore, the splitting should be done at a layer where the output size is minimal. Though, the latter work did not describe the partition strategy. The Shredder approach \cite{shredder_privacy} resolved this dilemma by considering the computation overhead during the noise injection process. The idea is to conduct an offline machine learning training to find the noise distribution that strikes a balance between privacy (i.e., information loss) and accuracy. In this way, the DNN model does not require retraining with the noisy data and the network can be cut at any point to apply directly the noise distribution. The partitioning decision is based on the communication and computation cost. A higher privacy level and lower communication overhead are guaranteed when the split is performed at deep layers; however, the allocation at the end-device becomes less scalable. Adding noise or extracting task-specific data can be included under the umbrella of differential privacy, which at a high level ensures that the model does not reveal any information about the private input data, while still presenting satisfactory classification. The performance of differential privacy is assessed by a privacy budget parameter $\epsilon$ that denotes the level of distinguishability. Authors in \cite{diffPrivacy} conducted theoretical analysis  to minimize $\epsilon$, while considering  accuracy and the communication overhead to offload the intermediate features among fog participants. 

\paragraph{Cryptography}
Cryptography is another technique that can be used to protect the distributed inference. The main idea is to encrypt the input data and process it using a model trained on encrypted dataset, in a way the intermediate data cannot be used by a malicious participant. Little research, including \cite{dist_privacy2}, investigated the encrypted DNN distribution, as this approach suffers from a prohibitive computation and communication overhead that exacerbates the complexity of the inference process, particularly when executed in resource-constrained devices.

\paragraph{Distribution for privacy}
All the previous techniques applied additional tasks to secure the shared data, e.g., feature extraction, adding noise, and encryption, which overloads the pervasive devices with computational overhead. Different from previous works, DistPrivacy \cite{DistPrivacy} used the partitioning scheme to guarantee privacy of the data. In fact, all the existing privacy-aware approaches adopted the per-layer distribution of the DNN model. This partitioning strategy incurs an intermediate shared information that can be reverted easily using adversarial attacks. The main idea in \cite{DistPrivacy} is to divide the data resulting from each layer into small segments and distribute it to multiple IoT participants, which contributes to hiding the proprieties of the original image as each participant has only small amount of information. Particularly, the authors adopted the filter splitting strategy, in such a way that each device computes only a part of the feature maps. However, as stated in section \ref{profiling}, this partitioning strategy results in large data transmission between participants. Therefore, the authors formulated an optimization that establishes a trade-off between privacy and communication overhead.

Table \ref{tab:privacy} illustrates different privacy-aware strategies for distributed inference existing in the literature and shows their performance. We can see that choosing the adequate strategy depends on the requirements of the pervasive computing system, as multiple trade-offs need to be established, such as the security level and accuracy, or the computation and communication loads.
\subsection{Lessons Learned}
\begin{itemize}
\item 
Current works addressing pervasive AI privacy proved their efficiency while trying to maintain an acceptable accuracy. However, some of these efforts incur significant extra communication and computation costs, while others incorporate new hyper-parameters that not only affect the accuracy but also
distress the communication.
    \item Most of the efforts in the literature explored the attacks that target the federated learning and the possible defenses. However, only little research investigated the threats facing distributed inference. More specifically, changing the intermediate data or injecting malicious information to mislead the prediction is not studied yet.   
    \item Limited efforts have investigated the privacy and security in multi-agent reinforcement learning. This could be attributed to the satisfactory performance of the well-known defense mechanisms (e.g., DP.) when applied to multi-agent systems without many modifications. It is also worth to mention that these privacy and security protocols add additional communication and computation requirements, which are already high in the case of multi-agent learning. Thus, most works assume that the agents are trusted and focus on minimizing communication and computational resource utilization.
\end{itemize}
\begin{figure*}[!h]
\centering
	\includegraphics[scale=0.43]{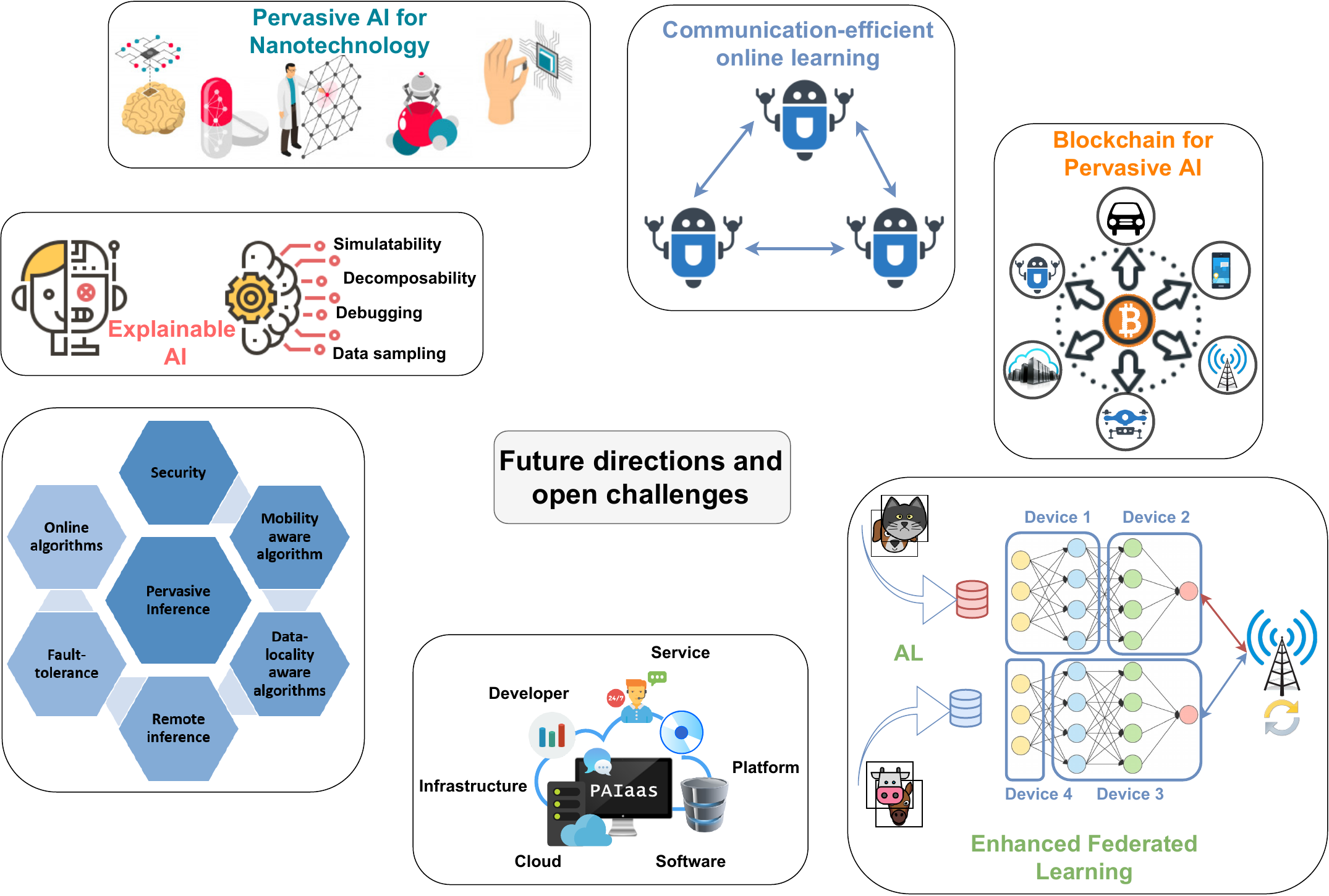}
	\caption{Future directions and open challenges.}
	\label{future_illustrated}
\end{figure*}
\section{Future directions and open challenges}\label{future}
In this section, we present a list of open challenges and issues facing the pervasive AI systems, and we propose some promising ideas to mitigate these issues. Specifically, we introduce the opportunities to integrate the pervasive AI in emerging systems, and we suggest some future directions for efficient distributed inference and enhanced federated learning algorithms. Finally, we present some innovative ideas related to the new concepts of multi-agent reinforcement learning. Fig. \ref{future_illustrated} presents an illustration of the proposed directions.
\subsection{Deployment of Pervasive AI in emerging systems}
\subsubsection{Pervasive AI-as-a-service}
While the 5G main goal is to provide high speed mobile services, the 6G pledges to establish next-generation softwarization and improve the network configurability in order to support pervasive AI services deployed on ubiquitous devices. However, the research on 6G is still in its infancy, and only the first steps are taken to conceptualize its design, study its implementation, and plan for use cases. Toward this end, academia and industry communities should pass from theoretical studies of AI distribution to real-world deployment and standardization, aiming at instating the concept of Pervasive AI-as-a-service (PAIaas). PAIaas allows the service operators and AI developers to be more domain-specific and focus on enhancing users’ quality of experience, instead of worrying about tasks distribution. Moreover, it permits to systemize the mass-production and unify the interfaces to access the joint software that gathers all participants and applications. 
Some recent works, including \cite{10.1145/3426745.3431337} and \cite{s20205796}, started to design distributed AI services. However, authors did not present an end-to-end architecture enclosing the whole process, neither they have envisaged an automated trusted management of the service provisioning.

\subsubsection{Incentive and trusty mechanism for distributed AI using blockchain}
The distribution of heavy and deep neural networks on ubiquitous and resource-limited devices contributes to minimizing the latency of the AI task and guarantees the privacy of the data. However, even though pervasive systems are composed of computing units existing everywhere, anytime, and not belonging necessarily to any operator, the distribution is based on the assumption that pervasive devices are consenting to participate in the collaborative system. In this context, several considerations should be examined first: (1) The design of an incentive mechanism to motivate different nodes to take over AI tasks and sacrifice their memory, energy, communication, and computation resources to gain some rewards (e.g., monetary remuneration and free access to services); (2) In addition to the security of the private data, the security of the participants’ information should also be guaranteed (e.g., locations, identifiers, and capacities). Recently, blockchain \cite{blockchain} \cite{9573346} has gained large popularity as a decentralized dataset managing transaction records across distributed devices, while ensuring  trusty communication. Moreover, the aforementioned incentivizing mechanism can also be handled by blockchain systems. More specifically, all source devices and pervasive nodes have to first register to the blockchain system to benefit from the distributed AI or to participate in the computation. Then, data-generating devices request help to accomplish a task and submit at the same time a transaction application to the blockchain with a reward. Next, when the joining devices complete the offloaded tasks, they return the results to the source device and validate the completion of the transaction. Finally, the recorded participants are awarded according to their contribution to the blockchain transaction. The edge-based blockchain has a promising potential to prevent the security threats of transferring data between heterogeneous, decentralized, and untrusted devices. However, this approach is still in its infancy. Particularly, deploying it in resource-constrained devices is challenging due to the huge energy and computation load of blockchain mining \cite{mining}.
\subsubsection{Explainable AI (XAI)}
The AI-based applications are increasingly involved in many fields, where the decisions are very critical to lives and personal wellness, such as smart health applications and autonomous drones used during wars. However, most of the users do not have visibility on how the AI is making decisions. This lack of explainability prevents us to fully trust the predictions generated by AI systems. Finding reasons and logical explanations for decisions made by AI is called Explainable AI (XAI) \cite{XAI1,XAI3}.  XAI is an emerging field that is expected to answer some questions, including: Why are some predictions  chosen, and why others not? When does an AI model succeed in taking the right decision, and when it fails?

Various techniques are used to explain the AI:  (1) One of these techniques is decomposability, which stands for the ability to describe each part of the model, extract features of the output, and analyze them using clustering methods. The pervasive AI system is the most adequate environment to empower XAI by improving the ability to interpret, understand, and explain the behavior of the model. More specifically, by distributing the inference, the model becomes algorithmically transparent, and each segment can be interpreted and clustered by its importance for the prediction. (2) Moreover, among the most important directions supporting the XAI is model debugging. Debugging a DNN allows to detect errors and understand their sources and their influence on misleading the prediction. A distributed model produces fine-grained outputs, that help to follow the inference process and localize the errors before reaching the prediction layer. (3) A third direction to explain the AI is the extraction of data samples that are highly correlated with the results generated by the model. In fact, similarly to human behaviors when trying to understand some processes, examples are analyzed to grasp the inner correlation that is derived by the AI model. Federated learning is based on clustering data entries and training local models. This technique permits us to narrow the examples search and enables the detection of the most influencing inputs on the model behavior. Research on XAI is still in its infancy, and pervasive DNN computing looks like a promising environment to track the AI process and interpret the results.
\subsection{Efficient algorithms for pervasive inference}
\subsubsection{Online resource orchestration}
The pervasive computing systems are characterized by an extremely dynamic environment, where the available computing resources are volatile, and the load of requests may follow some statistical distributions. 
Additionally, the quality of the collected data may affect the depth of the adopted DL and consequently the computation requirements of the tasks. As an example, capturing high quality images allows to have a good prediction using smaller networks. In this scenario, early-exit techniques or squeezed models can be adopted. 

Because of these network’s dynamics, the pervasive systems deploying distributed inference need a well-designed online resource orchestration and participants selection strategy to support the large number of AI services with minimum latency. Meanwhile, heterogeneous and limited resources, and high dimensional parameters of the DNN should be taken into consideration. In section \ref{PI}, we have introduced existing theoretical approaches to split different DNN networks and distribute the resultant segments into pervasive devices to optimize pervasive computing \cite{FullyDistribution, IoTInferencing,DeepThings,MoDNN}. Nonetheless, most of them focused on how to partition the model in order to maximize the model parallelization and minimize the dependency between participants. Yet, there is no relevant work that deeply studied the performance of inferences distribution and reported the bottleneck and gain of such an approach in long-term online resource orchestration, with different loads of requests and a dynamic behavior of participants and sources. In other words, the data parallelization is not well investigated in the literature, where sources can generate multiple requests at the same time and offload them to neighboring devices. In this scenario, the critical decision of each device is to choose whether to process the same task from different requests while minimizing the memory to store the filters’ weights or to compute sequential tasks from the same request while reducing the transmission among participants. Furthermore, the age-aware inference is an important factor that can be foreseen in online data parallelization. In fact, some requests are highly sensitive to delays and need to be processed timely, such as self-driving cars, whereas others are less critical, including machine translation and recommendation systems. Thus, prioritizing urgent tasks and  assigning better resources and intensive data parallelization to them is of high importance. We believe that pervasive AI computing should focus more on the online configuration to implement the above vision.
\subsubsection{Privacy-aware distributed inference}
Guaranteeing the privacy of the data shared between collaborative devices is one of the main concerns of pervasive computing systems, since untrusted participants may join the inference and observe critical information.  Because of this heterogeneity of ubiquitous devices, the trained models are subject to malicious attacks, such as black-box and white-box risks, by which the original inputs may be in jeopardy. In this case, privacy-aware mechanisms should be enhanced to ensure the security of the distributed inference process. Many efforts have been conducted in this context, such as noise addition and cryptography. Even though these techniques succeeded in hiding features of the data from untrusted devices, most of them suffer from computation overhead and incompatibility with some end-devices or DNNs. More specifically, noisy or encrypted data need to be re-trained to preserve the accuracy of the prediction, and each input has to be obfuscated, which adds a computation overhead. Moreover, encryption may not be applicable for all DNN operations nor possible in some end-devices due to the crypto key management requirements. A notable recent work in \cite{DistPrivacy} and \cite{9750858} proposed to use the distribution for data privacy, without applying any additional task requiring computation overhead. In fact, per-segment splitting leads by design to assigning only some features of the input data to participants. Authors, of this work, applied filter partitioning and conducted empirical experiments to test the efficiency of black-box attacks on different segments’ sizes (i.e., number of feature maps per device). The lower the number of feature maps per device, the higher the privacy. However, filter partitioning incurs high communication load and dependency between devices. This study is still immature. Other partitioning strategies (e.g., channel and spatial.) can be examined to identify the optimal partitioning and distribution that guarantee satisfactory privacy and minimum resource utilization per participant.
\subsubsection{Trajectory optimization of moving robots for latency-aware distributed inference}
The usage of robots (e.g., UAVs) proved its efficiency to improve services in critical and hard-reaching regions. Recently, moving robots have been used for real-time image analysis, such as highway inspection, search and rescue operations, and border surveillance missions. These devices have numerous challenges, including energy consumption and unstable communication with remote servers. Recent works, e.g., \cite{robots_power,robots}, proposed to avoid remote AI inferences and leverage the computation capacity of ground robots to accomplish the predictive tasks. However, only few works covered the distribution of the inference among flying drones, characterized by their faster navigation, higher power consumption, and ability to reach areas with high interferences (e.g., high-rise buildings) compared to ground devices \cite{9498967, 9428021}. Moreover, recent efforts did not cover the path planning for different moving robots to complete their assigned missions, while performing latency-aware predictions. More specifically, the time period between capturing the data to the moment when tasks from all the points are collected, should be minimized by optimizing the devices’ trajectories, and planning close paths for participants handling subsequent segments. Furthermore, the trajectories of devices with available resources should cross the paths of the nodes that need to offload the tasks, because of resource constraints.
\subsubsection{Remote inference of non-sequential DNN models}
A major part of pervasive inference literature analyzes the remote collaboration, where the source device computes the shallow layers of the model, while the cloud handles the deep layers \cite{Neurosurgeon,Boomerang}. In this context, the split point is chosen based on the size of the shared data, the resource capability of the end-device, and the network capacity. This DNN partitioning approach may work well for the standard sequential model, where filters are sequentially reducing the size of the intermediate data. However, state-of-the-art networks do not only include sequential layers with reduced outputs. Indeed, generative models (GAN) \cite{GAN} proved their efficiency for image generation, image quality enhancement, text-to-image enhancement, etc. Auto-encoders also showed good performance for image generation, compression, and denoising. These types of networks have large-sized inputs and outputs. Hence, despite the reduced intermediate data, the cloud servers have to return the high-sized results to the source device, which implies high transmission overhead. Another family of efficient neural networks is the RNN (see section \ref{DRN}) \cite{RNN}, used mostly for speech recognition and natural language processing. These networks  include loops in their structures and multiple outputs of a single layer, which imposes multiple communications with remote servers in case of partitioning. Other complex DNN structures prevent remote collaboration wisdom, such as the randomly wired networks and Bolzman Machines (BM) having a non-sequential dependency. Keeping up with ever-advancing deep learning designs is a major challenge for per-layer splitting, particularly for remote collaboration. Based on these insights, the scheduling of DNN partitioning should have various patterns depending on the model structure.
\subsubsection{Fault-tolerance of distributed inference}
When a deep neural network is split into small segments and distributed among multiple physical devices, the risk of nodes failure is increased, which leads to performance drop and even inference abortion. The typical networking wisdom resorts to re-transmission mechanisms along with scheduling redundant paths. These failure management techniques inevitably consume additional bandwidths. The DNNs are characterized by a unique structure that may enclose skip connections, convolutional neural connections, and recurrent links. These features of state-of-the-art networks implicitly increase the robustness and resiliency of the joint inference. More specifically, skip blocks allow receiving information from an intermediate layer in addition to the data fed from the previous one. These connections, serving as a memory for some DL models (e.g., ResNet), can play the role of fault-tolerant paths. If one of the devices fails or leaves the joint system, information from a prior participant can still be propagated forward to the current device via the skip blocks, which adds some failure resiliency. The skip connections proved an unprecedented ability to enhance the accuracy of deep models, in addition to its potential to strengthen the fault-tolerance of  pervasive computing. However, transmission overheads are experienced, particularly for failure-free systems. Thus, a trade-off between accuracy, resilience, and resource utilization should be envisaged. Another vision to be investigated is to train the system without skip connections and use them only in case of failures. This idea is inspired from the Dropout \cite{dropout} technique that is used to reduce the data overfitting problem. It is based on randomly dropping some neurons during the training and activating them during the inference. Studying the impact of cutting off some transmissions during the inference for different splitting strategies while re-thinking the dropout training is interesting to strengthen the fault-tolerance of pervasive computing. Very recent works \cite{FailureDis,resilinet} started to discuss such insights; however, they are still immature. 
\subsubsection{Data-locality-aware algorithms}
Most of the efforts, studying the pervasive inference, focus on splitting and parallelizing the DNN tasks related to a predictive request. Next, based on the resource requirements and their availability in the joint system (e.g., computation and energy), tasks are distributed and assigned to the participants. However, in terms of memory, only the weight of the input data is considered, whereas the weights to store the DNN structure are never taken into account.
For example, VGG-16 model has 138 M parameters and requires 512 Mb to store its filters \cite{VGG}. What worsens the situation is that some partitions impose copying the filters to all participants (e.g., spatial splitting.). Moreover, if the intelligent application is led by multiple DNN models and different segments are assigned to each device, a huge memory burden is experienced. Therefore, data-locality-aware algorithms should be designed. More specifically, the distribution system has to account for the past tasks assigned to each participant and try to maximize the re-usability of previously-stored weights, with consideration to the capacity of the devices. Minimizing the number of weights assigned to each participant, not only contributes to reduce the memory usage, but also guarantees the privacy of the structure against white-box attacks \cite{emna_white_box}.
\subsubsection{Pervasive inference for nanotechnology applications}
Nanotechnology is the field of innovation and research that focuses on creating particles (e.g., devices and materials) in the scale of atoms. These particles can be used in multiple domains, such as nanomedicine that studies new ways of detecting and curing diseases. One of the interesting examples of nanomedicine is the detection of diabetes through analyzing human's breaths.
Before nanotechnology, it was not possible to precisely detect nano biomarkers. Nowadays, intelligent and invisible nano-sensors can be trained to sniff human breath and analyse the concentration of specific particles. Still, reaching the full potential of nanomedicine (e.g., drug delivery systems and precision cancer medicine) is still yet to be fully realized. 

To guarantee that Nano particles achieve the targeted objectives, large amount of data and computational analysis is expected. While the traditional techniques opt for an in-depth understanding of biological and chemical knowledge, the AI only requires data training. Thus, it is highly interesting to integrate the AI to evaluate and formulate the nanoscale particles \cite{nano1,nano2}. However, these particles suffer from small energy capacity that limits their communication with remote devices (e.g., handheld mobiles and computers). Hence, the distribution of inference within the nano-sensors can provide localized processing and minimize the data transmission. In this context, new partitioning strategies should be envisaged, as the existing ones do not fit the extremely limited computational resources of the particles. Particularly, even neuron, spatial, or filter splitting involving numerous multiplications are considered complex tasks. Thus, per-multiplication partitioning and the related dependency between millions of nano-participants have to be investigated to ensure the practicality of this futuristic convergence between pervasive AI and nanotechnology.

\subsection{Enhanced federated learning algorithms}
\subsubsection{Active Federated Learning}  Given the main limitations of FL in terms of communication overheads and slow convergence,  combining AL concept with emerging FL schemes would be of great interest. Since most of the existing schemes for FL suffer from slow convergence, a novel active FL solution would be needed, which exploits the distributed nature of FL, while coping with  highly dynamic environments and ensuring adequately fast convergence. Indeed, heterogeneity of the local training data at distributed participating nodes and considering all nodes in the FL process can significantly slow down the convergence. Full nodes participation renders the centralized server to wait for the stragglers. Thus, we envision that: (1) exchanging some side information between different participating nodes (e.g., the unique data samples or class distribution) can significantly help in tackling the data heterogeneity problem; (2) considering partial node participation by proposing efficient user selection schemes can play an important role in decreasing communication overheads and accelerating the convergence. A preliminary study of this approach can be found in \cite{AFL}.
\subsubsection{Blending inter and intra data  parallelism for federated learning}
Deep neural networks require intensive memory and computational loads.  This challenge is compounded, when the model is larger and deeper, as it becomes infeasible to acquire training results from a single resource-limited device. Triggered by this challenge, federated learning is proposed to train deep models over tens and even hundreds of CPUs and GPUs, by taking advantage of \textit{inter-data parallelism} \cite{surveyV13}. At present, federated learning techniques split the data to be trained among pervasive nodes while copying the whole DL model to all of them. Still, small devices cannot participate in such a process due to their limited capacities. Hence, blending the \textit{inter-data parallelism} where the trained data is distributed and the \textit{intra-data parallelism} where the intermediate data of the model are partitioned, can be a feasible solution to enable training within non-GPU devices. Certainly, the practicality, gains and bottleneck of such an approach are to be examined and studied, as the backpropagation characterizing the training phase imposes huge dependency and communication between devices. 
\subsection{Communication-efficient multi-agent reinforcement learning}
    \subsubsection{Demonstrated applications} Since most of the MAB algorithms discussed in this paper are recent  \cite{chawla2020gossiping,agarwal2021multi,zhu2021federated,shi_federated_2021}, it remains interesting to see their implications on practical applications, for example, quantifying the effect of bounded communication resources or energy used in wearable devices and congestion between edge nodes. Similarly, quantifying the improvement in regret bounds on actual and  Quality of Experience (QoE) metrics can be promising. 
    
    \subsubsection{More general forms of MABs} The state of art algorithms in the distributed and federated setup adapt the finite-actions, and the stochastic case of the multi-agent settings. However, there exist many more general forms of the bandit problem that are yet to be studied under the multi-agent settings. These include but are not limited to adversarial-bandits, linear bandits, pure exploration, and non-stationary bandits \cite{lattimore_szepesvari_2020}. Investigating potential regret improvements and communication resource utilization in the MAB settings of non-stochastic and infinite-action bandits remains to be tackled. 
    
    \subsubsection{Heterogeneity of Bandit agents} In MAB settings, agents might not only differ in the instances they are trying to solve, but also in their computational capabilities. Different computational capabilities mean that agents interact with their environments at different rates, collecting an additional amount of samples and hence having different quality estimates. While the effect of this computational heterogeneity is heavily studied in supervised federated learning \cite{arxiv.2205.12493}, it is not yet investigated either in distributed or federated bandits. 
    
    \subsubsection{ MARL performance/communication trade-off} Methods that train in a logically centralized server and then execute in a decentralized manner (CTDE) are able to communicate less (even not at all) at execution while being able to learn good joint policy due to the central training phase, as illustrated earlier. However, their adaptability is not guaranteed when dealing with a non-stationary environment, and they might require re-training again to adapt to the new environment. On the other hand, fully decentralized agents can continue the learning throughout their deployment but need to communicate more often to reason about their joint action. Otherwise, learning can be challenging and might diverge \cite{du2020survey}. A natural goal is to design adaptable methods that communicate conservatively, which is the main motivation behind scheduling in learned communication. Thus, more work is needed to address the question of adaptable and communication-cognizant MARL.
    
    \subsubsection{MARL under networking constraints} 
    Several communication characteristics have not been investigated under the MDP and POMG settings. For example, while delay, noise, failure, and time-varying topologies are vital factors in today's practical networks, they were not considered in most of MARL papers. These factors were, however, considered in other optimization frameworks like multi-agent (distributed) convex optimization \cite{MAL-051}. Some of the works started to study bandwidth and multiple-access aspects \cite{wang_learning_2020,mao_learning_2020}. Yet, it is important to study the performance of emerging policies of MARL under realistic networking constraints.

\section{Conclusion}\label{conclusion}
Recently, AI and pervasive computing have drawn the attention of academia and industrial verticals, as their confluence has proved a high efficiency to enhance human’s productivity and lifestyle. Particularly, the  computing capacities offered by the massive number of ubiquitous devices open up an attractive opportunity to fuel the continuously advancing and pervasive IoT services, transforming all aspects of our modern life.  In this survey, we presented a comprehensive review of the resource allocation and communication challenges of pervasive AI systems, enabling to support a plethora of latency-sensitive applications. More specifically, we first presented the fundamentals of AI networks, applications and performance metrics, and the taxonomy of pervasive computing and its intersection with AI. Then, we summarized the resource management algorithms for the distributed training and inference. In this context, partitioning strategies, architectures, and communication issues and solutions were extensively reviewed. Additionally, relevant use cases were described and futuristic applications were discussed. The challenges encountered in this paper revolve around 
choosing the categorization of different AI distribution strategies, as for example the MARL can be classified under the umbrella of pervasive training, pervasive decision-making, or simply pervasive online learning. 

Multiple challenges remain to be addressed, to further improve the performance, as well as the resource management, privacy, and avant-garde applications. Therefore, we presented our vision of technical challenges and directions that may emerge in the future, along with some opportunities for innovation. We hope that this survey will elicit fruitful discussion and inspire new promising  ideas. 

\section*{Acknowledgment}
This work was made possible by NPRP grant  NPRP12S-0305-190231 and NPRP13S-0205-200265 from the Qatar National Research Fund (a member of Qatar Foundation). The findings achieved herein are solely the responsibility of the authors.    

\ifCLASSOPTIONcaptionsoff
  \newpage
\fi
\bibliographystyle{IEEEtran}
\bibliography{References}
\balance

\end{document}